\documentclass[twocolumn]{aastex63}
\usepackage{lineno}

\usepackage{xcolor}
\definecolor{mygreen}{rgb}{0.13,0.54,0.13}
\usepackage[normalem]{ulem}

\usepackage{natbib}
\usepackage{color}
\usepackage{mathtools}
\usepackage{hyperref} 

\def    \apjl  		{\rm {ApJL}}

\def    \apjl  		{\rm {ApJL}}

\def	\cm		{\,{\rm {cm}}}
\def	\K		{\,{\rm K}}
\def	\g		{\,{\rm {g}}}
\def	\mum	{\,{\mu \rm{m}}}

\def \bea {\begin{eqnarray}}
\def \ena {\end{eqnarray}}                  

\def    \ba     {\bf  a}

\def	\ba	{{\bf a}}

\def	\B	{{\rm B}}

\def	\bF	{{\bf F}}

\def	\bJ	{{\it \bf J}}
\def	\bk	{{\it \bf k}}

\def    \bmu    {{\hbox{\boldsym\char'026}}}	

\def	\br	{{\bf r}}
\def	\bv	{{\bf v}}

\def	\C	{{\rm C}}

\def	\cm	{\,{\rm cm}}
\def	\km	{\,{\rm km}}

\def	\max	{\,{\rm max}}
\def	\d	{{\rm d}}

\def	\D	{{\rm D}}
\def	\eff	{{\rm eff}}

\def	\erg	{\,{\rm erg}}

\def	\g	{\,{\rm g}}
\def	\gas	{\,{\rm gas}}
\def	\G	{{\rm G}}

\def	\H	{{\rm H}}

\def	\N	{{\rm N}}

\def	\s	{\,{\rm s}}
\def	\sp	{{\rm sp}}

\def	\AU	{\,{\rm au}}

\def	\V	{{\rm V}}
\def	\Bar	{{\rm Bar}}
\def	\rad	{{\rm rad}}
\def	\yr	  {\,{\rm yr}}

\def \St {{\rm St}}

\def	\ahat		{\hat{\bf a}}

\def    \Bv     	{{\it \bf  B}}

\def    \kv     	{{\bf  k}}

\def	\Rv			{{\bf R}}
\def	\bu			{{\bf u}}
\def    \br     	{{\bf  r}}
\def	\ba			{{\bf a}}

\def    \gas     	{{\rm gas}}

\font\mib=cmmib10

\def\bOmega{\hbox{\mib\char"0A}}
\def\bmu{\hbox{\mib\char"16}}

\usepackage{multirow}
\usepackage{rotating}
\usepackage{longtable}

\begin{document}
\shorttitle{Grain alignment in protostellar environments}
\shortauthors{Hoang et al.}
\title{On Internal and External Alignment of Dust Grains in Protostellar Environments}

\author{Thiem Hoang}
\affiliation{Korea Astronomy and Space Science Institute, Daejeon 34055, Republic of Korea, \href{mailto:thiemhoang@kasi.re.kr}{thiemhoang@kasi.re.kr}}
\affiliation{Korea University of Science and Technology, 217 Gajeong-ro, Yuseong-gu, Daejeon, 34113, Republic of Korea}

\author{Le Ngoc Tram}
\affiliation{Max-Planck-Institut f\"ur Radioastronmie, Auf dem H\"ugel 69, 53-121, Bonn, Germany}

\author{Vo Hong Minh Phan}
\affiliation{Institute for Theoretical Particle Physics and Cosmology (TTK), RWTH Aachen University, 52056 Aachen, Germany} 

\author{Nguyen Chau Giang}
\affiliation{Korea Astronomy and Space Science Institute, Daejeon 34055, Republic of Korea}
\affiliation{Korea University of Science and Technology, 217 Gajeong-ro, Yuseong-gu, Daejeon, 34113, Republic of Korea}

\author{Nguyen Thi Phuong}
\affiliation{Korea Astronomy and Space Science Institute, Daejeon 34055, Republic of Korea}
\affiliation{Department of Astrophysics, Vietnam National Space Center, Vietnam Academy of Science and Technology, 18, Hoang Quoc Viet, Nghia Do, Cau Giay, Ha Noi, Vietnam}

\author{Nguyen Duc Dieu}
\affil{Université de Lyon 1, ENS de Lyon, CNRS, Centre de Recherche Astrophysique de Lyon (CRAL) UMR5574, F-69230 Saint-Genis-Laval, France}

\begin{abstract}
Multiwavelength observations toward protostars reveal complex properties of dust polarization, which is challenging to interpret. 
Here we study the physical processes inducing the alignment of the grain axis of maximum inertia moment with the angular momentum ($\bJ$, i.e., internal alignment) and of $\bJ$ with the magnetic field (i.e., external alignment) of very large grains (VLGs, of radius $a>10\mum$) using the alignment framework based on radiative torques (RATs) and mechanical torques (METs). We derive analytical formulae for critical sizes of grain alignment, assuming grains aligned at low$-J$ and high$-J$ attractors by RATs (METs). For protostellar cores, we find that super-Barnett relaxation induces efficient internal alignment for VLGs with large iron inclusions, but inelastic relaxation is efficient for VLGs regardless of composition aligned at high$-J$ attractors by RATs (METs). For external alignment, VLGs with iron inclusions aligned at high$-J$ attractors have magnetic alignment by RATs ($B-$RAT) or METs ($B-$ MET), enabling dust polarization as a reliable tracer of magnetic fields in dense regions. Still, grains at low$-J$ attractors or without iron inclusions have alignment with $\bJ$ along the radiation direction ($k-$RAT) or gas flow ($v-$MET). For protostellar disks, we find that super-Barnett relaxation is efficient for grains with large iron inclusions in the outer disk thanks to spin-up by METs, but inelastic relaxation is inefficient. VLGs aligned at low-J attractors can have $k-$RAT ($v-$MET) alignment, but grains aligned at high$-J$ attractors have likely $B-$RAT ($B-$MET) alignment. We also find that grain alignment by METs is more important than RATs in protostellar disks. 

\end{abstract}

\section{Introduction}
Dust and magnetic fields are ubiquitous in the Universe and play important roles in many astrophysical processes, including star formation and evolution of the interstellar medium (ISM). Dust grains are the building blocks of planets and catalytic surfaces for the formation of water and complex molecules. Dust absorbs optical-UV starlight and reemits in infrared, which is a powerful tracer of modern astrophysics (\citealt{2011piim.book.....D}). Interstellar dust grains are known to have non-spherical shapes and are aligned with interstellar magnetic fields, as demonstrated through the polarization of starlight (\citealt{Hall:1949p5890,Hiltner:1949p5856}) as well as polarized thermal dust emission \citep{Hildebrand:1988p2566,2015A&A...576A.104P}. Therefore, dust polarization induced by grain alignment become the popular technique to observe interstellar magnetic fields (see \citealt{2019FrASS...6...15P} for a recent review) and extragalactic magnetism \citep{Lopez-Rodriguez.2022}.

The process of grain alignment in general includes (1) the alignment of the axis of maximum moment of inertia ($\ahat_{1}$) with the angular momentum ($\bJ$, so-called internal alignment) and (2) the alignment of $\bJ$ with a preferred direction in space (the magnetic field, the anisotropic radiation, or the gas flow; so-called external alignment, see \citealt{2015ARA&A..53..501A,LAH15}). 

The leading process for the internal alignment of interstellar grains is the Barnett relaxation effect (\citealt{1979ApJ...231..404P}, hereafter P79). The Barnett relaxation arises from the dissipation of grain rotational energy into heat due to rotating Barnett magnetization within a paramagnetic grain rotating around a non-principal axis. For the ISM (e.g., the diffuse ISM and molecular clouds (MCs)) where grains are essentially small (of radius of $a<1\mum)$\footnote{In this paper, small grains are defined by their sizes of $a<1\mum$, large grains of $a\sim 1-10\mum$, and very large grains of sizes $a\gtrsim 10\mum$.}, the Barnett relaxation is usually faster than the randomization of grain orientation by gas collisions, resulting in the perfect internal grain alignment with $\ahat_{1}\|\bJ$ (e.g., \citealt{Hoang.2022}). For external alignment, the Larmor precession of the grain magnetic moment aligned with $\bJ$ around the magnetic field ($\Bv$) is much faster than the gas randomization, so that $\Bv$ is the axis of grain alignment (e.g., \citealt{Hoang.2022}). Moreover, the angular momentum $\bJ$ can be aligned with $\Bv$ by radiative torques (RATs; \citealt{1976Ap&SS..43..291D,1997ApJ...480..633D,2007MNRAS.378..910L,Hoang:2008gb}), mechanical torques (METs; \citealt{2007ApJ...669L..77L,2018ApJ...852..129H}), and/or enhanced paramagnetic relaxation \citep{1951ApJ...114..206D,2008ApJ...676L..25L}. The first two mechanisms are most efficient, which are known as the $B-$RAT and $B-$MET alignments, whereas the third one is to enhance the efficiency of $B-$RAT and $B-$MET alignment \citep{2016ApJ...831..159H}. Within the RAT or MET paradigm, the effect of paramagnetic relaxation can enhance the alignment degree beyond the level induced by RATs or METs alone. Thus, this effect is called Magnetically enhanced RAT alignment (MRAT, \citealt{2016ApJ...831..159H}) or Magnetically enhanced MET alignment (\citealt{2018ApJ...852..129H}). The resulting polarization of starlight and of thermal dust emission is parallel and perpendicular to the magnetic field, respectively. Therefore, dust polarization is a reliable tool to trace interstellar magnetic fields.

Magnetic fields are usually thought to play an important role in the formation of stars and protostellar disks (see \citealt{2012ARA&A..50...29C}; \citealt{2019FrASS...6...15P} for reviews). Recent advance in interferometric polarimetry allows us to conduct high-spatial resolution observations of polarized thermal dust emission toward very dense regions where young stars are forming (hereafter star-forming regions or SFRs) (\citealt{Hull:2019hw}). For instance, dust polarization observations toward a large sample of protostars (Class 0 YSOs) using Atacama Large Millimeter Array (ALMA) are reported by \cite{2018ApJ...855...92C} (for Perseus MC), \citealt{Sadavoy.2019} (for
Ophiuchus cloud), and \cite{Liu.2021fkn} (for OMC-3). ALMA dust polarization observations toward small scales of $\sim 100\AU$ within protostellar disks have also reported (\citealt{Stephens:2017ik,Kataoka:2017fq}). We note that on the disk-scales of $\sim 100$ au, dust polarization can arise from scattering of radiation emitted by dust itself, a process known as self-scattering \citep{2015ApJ...809...78K}. The key features of sub(mm) dust polarization observed toward protostars include the complex variation of dust polarization (in both pattern and fraction) from the protostellar envelope to the inner disk region ($\lesssim 100$ au) and the variation of the polarization pattern with the wavelength (e.g., \citealt{Stephens:2017ik,Kataoka:2017fq,Liu.2021fkn}). As a result, the question of whether dust polarization can trace magnetic fields in protostellar environments is crucially important, yet remains unclear. To answer this question, we need first to understand how and where dust grains can align and what are the polarization properties produced by aligned grains in these environments. 

Incidentally, very large grains (VLGs) of radius $a> 10\mum$ are often reported in protostellar environments, including the envelope (\citealt{2014A&A...567A..32M}), protostellar disks around Class 0 protostars (\citealt{2009ApJ...696..841K,Galametz:2019fj}), and protoplanetary disks (e.g., \citealt{2014arXiv1402.1354T,Mauco.2021}). Therefore, the key difference in the physical properties of the ISM and protostellar environments is the huge difference in grain sizes and the gas density by several orders of magnitudes. This indicates that the leading mechanisms for grain alignment in the ISM cannot be directly applied to the protostellar conditions.

Extensive efforts have been done to model and interpret sub(mm) dust polarization from protostellar environments (see e.g., \citealt{Valdivia.2022}). Nevertheless, previous studies usually make simplified assumptions on grain alignment physics for modeling dust polarization. For example, grains are usually assumed to be aligned with the shortest axis ($\ahat_{1}$) perfectly aligned with the magnetic field or radiation direction (\citealt{Ohashi:2018gl,Ohashi:2020tb,Kataoka.2019,Yang.2018,Harrison.2019}). That assumption disregarded the important effect of internal alignment (e.g., \citealt{Mori.2021,Lin.2021hoj,kk7}), which does not ensure that the inferred magnetic fields are accurate. 

A detailed modeling of thermal dust polarization by aligned grains due to RATs in a protoplanetary disk was presented in \cite{2017ApJ...839...56T}. They found that VLGs can be aligned with $\bJ$ along the radiation direction ($\kv$) instead of along the $\Bv$, which is known as $k-$RAT, as predicted in \cite{2007MNRAS.378..910L}. However, they assumed the perfect internal alignment for all grains rotating suprathermally at an attractor point with the angular momentum much greater than the thermal value (i.e., high$-J$ attractors; \citealt{Hoang:2008gb,2014MNRAS.438..680H}). Such an assumption is not always justified because the perfect internal alignment is only achieve when internal relaxation is faster than the gas randomization. Note that internal relaxation depends not only on the rotation rate, but also on the grain size, the magnetic susceptibility (for Barnett relaxation; \citealt{Hoang.2022}), and shear modulus of grain material (for inelastic relaxation; P79; \citealt{1999MNRAS.303..673L} (LE99)). Moreover, before grains reaching a high$-J$ attactor by RATs/METs, they experience a period of thermal rotation during which efficient internal alignment is not satisfied \citep{Hoang:2008gb,2016ApJ...831..159H}. Since the net dust polarization is dominated by grains at high$-J$ attractors \citep{2014MNRAS.438..680H,2016ApJ...831..159H}, a detailed study of the dependence of internal relaxation on the grain angular velocity is required to accurately determine how internal alignment occurs at high$-J$ attractors. Here we refer to the $k-$RAT with the perfect internal alignment of $\ahat_{1}\|\bJ$ as {\it right} $k-$RAT. Subsequent studies for HL Tau (\citealt{Stephens:2017ik}; \citealt{Kataoka:2017fq}) interpret the azimuthal polarization patterns as the evidence of {\it right} $k-$RAT, but the details of grain alignment are not discussed. \cite{Yang.2018} attempted to interpret dust polarization observed toward HL Tau and suggested that the Gold mechanism (\citealt{1952MNRAS.112..215G}) can produce the elliptical polarization pattern, compared to the circular pattern predicted by the {\it right} $k-$RAT. As shown by \cite{2007MNRAS.378..910L} and \cite{2018ApJ...852..129H}, mechanical torques are far more efficient than the Gold mechanism. Later, \cite{Kataoka.2019} discussed dust polarization produced by grain alignment by mechanical torques, but the perfect internal alignment is also assumed. Recently, \cite{Draine.2022} suggested that circular polarization of dust emission could be detected in protoplanetary disks, assuming that the grain has efficient internal alignment so that the shortest axis is parallel to $\bJ$.

We note that for paramagnetic and super-paramagnetic grains (i.e., grains with iron inclusions) in the ISM, which are small of radius $a\lesssim 1\mum$ (see e.g., \citealt{2016ApJ...831..159H}), internal alignment is expected to be efficient thanks to fast Barnett relaxation and nuclear relaxation compared to the randomization of grain orientation by gas collisions (\citealt{2009ApJ...697.1316H}; \citealt{Hoang.2022}). Recently, \cite{Hoang.2022} studied the internal alignment by Barnett and nuclear relaxation effects for dust grains in dense clouds (DCs) and found that micron-sized grains can have efficient internal alignment if grains have embedded iron inclusions (see also \citealt{2016ApJ...831..159H}). Indeed, observations reveal that about $90\%$ of Fe is locked in dust \citep{2009ApJ...700.1299J,2016ApJ...825..136D}. Although the form of Fe in dust is unclear, one expect that a fraction of Fe is in the forms of metallic Fe or iron oxides (FeO, Fe$_2$O$_3$, Fe$_{3}$O$_{4}$) nanoparticles.
Thus, the existence of iron inclusions is expected for large dust grains in protostellar environments which are grown by grain-grain collisions, during which iron nanoparticles are most likely incorporated into large dust grains.  Moreover, iron nanoparticles are reported to be present in local interstellar dust from the {\it Cassini} mission \citep{Altobelli:2016dl} and in primitive interplanetary dust \citep{Hu.2021yit}. Because physical conditions (i.e., gas density, radiation field, and gas kinematic) of protostellar environments are much different from those of DCs, the question whether VLGs can have efficient internal alignment remains uncertain and will be addressed in this paper. For our study, we consider composite grains with iron inclusions, which is the most likely form of dust in protostellar environments.

Moreover, the problem of external alignment of VLGs in protostellar environments is not yet studied in details. External alignment of grains in protostellar environments is expected to be complicated due to dynamical nature of star-forming regions. While grains in the diffuse ISM and MCs are most likely aligned with $\bJ$ along the magnetic field via RATs/METs (so-called $B-$RAT/$B-$MET alignment), grains in protostellar environments may align with $\bJ$ along the radiation direction/gas flow (so-called $k-$RAT/$v-$MET alignment, \citealt{2019ApJ...883..122L}). Understanding in what conditions the $B-$RAT/$B-$MET alignment can still occur is crucially important for interpreting dust polarization and tracing magnetic fields using dust polarization. Therefore, the main challenge for modeling and interpreting dust polarization and for inferring magnetic fields in protostellar environments lies in the lack of detailed studies of grain alignment for these complex regions. The physics of grain alignment (both internal and external alignment processes) previously developed and tested for the ISM (see \citealt{2015ARA&A..53..501A}; \citealt{LAH15} for reviews) must be revised for protostellar environments.

For the DC conditions, as shown in \cite{Hoang.2022}, Barnett relaxation can be efficient only when grains have iron inclusions, but the maximum size for internal alignment is less than $10\mum$. The efficiency of internal relaxation can be enhanced by increasing its grain angular momentum to above the thermal value (i.e., grains having suprathermal rotation; \citealt{Hoang.2022}). Suprathermal rotation can be achieved by RATs due to protostellar radiation \citep{Hoang.2021} or METs due to grain drift relative to the gas \citep{2007ApJ...669L..77L,2018ApJ...852..129H}. Therefore, the most promising mechanism that can induce internal alignment is inelastic relaxation due to its rapid increases with the angular velocity (P79; LE99), which will be quantified in this study.

The main goal of this paper is to discuss in detail the main processes of internal and external alignment for very large grains in protostellar environments and determine the range of grain sizes with efficient alignment  using the framework of the RAT (MET) alignment. Such a range of grain sizes with efficient grain alignment is required for physical forward modeling of dust polarization and interpreting observational polarization data. Grain alignment is also found to be important for grain growth in protostellar environments and leaves imprints in the internal structure of cometary dust aggregates that can be observed via in-situ spacecraft (\citealt{Hoang.2022}).

The paper is structured as follows. In Section \ref{sec:grainmodel}, we describe the grain model and magnetic properties. In Section \ref{sec:paradigm}, we review the essential components of the modern paradigm of interstellar grain alignment based on RATs and METs. In Sections \ref{sec:IntAlign} and \ref{sec:extalign}, we study the leading physical processes that induce internal alignment and external alignment, and then derive the basic formulae for quantifying grain alignment in protostellar environments. In Section \ref{sec:disk}, we apply our theoretical framework obtained from the previous section to study grain alignment within a protostellar disk. An extended discussion on implications of our results for dust polarization observations is presented in Section \ref{sec:discuss}. We summarize our main findings in Section \ref{sec:summary}.

\section{Grain model and assumptions}\label{sec:grainmodel}
\subsection{Grain geometry}
\begin{figure}
\includegraphics[width=0.45\textwidth]{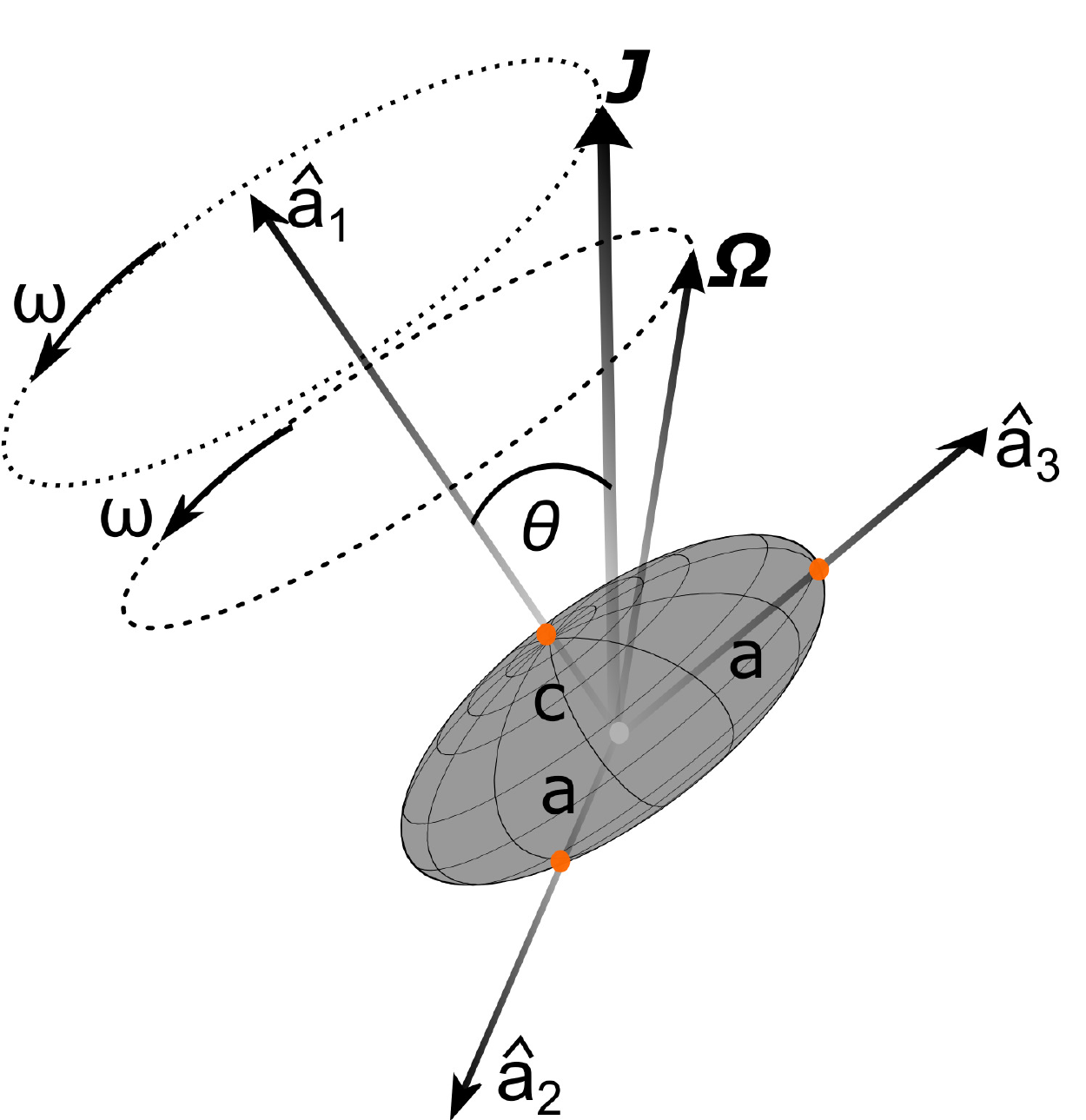}
\caption{Illustration of torque-free motion of an oblate spheroidal grain with semi-major and semi-minor axes of $a$ and $c$. In the grain's body frame, the angular momentum ($\bJ$) and angular velocity ($\bOmega$) are both precessing around the axis of maximum moment of inertia $\hat{\ba}_{1}$ with the angular rate $\omega$.}
\label{fig:torque-free}
\end{figure}

Astrophysical dust grains are expected to have an irregular (non-spherical) shape to produce the polarization of starlight and thermal dust emission. A triaxial irregular shape is described by the principal axes $\ahat_{1},\ahat_{2},\ahat_{3}$. Let $I_{1}> I_{2}\ge I_{3}$ be the principal moments of inertia along the principal axes, respectively. However, for the sake of convenience of numerical estimates, throughout this paper, we assume an oblate spheroidal shape for dust grains, such that the principal moments of inertia along the principal axes are $I_{1}> I_{2}=I_{3}$. Let us denote $I_{\|}\equiv I_{1}$ and $I_{\perp}\equiv I_{2}$ for simplicity, and $h=I_{\|}/I_{\perp}$ be the ratio of the principal moments of inertia. The length of the symmetry (semi-minor) axis is denoted by $c$ and the lengths of the semi-major axes are denoted by $a$ and $b$ with $a=b$. Our notations are summarized in Table \ref{tab:notation}. The assumed grain model is illustrated in Figure \ref{fig:torque-free}.

For a general case, the grain angular momentum ($\bJ$) and angular velocity ($\bOmega$) are not parallel to the axis of maximum inertia, $\ahat_{1}$. So let $\theta$ be the angle between $\bJ$ and $\ahat_{1}$. The angular velocity component projected onto $\ahat_{1}$ is $\Omega_{1}=J_{1}/I_{1}=J\cos\theta/I_{1}\equiv \Omega_{0}\cos\theta$ with $\Omega_{0}=J/I_{1}$ (see Figure \ref{fig:torque-free}). In the grain's body frame, the tips of $\bJ$ and $\bOmega$ precess around $\ahat_{1}$ with the same angular rate of $\omega=(h-1)\Omega_{1}=(h-1)\Omega_{0}\cos\theta$, which is evaluated at $\theta=\pi/4$ for numerical estimates (e.g., \citealt{1979ApJ...231..404P}; \citealt{Hoang:2010jy}).

The principal moments of inertia for the rotation parallel and perpendicular to the grain symmetry axis are given by
\bea
I_{\|}&=&\frac{8\pi}{15}\rho a^{4}c=\frac{8\pi}{15}\rho sa^{5},\label{eq:Ipar}\\
I_{\perp}&=&\frac{4\pi}{15}\rho a^{2}c\left(a^{2}+c^{2}\right)=\frac{4\pi}{15}\rho sa^{5}\left(1+s^{2}\right),\label{eq:Iparperp}
\ena 
where $\rho$ is the mass density of the grain, $s=c/a<1$ is the axial ratio, and $h=I_{\|}/I_{\perp}=2/(1+s^{2})>1$ for $s<1$ (see also \citealt{2014MNRAS.438..680H}). The effective grain size $a_{\eff}$ is defined as the equivalent sphere of the same volume, so $a_{\eff}=(ca^{2})^{1/3}=s^{1/3}a$. For our formulae and numerical estimates, we use the semi-major length $a$ and adopt a fixed axial ratio of $s=1/2$.

\subsection{Grain rotation and rotational damping}
The rotational kinetic energy of an axisymmetric grain is given by
\bea
E_{\rm rot}(J,\theta)&=&\sum_{i=1}^{i=3}\frac{I_{i}\Omega_{i}^{2}}{2}=\sum_{i=1}^{i=3}\frac{J_{i}^{2}}{2I_{i}}=\frac{J_{1}^{2}}{2I_{1}}+\frac{J_{\perp}^{2}}{2I_{2}}
\nonumber\\
&=&\frac{J^{2}}{2I_{1}}\left[1+(h-1)\sin^{2}\theta\right],\label{eq:Erot}
\ena
where $J_{\perp}^{2}=J_{2}^{2}+J_{3}^{2}=J^{2}\sin^{2}\theta$.

In the ISM, dust grains interact with gas species through random collisions and can achieve the energy balance if we disregard other interaction processes such as with interstellar radiation and cosmic rays. Let $\Omega_{T}$ be the thermal angular velocity of the grain rotation along the symmetry axis in the gas of temperature $T_{\gas}$, which is defined by $I_{\|}\Omega_{T}^{2}/2=kT_{\rm gas}/2$. 
Then, $\Omega_{\rm T}=(kT_{\rm gas}/I_{\|})^{1/2}=5.2\times 10^{4}\hat{\rho}^{-1/2}T_{\gas,1}^{1/2}s^{-1/2}a_{-5}^{-5/2}\rm rad\s^{-1}$ with $T_{\gas,1}=T_{\gas}/10\K$ and $\hat{\rho}=\rho/(3\g\cm^{-3})$. Throughout this paper, we all values are given in cgs units, unless stated otherwise. The grain thermal angular momentum can be defined as $J_{T}=I_{\|}\Omega_{T}$.

Dust grains can be spun-up to an angular momentum greater than its thermal value (i.e., {\it suprathermal} rotation) by various grain surface processes \citep{1979ApJ...231..404P}, interaction with radiation (RATs) and gas flow (METs) (see \citealt{2020Galax...8...52H} for a review). To describe the grain suprathermal rotation, we introduce a dimensionless parameter, $\St=J/J_{T}=\Omega_{0}/\Omega_{T}$, which is referred to as the {\it suprathermal rotation number}.


Evaporation of gas species from the grain surface (e.g., those species that stick to the grain surface upon collisions) carries away some the grain's angular momentum and induces the damping of the grain rotation. The characteristic timescale of the gas rotational damping is \citep{1993ApJ...418..287R}
\bea
\tau_{\gas}&=&\frac{3}{4\sqrt{\pi}}\frac{I_{\|}}{1.2n_{\rm H}m_{\rm H}
v_{\rm T}a^{4}\Gamma_{\|}}=\frac{\sqrt{\pi}\rho sa }{3n_{\H}m_{\H}v_{T}\Gamma_{\|}}\nonumber\\
&\simeq& 0.083\hat{\rho}\left(\frac{sa_{-5}}{n_{8}T_{\gas,1}^{1/2}\Gamma_{\|}}\right)~{\rm yr},\label{eq:tgas}
\ena
where $a_{-5}=a/(10^{-5}\cm)$, $n_{8}=n_{\H}/(10^{8}\cm^{-3}$), $v_{T}=(2kT_{\gas}/m_{\H})^{1/2}$, and the gas density is normalized using the typical value of the protostellar core with $n_{\H}=10^{8}\cm^{-3}$. 

Above, $\Gamma_{\|}$ is the geometrical factor given by \citep{1993ApJ...418..287R}
\bea
\Gamma_{\parallel}(e_{m}) &=& \frac{3}{16}\left(3+4(1-e_{m}^{2})g(e_{m})\right)\nonumber\\
&& \frac{3}{16}\left(-e_{m}^{-2}\left[1-(1-e_{m}^{2})^{2}g(e_{m})\right]\right),~~~\label{eq:Gamma_par}
\ena
where $e_{m}=\sqrt{1-s^{2}}$ is the grain shape eccentricity, and
$g(e_{m})$ is given by
\bea
g(e_{m})=\frac{1}{2e_{m}}\mbox{ln}\left(\frac{1+e_{m}}{1-e_{m}}\right).\label{eq31} 
\ena
One has $\Gamma_{\|}=1$ for spherical grains of $s=1$, and $\Gamma_{\|}=0.51, 0.62, 0.74$ for $s=1/3,1/2,2/3$, respectively.

Subject to a radiation field of energy density $u_{\rm rad}$, dust grains are heated to high temperatures and subsequently cool down by infrared (IR) emission. The IR emission also results in the rotational damping of the grain due to the loss of angular momentum carried away by photons (see \citealt{1998ApJ...508..157D}). For a grain in thermal equilibrium of equilibrium temperature $T_{d}$, the IR damping rate is $\tau_{\rm IR}^{-1}=F_{\rm IR}\tau^{-1}_{\gas}$ with $F_{\rm IR}$ being the dimensionless IR damping parameter,
\bea
F_{\rm IR}\simeq \left(\frac{3.8\times 10^{-3}}{s^{1/3}a_{-5}}\right)\left(\frac{U_{6}^{2/3}}{n_{8}T_{\gas,1}^{1/2}}\right),\label{eq:FIR}
\ena 
where $U_{6}=U/10^{6}$ with $U=u_{\rm rad}/u_{\rm MMP83}$ the strength of the radiation field where $u_{\rm MMP83}$ is the energy density of the radiation field in the solar neighborhood from \cite{1983A&A...128..212M}. For VLGs in protostellar environments of high gas density, $F_{\rm IR}\ll 1$, so the rotational damping is dominated by gas collisions.

\subsection{Magnetic susceptibility and magnetic moment}
Iron is among the most abundant elements in the Universe. Observations reveal that about $90\%$ of Fe is locked in dust \citep{2009ApJ...700.1299J,2016ApJ...825..136D}. The presence of Fe atoms with unpaired electrons makes dust a natural paramagnetic material (PM). Let $f_{\rm p}$ is the fraction of atoms which are paramagnetic (e.g., Fe) in the dust. The number density of paramagnetic atoms in the dust grain is then $n_{\rm p}=f_{\rm p}n$ with $n$ the total number density of atoms.

The zero-frequency susceptibility of a paramagnetic grain at rest with a dust temperature $T_{d}$ is described by the Curie's law:
\bea
\chi_{\rm p}(0)&=&\frac{n_{\rm p}\mu_{\rm p}^{2}}{3k_{\B}T_{\d}},\label{eq:curielaw}
\ena
which corresponds to
\bea
\chi_{\rm p}(0)\simeq 0.06f_{{\rm p}}n_{23}\hat{p}^{2}\left(\frac{10\K}{T_{d}}\right),\label{eq:chi_para}
\ena
where $n_{23}=n/10^{23}\cm^{-3}$, $\mu_{p}=p\,\mu_{\rm B}$ is the effective magnetic moment per iron atom with $g_{e}$ the g-factor and $\hat{p}=p/5.5$, $\mu_{B}=e\hbar/2m_{e}c$ the Bohr magneton. For silicate of MgFeSiO$_{4}$ structure, one has $f_{\rm p}=1/7$ and $p\approx 5.5$ (see \citealt{2016ApJ...831..159H}). A smaller value of $f_{\rm p}$ is expected because iron may be present in other structures.

Fe atoms are likely incorporated in the dust in the form of iron clusters, which makes dust a superparamagnetic material (SPM, e.g., \citealt{Jones:1967p2924}). Iron nanoparticles are an essential constituent in the composite Astrodust model \citep{Draine.2021b1e}. Let $N_{\rm cl}$ be the number of iron atoms per cluster and $\phi_{\rm sp}$ be the volume filling factor of iron clusters. Following \cite{2016ApJ...831..159H}, the zero-frequency superparamagnetic susceptibility is described by
\bea
\chi_{\rm sp}(0)\simeq 0.052N_{\rm cl}\phi_{\rm sp}\hat{p}^{2}\left(\frac{10\K}{T_{d}}\right),\label{eq:chi_sp}
\ena
where possible value of $N_{\rm cl}$ spans from $\sim 20$ to $10^{5}$ (\citealt{Jones:1967p2924}), and $\phi_{\rm sp}\sim 0.3$ if $100\% $ of Fe abundance present in the dust in form of iron clusters (\citealt{2016ApJ...831..159H}). For a given $\phi_{\rm sp}$, the superparamagnetic susceptibility is larger for grains with larger iron clusters. 

Comparing Equations (\ref{eq:chi_sp}) with (\ref{eq:chi_para}), one can see that $\chi_{\rm sp}(0)\sim \chi_{\rm p}(0)$ for $N_{\rm cl}\sim 1$, assuming $\phi_{\rm sp}\sim f_{\rm p}$. Therefore, in the following, we adopt $\chi_{\rm sp}$ to describe the magnetic susceptibility of magnetic dust for a general case, and $N_{\rm cl}\sim 1$ is implicitly intended for the PM case. 

The imaginary part of the complex magnetic susceptibility is a function of frequency and usually represented as $\chi_{2}(\omega)=\omega K(\omega)$ where $K(\omega)$ is the function obtained from solving the magnetization dynamics equation. For superparamagnetic material, $K_{\rm sp}(\omega)$ is given by (see \citealt{2016ApJ...831..159H}):
\bea
K_{\sp}(\omega) &=& \frac{\chi_{\sp}(0)\tau_{\sp}}{[1+(\omega \tau_{\sp}/2)^{2}]^{2}},\nonumber\\
&\simeq & 5.2\times 10^{-11}N_{\rm cl}\phi_{\rm sp}\hat{p}^{2}\left(\frac{k_{\rm sp}(\omega)}{T_{d,1}}\right)\s,~~~\label{eq:kappa_sp}
\ena
where $T_{d,1}=T_{d}/10\K$, $\tau_{\rm sp}$ is the timescale of remagnetization by thermal fluctuations given by
\bea
\tau_{\sp}\approx \nu_{0}^{-1} \exp\left(\frac{N_{\rm cl}T_{\rm act}}{T_{d}}\right)
\label{eq:tau_sp}
\ena
with $\nu_{0}$ the characteristic frequency of thermal fluctuations of iron clusters, and $T_{\rm act}$ the activation temperature of superparamagnetism (see e.g., \citealt{Jones:1967p2924}), which have the typical values of $\nu_0\approx 10^{9}\s^{-1}$ and $T_{\rm act}\approx 0.011\K$ from experiments (see \citealt{Morrish:2001vp}). Above,
\bea
k_{\rm sp}(\omega)= \exp\left(\frac{N_{\rm cl}T_{\rm act}}{T_{d}}\right)\left[1+\left(\frac{\omega\tau_{\rm sp}}{2}\right)^{2}\right]^{-2},\label{eq:gsp}
\ena
which increases with $N_{\rm cl}$ but decreases with increasing $T_{d}$ and $\omega$. 

The susceptibility $K_{\rm sp}(\omega)$ is constant for low frequency but decreases rapidly at high frequency for $\omega>2/\tau_{\rm sp}\sim 2\nu_{0}\sim 10^{9}\rad\s^{-1}$. 

A magnetic grain of zero-frequency susceptibility, $\chi(0)$, rotating with an angular velocity $\bOmega$ becomes magnetized via the Barnett effect \citep{Barnett:1915p6353} and acquires an instantaneous magnetic moment,
\bea
\bmu_{\rm Bar}= \frac{\chi(0)V}{\gamma_{e}} \bOmega,\label{eq:muBar}
\ena
where $V$ is the grain volume, $\gamma_{e}=-g_{e}\mu_{B}/\hbar$ the electron gyromagnetic ratio where $g_{e}\approx 2$, $\chi(0)=\chi_{\rm p}(0)$ and $\chi_{\rm sp}(0)$ for paramagnetic and superparamagnetic grains, respectively.

\section{Review of the interstellar grain alignment paradigm driven by RATs/METs}\label{sec:paradigm}
Here we first review the essential components (framework) of the modern paradigm of grain alignment for interstellar grains driven by RATs and METs. We then discuss that the RAT (MET) paradigm can be used to describe grain alignment in protostellar environments.

\subsection{RATs (METs) and its basic effects on grain dynamics: Spin-up/spin-down, precession and alignment}
\begin{figure}
\includegraphics[width=0.5\textwidth]{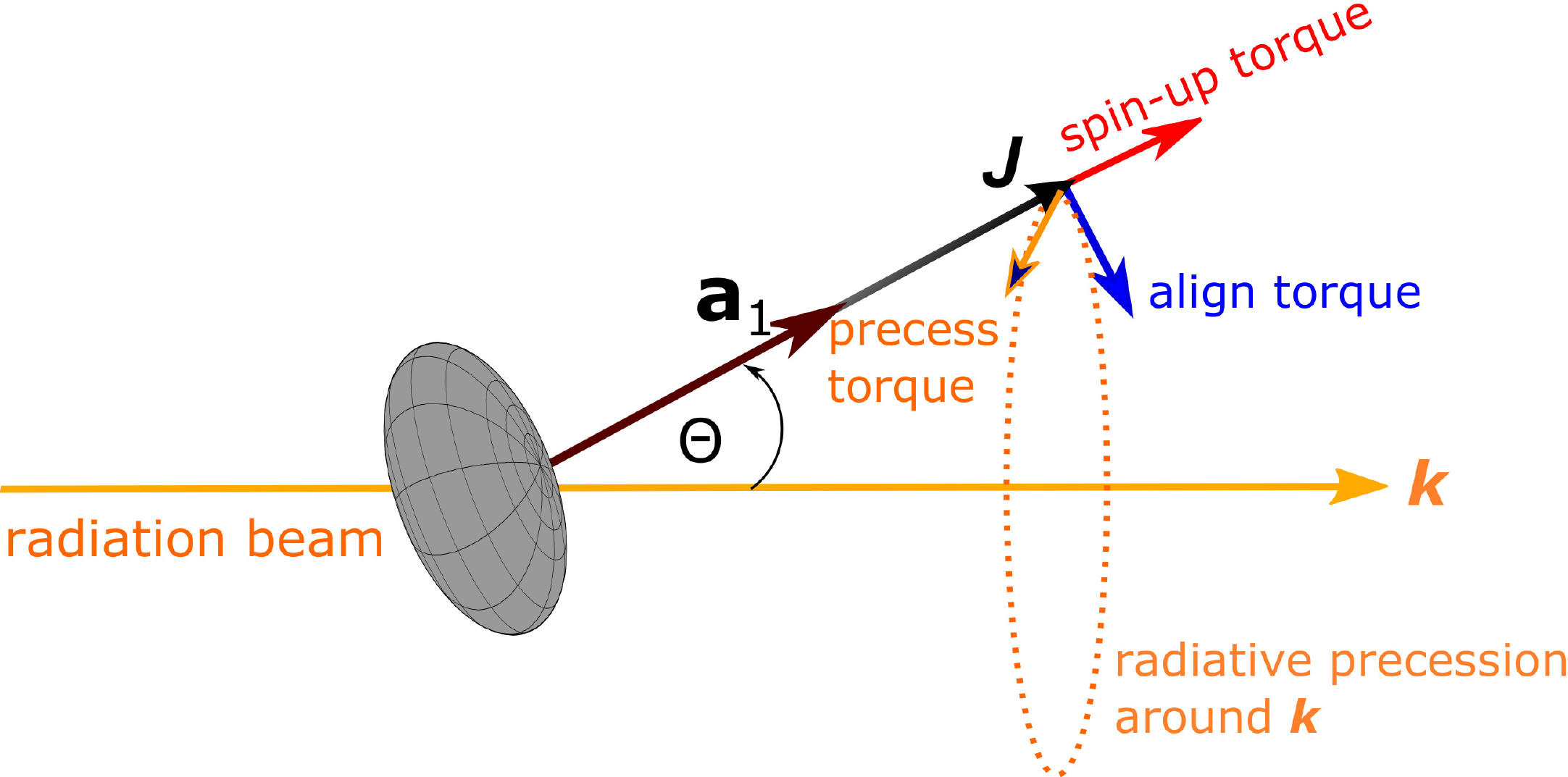}
\caption{Illustration of the three effects of RATs by an anisotropic radiation field, including the spin-up, alignment, and precession around $\kv$. The perfect internal alignment of the grain's shortest axis, $\ahat_{1}$, with ${\bf J}$ is assumed.}
\label{fig:kRAT}
\end{figure}

\cite{1976Ap&SS..43..291D} first suggested that
the interaction of an anisotropic radiation with a helical grain can induce RATs due to the differential scattering/absorption of left- and right-handed circularly polarized photons. Latter, RATs were numerically demonstrated in \cite{1996ApJ...470..551D} for three irregular grain shapes. \cite{2007MNRAS.378..910L} introduced an Analytical Model (AMO) of RATs which is based on a helical grain consisting of an oblate spheroid and a weightless mirror. The AMO is shown to reproduce the basic properties of RATs obtained from numerical calculations for realistically irregular grain shapes \citep{2007MNRAS.378..910L,Hoang:2008gb,Herranen.2021}, and enables us to make quantitative predictions for various conditions \citep{2014MNRAS.438..680H} and dust compositions \citep{2008ApJ...676L..25L,2009ApJ...695.1457H,2016ApJ...831..159H}. Many predictions were observationally tested (see \citealt{2015ARA&A..53..501A}). As shown in previous studies \citep{1997ApJ...480..633D,2007MNRAS.378..910L,Hoang:2008gb}, RATs in general can induce three fundamental effects on grain rotational dynamics, including (1) the grain precession around the radiation direction, (2) spin-up the grain to suprathermal rotation as well as spin-down to thermal rotation, (3) and align the grain with $\bJ$ along the radiation $\kv$ (see Figure \ref{fig:kRAT}).

The interaction of an irregular grain with a gas flow results in METs on the dust grain \citealt{2007ApJ...669L..77L,2018ApJ...852..129H}). Due to the equivalence of frame of reference, METs arising from the grain at rest bombarded by a gas (mechanical) flow of speed $v_{d}$ are the same as those induced by the grain drifting through the ambient gas with the same speed. Similar to RATs, METs have a component that causes the spin-up and spin-down, a torque component that aligns the grain with the gas flow, and another component that causes the grain to precess around the flow direction, $\bv_{d}$ \citep{2007ApJ...669L..77L}.

\subsection{Spin-up and suprathermal rotation by RATs}
Grain suprathermal rotation due to the spin-up effect of RATs is crucially important for modeling grain alignment and rotational disruption (see \citealt{2020Galax...8...52H} for a review).

Let $\gamma_{\rm rad}$ and $\bar{\lambda}$ be the anisotropy degree and the mean wavelength of the radiation field. Following \cite{Hoang.2021z5d} (see also \citealt{Hoang.2021}), an irregular grain of effective size $a_{\rm eff}$ subject to a protostellar radiation field can be spun up by RATs to a maximum angular velocity given by,
 \bea
\Omega_{\rm RAT}&=& \frac{3\gamma_{\rm rad} u_{\rm rad}a_{\rm eff}\bar{\lambda}^{-2}}{1.6n_{\rm H}\sqrt{2\pi m_{\rm H}kT_{\rm gas}}}\left(\frac{1}{1+F_{\rm IR}}\right)\nonumber\\
&\simeq &9.4\times 10^{7} s^{1/3}a_{-5}\left(\frac{\bar{\lambda}}{1.2\mum}\right)^{-2}
\left(\frac{\gamma_{\rm rad} U_{6}}{n_{8}T_{\gas,1}^{1/2}}\right)\nonumber\\
&&\times\left(\frac{1}{1+F_{\rm IR}}\right)\rad\s^{-1},\label{eq:omega_RAT1}
\ena
for $a_{\rm eff}\lesssim a_{\rm trans}$ with $a_{\rm trans}=\bar{\lambda}/2.5$ being the transition size at which the average RAT efficiency changes the slope.\footnote{\cite{Hoang.2021} omitted the subscript in the effective size $a_{\rm eff}$, which is written explicitly in this paper.}

For large grains with $a_{\rm eff}> a_{\rm trans}$, one has
\bea
\Omega_{\rm RAT}&=&\frac{1.5\gamma_{\rm rad} u_{\rm rad}\bar{\lambda}a_{\rm eff}^{-2}}{12n_{\rm H}\sqrt{2\pi m_{\rm H}kT_{\rm gas}}}\left(\frac{1}{1+F_{\rm IR}}\right)\nonumber\\
&\simeq& 8.1\times 10^{9}s^{-2/3}a_{-5}^{-2}\left(\frac{\bar{\lambda}}{1.2\mum}\right) \left(\frac{\gamma_{\rm rad} U_{6}}{n_{8}T_{\gas,1}^{1/2}}\right)\nonumber\\
&&\times\left(\frac{1}{1+F_{\rm IR}}\right)\rad\s^{-1}.\label{eq:omega_RAT2}
\ena



The suprathermal rotation number for the grain spin-up by RATs is then,
\bea
\St_{\rm RAT}&=&\frac{\Omega_{\rm RAT}}{\Omega_{T}}\nonumber\\
&\simeq &1800 \hat{\rho}^{1/2}s^{5/6}a_{-5}^{7/2}\left(\frac{\bar{\lambda}}{1.2\mum}\right)^{-2}
\left(\frac{\gamma_{\rm rad} U_{6}}{n_{8}T_{\gas,1}}\right)\nonumber\\
&\times&\left(\frac{1}{1+F_{\rm IR}}\right),\label{eq:S_RAT1}
\ena
and
\bea
\St_{\rm RAT}&\simeq& 2.1\times 10^{5}\hat{\rho}^{1/2}s^{-1/6}a_{-5}^{1/2}\left(\frac{\bar{\lambda}}{1.2\mum}\right) \left(\frac{\gamma_{\rm rad} U_{6}}{n_{8}T_{\gas,1}}\right)\nonumber\\
&\times&\left(\frac{1}{1+F_{\rm IR}}\right),\label{eq:S_RAT2}
\ena


\noindent
which reveal that the suprathermal rotation number increases rapidly with the grain size as $a^{7/2}$ for $s^{1/3}a<a_{\rm trans}$ and as $a^{1/2}$ as $s^{1/3}a>a_{\rm trans}$.

We note that for protostellar conditions of high gas density, $1+F_{\rm IR}\approx 1$ because $F_{\rm IR}\ll 1$ (Eq. \ref{eq:FIR}), so that Equations (\ref{eq:S_RAT1}) and (\ref{eq:S_RAT2}) yield $\St_{\rm RAT}\propto U/(n_{\H}T_{\gas})$. However, for the condition of strong radiation fields and low gas density, $F_{\rm IR}\gg 1$ (see Eq. \ref{eq:FIR}), using $1+F_{\rm IR}\approx F_{\rm IR}$, one obtains $\St_{\rm RAT}\propto U^{1/3}$.

\subsection{Radiative precession around the radiation direction and k-RAT alignment}
In the presence of an anisotropic radiation field, the irregular grain experiences radiative precession due to RATs, with the grain angular momentum precessing around the radiation direction ($\bk$). For a grain with angular momentum $J$, the radiative precession time is given by (\citealt{2007MNRAS.378..910L}; \citealt{2014MNRAS.438..680H}),
\bea
\tau_{k}&=&\frac{2\pi}{|d\phi/dt|}\approx\frac{2\pi J}{\gamma_{\rm rad} u_{\rad}\lambda a_{\rm eff}^{2}Q_{e3}},\nonumber\\
&\simeq& 56.8
\hat{\rho}^{1/2}T_{\gas,1}^{1/2}\hat{s}^{-1/6}a_{-5}^{1/2}\St\left(\frac{1.2\mum}{\gamma_{\rm rad}\bar{\lambda}\hat{Q}_{e3}U}\right)
\yr,~~~\label{eq:tauk}
\ena 
where $\hat{Q}_{e3}=Q_{e3}/10^{-2}$ with $Q_{e3}$ the third component of RATs that induces the grain precession around $\bk$ and the normalization is done using the typical value of $Q_{e3}$ (see \citealt{2007MNRAS.378..910L}). 

Equation (\ref{eq:tauk}) shows the linear dependence of radiative precession on the instantaneous angular momentum (or suprathermal rotation number $\St$) of the grain. This is an important feature that needs to take into account when studying the RAT alignment. 

When the radiative precession is faster than gas damping, then, RATs can cause the grain angular momentum ($\bJ$) to be coupled with $\kv$. In this case, $\kv$ becomes an axis of grain alignment by RATs, which is referred to as $k-$RAT alignment \citep{2007MNRAS.378..910L}. 

\subsection{Spin-up and suprathermal rotation by METs}
As RATs, the suprathermal rotation by METs determine the efficient grain alignment and rotational disruption (see \citealt{2020Galax...8...52H} for a review). Following \cite{2018ApJ...852..129H}, the magnitude of METs induced by the grain drift through the gas of density $n_{\H}$ with velocity $v_{d}$ that acts to spin-up the irregular grain is given by
\bea
\Gamma_{\rm MET}=n_{\H}m_{\H}v_{d}^{2} \pi a_{\rm eff}^{3}Q_{\rm spinup},
\label{eq:Gamma_MET}
\ena
where $Q_{\rm spinup}$ is the spin-up efficiency which physically depends on the grain shape, structure, and scattering properties \citep{2018ApJ...852..129H}. 
For a spherical shape, $Q_{\rm spinup}$ is zero. For irregular shapes, numerical calculations in \cite{2018ApJ...852..129H} report the range of $Q_{\rm spinup}\sim 10^{-6}-10^{-3}$ (see the spin-up component in their Figure 5). Recent Monte-Carlo simulations in \cite{Reissl.2022} show that the MET efficiency $Q_{\rm spinup}$ changes significantly, upto three orders of magnitude, with grain shapes. Therefore, for an arbitrary shape, we treat $Q_{\rm spinup}$ as a free parameter, with the conservative value of $Q_{\rm spinup}=10^{-3}$.

The maximum angular velocity of the dust grain spun-up by METs is given by \citep{2018ApJ...852..129H}
\bea
\Omega_{\rm MET}&=&\frac{\Gamma_{\rm MET}\tau_{\rm gas}}{I_{\|}}\nonumber\\
&=&n_{\H}m_{\H}v_{d}^{2}Q_{\rm spinup}\pi sa^{3}\times \frac{\sqrt{\pi}\rho sa}{3n_{\H}m_{\H}v_{T}\Gamma_{\|}I_{\|}}\nonumber\\
&=&\left(\frac{s_{d}^{2}v_{T}}{a}\right)\left(\frac{5\sqrt{\pi}sQ_{\rm spinup}}{8\Gamma_{\|}}\right)\nonumber\\
&=&4.5\times 10^{4}s_{d,-1}^{2}\frac{sT_{\gas,1}^{1/2}}{a_{-5}}\frac{Q_{\rm spinup,-3}}{\Gamma_{\|}}~\rad\s^{-1},
\label{eq:omega_MET}
\ena
where $s_{d}=v_{d}/v_{T}$, $s_{d,-1}=s_{d}/0.1$ and $Q_{\rm spinup,-3}=Q_{\rm spinup}/10^{-3}$. Above, we disregard the contribution of IR damping because $F_{\rm IR}\ll 1$ in high-density regions for simplicity.


The suprathermal rotation number for the grain spin-up by METs is then,
\bea
\St_{\rm MET}&=&\frac{\Omega_{\rm MET}}{\Omega_{T}}=\left(\frac{s_{d}^{2}v_{T}}{a}\right)\left(\frac{5s\sqrt{\pi}Q_{\rm spinup}}{8\Gamma_{\|}}\right)\times \left(\frac{I_{\|}}{kT_{\rm gas}}\right)^{1/2}\nonumber\\
&=&s_{d}^{2}\left(\frac{5s\sqrt{\pi}Q_{\rm spinup}}{8\Gamma_{\|}}\right)\times \left(\frac{16\pi \rho sa^{3}}{15m_{\H}}\right)^{1/2}\nonumber\\
&\simeq &0.86\hat{\rho}^{1/2}s_{d,-1}^{2}a_{-5}^{3/2}\left(\frac{s^{3/2}Q_{\rm spinup,-3}}{\Gamma_{\|}}\right),\label{eq:S_MET}
\ena
which only depends on the grain properties and the drift speed, but does not depend on the gas density. This is the basic property of METs, which is different from RATs in which $\St_{\rm RAT}$ decreases rapidly with the gas density because the suprathermal rotation number is determined by spin-up by RATs and rotational damping by the gas. 

\begin{figure}
\includegraphics[width=0.5\textwidth]{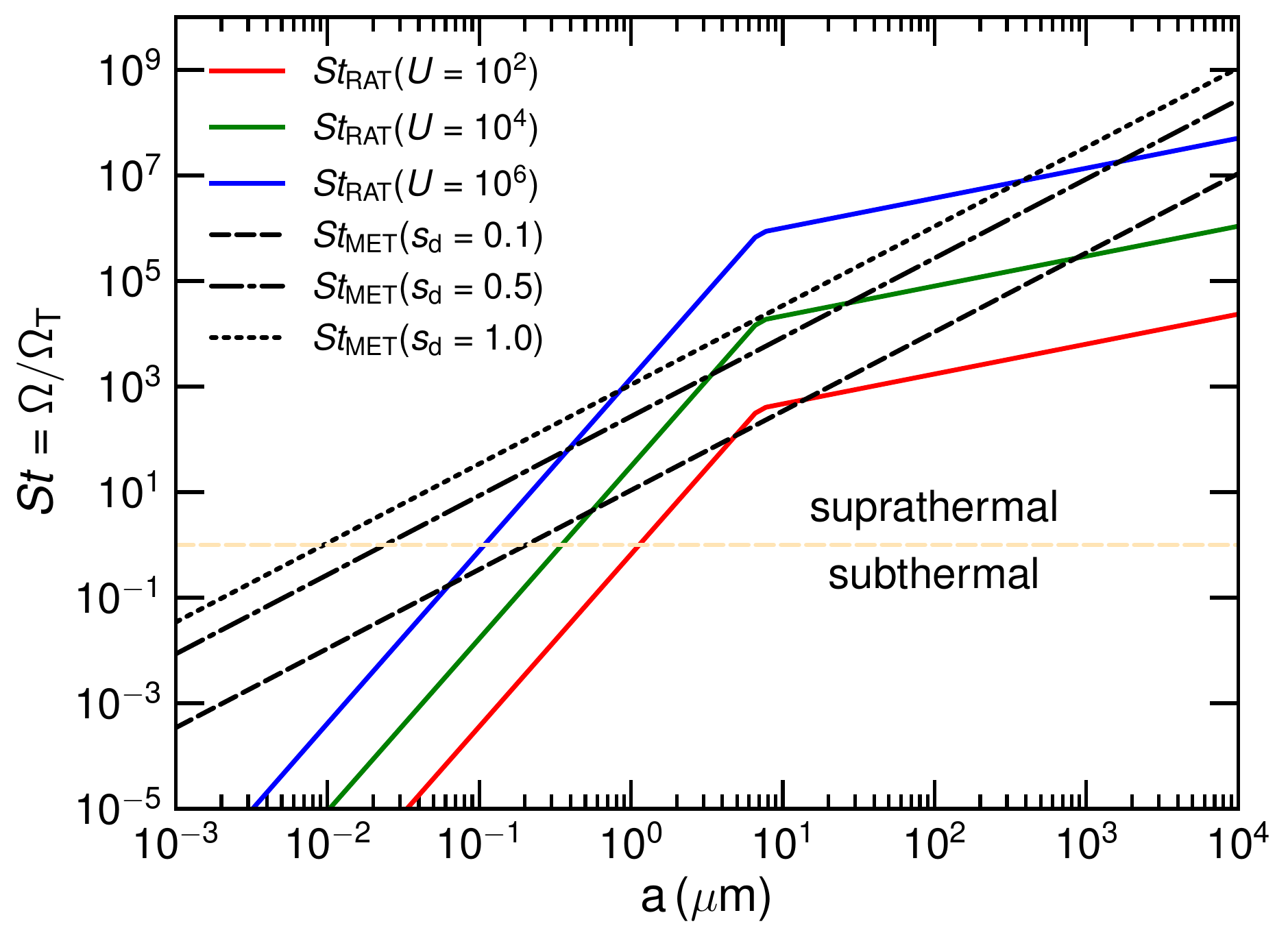}
\caption{Suprathermal rotation number ($\St$) of dust grains spun up by RATs and METs as functions of the grain size ($a$). The values of $\St_{\rm RAT}$ are shown by the solid lines for the different radiation strength, $U$, and $\St_{\rm MET}$ values are shown by the broken lines for the different drift parameter, $s_{\rm d}$. The dashed horizontal line indicates where $\St=1$. We assumed typical physical parameters with $n_{\H}=10^{8}\cm^{-3}$, $\bar{\lambda}=10\mum$, and $Q_{\rm spinup}=10^{-3}$ for METs.}
\label{fig:St}
\end{figure}

Figure \ref{fig:St} shows the suprathermal rotation numbers for grains spun up by RATs and METs for the different radiation strengths and drift velocities, assuming thermal equilibrium of gas and dust with $T_{\gas}=T_{d}\simeq 16.4\times a^{-1/15}_{-5} U^{1/6}\K$. For this particular condition, METs are more important than RATs for smallest and largest grains. For the radiation field of $U=10^{2}$, METs dominate for $s_{d}\ge 0.5$.

\subsection{Mechanical precession around the gas flow}

Following \citealt{2007MNRAS.378..910L} (see their Appendix A), the mechanical torque component that induces the grain precession around the gas flow arises dominantly from secular reflection of hydrogen atoms with the surface of the spheroid ellipsoid. From Eq. (A15) and (A19) from \citealt{2007MNRAS.378..910L}, one can derive the precession component of METs by replacing the radiation energy density $u_{\rm rad}$ by the kinetic energy density of the gas flow, $n_{\H}m_{\H}v_{d}^{2}$:
\bea
    \Gamma_{\rm MET,prec}&=&n_{\H}m_{\H}v_{d}^2 a^3 se(e^2-1)K(\Theta,e) \sin 2\Theta\nonumber\\
    & =&n_{\H}m_{\H}v_{d}^2 a^{3} Q_{\rm prec},
    \label{eq:MET_prec}
\ena
where $\Theta$ is an angle between the grain symmetry axis and the drift direction, $e$ is the eccentricity of the oblate grain and $K(\Theta, e)$ is a fitted function of order unity (see also \citealt{2019ApJ...883..122L}). To account for various shapes, we have introduced $Q_{\rm prec}=se(e^{2}-1)K(\Theta,e)\sin2\Theta$ to describe the MET efficiency of precession. For numerical estimates of the precession torque, we take $Q_{\rm prec}=0.1$ as a typical value.\footnote{The pre-factor $n_{\H}$ is missing in Eq.15 of \cite{2019ApJ...883..122L}, and we denote $a$ to be the semi-major length where \cite{2019ApJ...883..122L} uses $a$ for the semi-minor axis.}

For a grain rotating with angular velocity $\Omega$, the MET induces the precession around the drift direction at precession frequency:
\bea
\Omega_{\rm MET,prec}=\frac{d\phi}{dt}=\frac{\Gamma_{\rm MET,prec}}{I_{\|}\Omega},
\label{eq:omega_MET_prec}
\ena
so the mechanical precession timescale is
\bea
    \tau_{v}&=& \frac{2\pi}{\Omega_{\rm MET,prec}}=\frac{2\pi I_{\|}\Omega}{\pi a^{3}n_{\rm H}m_{\rm H}v_{d}^{2}Q_{\rm prec}}\nonumber\\     &\simeq& 1.7\times 10^{-2}\frac{\St}{a_{-5}^{1/2}Q_{\rm prec,-1}s_{d,-1}^{2}}\left(\frac{1}{n_{8}T_{\gas,1}^{1/2}}\right) \yr,~~~\label{eq:tau_v}
\ena
where $\St=\Omega/\Omega_{T}$ and $Q_{\rm prec,-1}=Q_{\rm prec}/0.1$.\footnote{The term $a_{-5}^{1/2}$ is missing in Eq. (16) of \cite{2019ApJ...883..122L}} Equation \ref{eq:tau_v} implies the decrease of the precession timescale with the gas density.

When the radiative precession is faster than gas damping, the grain alignment occurs with $\bJ$ along $\bv$, which is referred to as $v-$ MET alignment  \citep{2007ApJ...669L..77L}. 

\subsection{Larmor precession and $B-$RAT ($B-$MET) alignment}
\begin{figure}
\includegraphics[width=0.5\textwidth]{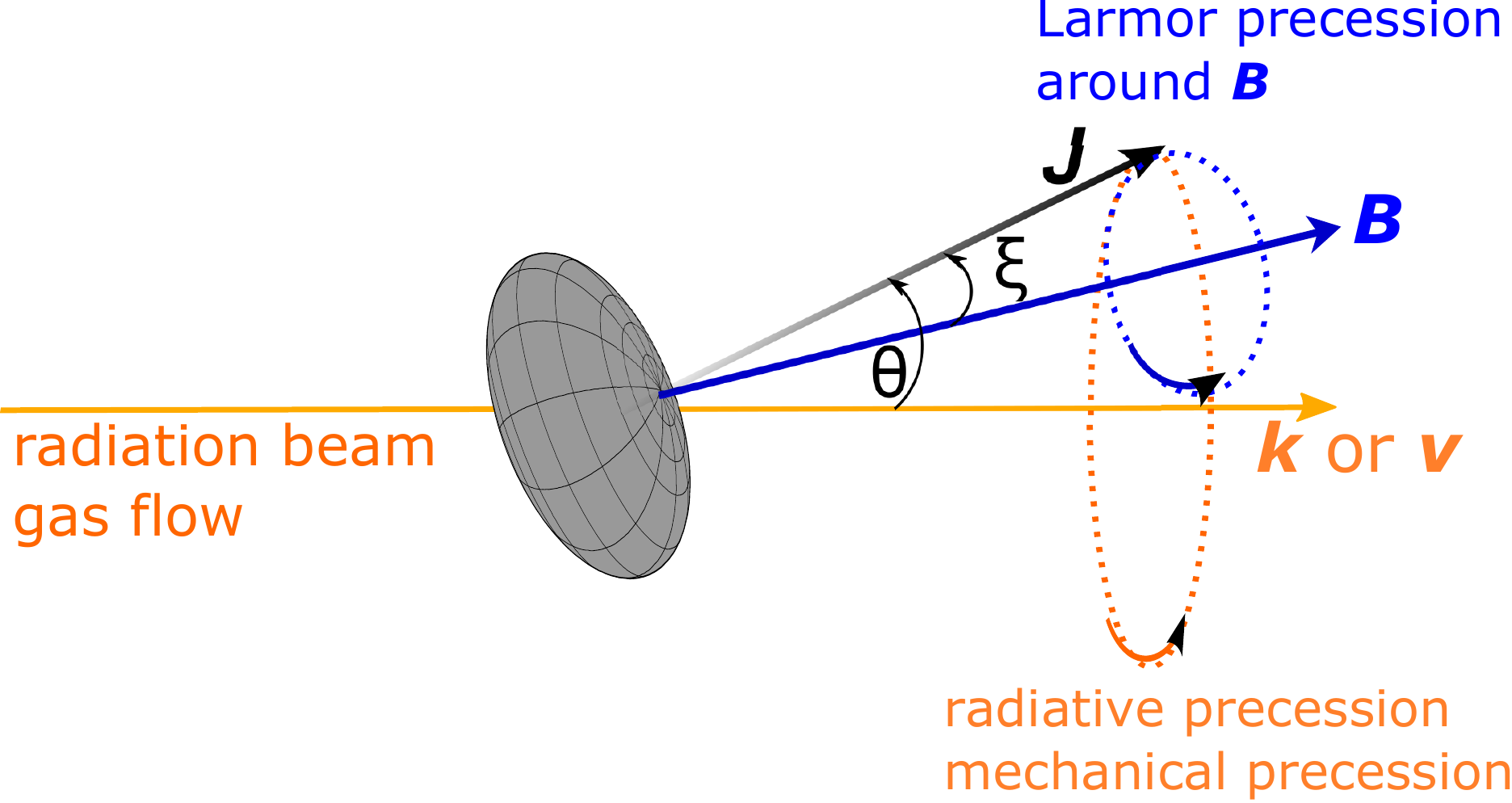}
\caption{Schematic illustration of the grain precession in the presence of a magnetic field and a radiation beam, or a gas flow. In addition to radiative (or mechanical) precession, the grain experiences the Larmor precession around the magnetic field (${\bf B}$). The fastest precession direction establishes the alignment axis by RATs (METs) in the presence of the magnetic field.}
\label{fig:KBE_RAT}
\end{figure}

The interaction of the grain magnetic moment produced by the Barnett effect with an external magnetic field causes the Larmor precession of the grain angular momentum ($\bJ$) around the magnetic field ($\Bv$) direction (see Figure \ref{fig:KBE_RAT}). For superparamagnetic grains, the rate of the Larmor precession is given by
\bea
\tau_{B}&=&\frac{2\pi}{|d\phi/dt|}=\frac{2\pi I_{\|}\Omega}{|\mu_{\Bar}|B}=\frac{2\pi |\gamma_{e}|I_{\|}}{\chi_{\rm sp}(0)VB},\nonumber\\
&=&\frac{2\pi g_{e}\mu_{B}}{\hbar}\frac{2\rho a^{2}}{5\chi_{\rm sp}(0)B}\simeq
8.1\times 10^{-4}\frac{\hat{\rho} T_{d,1}a_{-5}^{2}}{N_{\rm cl}\phi_{sp,-2}\hat{p}^{2}B_{3}} \yr.~~~~
\label{eq:tauB}
\ena
where $B_{3}=B/(10^{3}\mu G)$ is the normalized magnetic field strength.

Comparing Equations (\ref{eq:tauB}) with (\ref{eq:tgas}), one can see that $\tau_{B}\ll \tau_{\gas}$ for the conditions where the gas density is not very high (i.e., $n_{8}\ll 1$). Moreover, comparing with Equations (\ref{eq:tauk}) and (\ref{eq:tau_v}), one can also see that $\tau_{B}\ll \tau_{k}$ and $\tau_{B}\ll \tau_{v}$ for the typical parameters of the diffuse ISM to DCs with the gas density of $n_{8}<1$, the grain size of $a<10\mum$, the radiation field of $U\sim 1$, and the drift velocity of $s_{d,-1}\sim 1$. Therefore, the magnetic field is the preferred axis of grain alignment for PM and SPM dust in the ISM and MCs, which is referred to as the $B-$RAT ($B-$MET) alignment.

\subsection{A general model of grain alignment by RATs and METs}\label{sec:RATmodel}

In addition to the spin-up, spin-down, and precession effects, RATs (METs) have an aligning torque component that acts to align the grain angular momentum with $\kv$ (or $\bv$) at a high$-J$ attractor and a low$-J$ attractor \citep{2007MNRAS.378..910L, Hoang:2008gb,2016ApJ...831..159H}. 
Therefore, a fraction of grain ensemble can be aligned at the high$-J$ attractor, denoted by $f_{\rm high-J}$, and the $1-f_{\rm high-J}$ fraction of grains are aligned at the low$-J$ attractor when the grain randomization by gas collisions is disregarded \citep{2014MNRAS.438..680H}. This process can occur on a timescale shorter than the gas damping, which is so-called {\it fast} alignment \citep{2007MNRAS.378..910L,Lazarian.2020}.

Grains aligned at the high$-J$ attractor rotate suprathermally with $\Omega\sim \Omega_{\rm RAT}$ for RATs or $\Omega_{\rm MET}$ for METs, corresponding to the suprathermal number of $\St_{\rm high-J}>1$. Grains at the low$-J$ attractor rotate with $\Omega\sim \Omega_{T}$ or $\St_{\rm low-J}\sim 1$. When the gas randomization effect is taken into account, grains at low$-J$ attractors are randomized and they are eventually transported to the high$-J$ attractor after a timescale greater than the gas damping time, which is usually called {\it slow} alignment \citep{Hoang:2008gb,Lazarian.2020}. Without high$-J$ attractors, grains are cycling between the low$-J$ attractor and high$-J$ repellor point and have an average low degree of alignment \citep{Hoang:2008gb,2016ApJ...831..159H,Lazarian.2020}. The exact value of $f_{\rm high-J}$ depends on the grain properties (shape and size) and magnetic susceptibility. For grains with ordinary paramagnetic material (e.g., silicate), \cite{Herranen.2021} found that $f_{\rm high-J}$ can be about $10-70\%$ based on calculations of RATs for an ensemble of Gaussian random shapes. The presence of iron inclusions embedded in the grains increases grain magnetic susceptibility and superparamagnetic relaxation, which can produce the universal high$-J$ attractors (i.e., $f_{\rm high-J}\sim 100\%$) \citep{2016ApJ...831..159H,Lazarian.2020}. 

The above model of grain alignment with low$-J$ and high$-J$ attractors established for the ISM can be applied to protostellar environments where strong RATs/METs can rapidly drive the grain angular momentum to align with a preferred direction in space. Indeed, for grains with efficient internal relaxation, internal alignment occurs rapidly so $\ahat_{1}\|\bJ$, thus, the situation is similar to the case of interstellar grains. On the other hand, for VLGs with slow internal relaxation, internal alignment occurs much slower than external alignment \citep{2009ApJ...697.1316H}.\footnote{We disregard the situation where the internal alignment and external alignment processes occur on the similar timescale.}Therefore, in the following, we will use this idealized model of external alignment by RATs/METs to evaluate the efficiency of internal relaxation and grain alignment for protostellar environments.

\section{Internal Alignment by Internal Relaxation}\label{sec:IntAlign}
Internal alignment of the grain axis with the grain angular momentum is essential for modeling and interpreting dust polarization. Here, we study two leading processes that can induce internal alignment in protostellar conditions, including the Barnett relaxation and inelastic relaxation (P79; LE99). We determine the physical parameters for which internal relaxation is faster than gas randomization that induces efficient internal alignment for grains rotating thermally at the low$-J$ attractor and suprathermally at the high$-J$ attractor due to RATs (METs), as implied by the RAT (MET) paradigm.

\subsection{Barnett Relaxation}
A spinning paramagnetic grain acquires an instantaneous magnetic moment, $\bmu\propto \bOmega$, via the Barnett effect. In the grain's body frame, the magnetization vector has a component perpendicular to $\ahat_{1}$, rotating with respect to $\ahat_{1}$ at an angular rate $\omega$ (see Figure \ref{fig:torque-free}). The rotating magnetization component has some lag behind the grain material and induces the dissipation of the grain rotational energy, leading to the internal alignment of $\ahat_{1}$ with $\bOmega$ and $\bJ$  that corresponds to the minimum rotational energy state (\citealt{1979ApJ...231..404P}).

As shown in \cite{Hoang.2022}, Barnett relaxation for paramagnetic grains is inefficient for large grains in dense regions like protostellar cores and disks of density $n_{\H}\gtrsim 10^{6}\cm^{-3}$.
For superparamagnetic grains, the relaxation time by the Barnett effect (so-called super-Barnett relaxation) is given by
\bea
\tau_{\rm BR,sp}&=&\frac{\gamma_{e}^{2}I_{\|}^{3}}{VK_{\rm sp}(\omega)h^{2}(h-1)J^{2}}\nonumber\\
&\simeq &0.16 \hat{\rho}^{2}f(\hat{s})a_{-5}^{7}\left(\frac{1}{N_{\rm cl}\phi_{\rm sp,-2}\hat{p}^{2}}\right)\left(\frac{J_{d}}{J}\right)^{2}\nonumber\\
&&\times \left(\frac{T_{d,1}}{k_{\rm sp}(\omega)}\right) \yr,
\label{eq:tauBar_sup}
\ena
where $\hat{s}=s/0.5, f(\hat{s})=\hat{s}[(1+\hat{s}^{2})/2]^{2}$, and $J_{d}=\sqrt{I_{\|}k_{\B}T_{d}/(h-1)}$ is the dust thermal angular momentum (see also \citealt{2014MNRAS.438..680H}), $\phi_{\rm sp,-2}=\phi_{\rm sp}/0.01$. 



\subsection{Inelastic Relaxation}
Atoms and molecules within a precessing dust grain experience centrifugal acceleration (stress). The centrifugal force makes the material stretch out (strain), while the mutual attractive force between atoms tends to pull it. The stress causes the material deformation, so-called strain. Due to inelasticity of dust material, the deformation stress-strain process induces the dissipation energy of the rotation energy into the heat, resulting in the decrease of the rotational energy to its minimum state corresponding to the internal alignment of the axis of maximum moment of inertia ($\ahat_{1}$) with $\bJ$. This effect, frequently called inelastic relaxation, is first studied for asteroid (\citealt{Prendergast.1958}; \citealt{1973MNRAS.165..403B}) and then applied to interstellar dust (P79). Later, LE99 revisited the inelastic relaxation by taking into account the double-frequency contribution for the squared-prism grain shape. \cite{Molina.2003} studied inelastic relaxation for the oblate spheroidal shape using the LE99's approach.

The strain caused by centrifugal stress produces a potential energy stored within the grain material, denoted by $W$. Following \cite{Molina.2003}, the total strain energy for an oblate spheroidal grain is given by
\bea
W&=&\left[W^{(\omega)}+2W^{(2\omega)}\right]\nonumber\\
&=&\frac{32\pi \rho^{2}\Omega_{0}^{4}\sin^{4}\theta}{105\mu}\frac{a^{2}c^{5}{\rm cotan}^{2}\theta+a^{6}c/(1+\sigma)}{(1+s^{2})^{4}},\label{eq:W}
\ena
where $W^{(\omega)}$ and $W^{(2\omega)}$ are the amplitudes of the strain energies associated with the principal oscillation frequency $\omega$ and double frequency $2\omega$, respectively (see \citealt{Efroimsky:2000p5384}). Here, $\mu$ is the shear modulus (i.e., modulus of rigidity), and $\sigma$ is the Poisson ratio taken to be $\sigma=0.25$ \citep{Molina.2003}.

In principle, the rate of energy loss by inelastic relaxation is proportional to the potential energy and precession rate $\omega$. Thus, one can write 
\bea
\frac{dE_{\rm loss}}{dt}= \left(\frac{2\omega W}{Q}\right),\label{eq:dEloss_dt}
\ena
where $Q$ is the quality factor of grain material,\footnote{The quality factor $Q$ characterizes the material quality such that a higher (lower) value $Q$ corresponds to slower (faster) energy dissipation into heat via inelasticity.} which is assumed independent of the frequency, and the factor 2 accounts for the average of elastic energy (\citealt{Efroimsky:2000p5384}).

Using the grain rotational energy from Equation (\ref{eq:Erot}) and the energy conservation law, one obtains
\bea
\frac{dE_{\rm rot}}{dt}&=&-\frac{dE_{\rm loss}}{dt}\nonumber\\
&=&(h-1)\frac{64\pi \rho^{2}\Omega_{0}^{5}\sin^{4}\theta \cos\theta}{105\mu Q}\nonumber\\
&\times&\frac{a^{2}c^{5}{\rm cotan}^{2}\theta+a^{6}c/(1+\sigma)}{(1+s^{2})^{4}}.\label{eq:dErot_dt}
\ena

Using $dE_{\rm rot}/dt=I_{1}\Omega_{0}^{2}(h-1)\sin\theta\cos\theta d\theta/dt$ and the above equation, one obtains
\bea
\frac{d\theta}{dt}=\frac{8}{7}\frac{a^{2} \rho \Omega_{0}^{3}}{\mu Q}\sin^{3}\theta\frac{s^{4}{\rm cotan}^{2}\theta+1/(1+\sigma)}{(1+s^{2})^{4}}.\label{eq:dtheta_dt}
\ena

The characteristic time of inelastic relaxation can be estimated as
\bea
\tau_{\rm iER}=\left|\frac{1}{d\theta/dt}\right|_{\theta=\pi/4}\approx \frac{\mu Q}{\rho a^{2}\Omega_{0}^{3}}g(s),\label{eq:ti_LE}
\ena
where $g(s)$ is a geometrical factor that depends on the axial ratio as
\bea
g(s)=\frac{ 2^{3/2}7}{8}\frac{(1+s^{2})^{4}}{s^{4}+1/(1+\sigma)},\label{eq:gs}
\ena
which corresponds to $g(s)=7.0$ and $4.6$ for $s=1/2$ and $1/3$, respectively.

Representing the grain angular velocity as the suprathermal rotation number, $\Omega_{0}=\St \Omega_{T}$, Equation (\ref{eq:ti_LE}) becomes
\bea
\tau_{\rm iER}\approx \frac{\rho^{1/2}a^{11/2}\mu Q }{(kT_{\rm gas})^{3/2}\St^{3}}g'(s),\label{eq:tine_LE}
\ena
where $g'(s)\simeq 2.2s^{3/2}g(s)$. Using the same normalization throughout this paper, one obtains
\bea
\tau_{\rm iER}=0.034\hat{\rho}^{1/2}a_{-5}^{11/2}\frac{\mu_{8}Q_{3}}{T_{\gas,1}^{3/2}}\frac{g'(s)}{\St^{3}}~\yr,\label{eq:tau_iER}
\ena
where $\mu_{8}=\mu/(10^{8}\erg/\cm^{3})$, and $Q_{3}=Q/10^{3}$. The typical value $Q$ is about 100 for silicate rocks (see \citealt{Efroimsky:2000p5384}) and between 400 to 2000 for vitreous silica (see \citealt{1979ApJ...231..404P}). The value of $\mu$ depends on the grain structure, which is high for compact and low for composite/porous grains or dust aggregates (\citealt{Seizinger:2013ee}). Studies of cometary dust reveal $\mu \sim 3.6\times 10^{7}-3.46\times 10^{9}\erg\cm^{-3}$ (\citealt{Knapmeyer.2018}). Throughout this paper, we assume $\mu_{8}=1$ and $Q_{3}=1$ for numerical results, unless stated otherwise.

For the thermal rotation of $\St=1$, the inelastic relaxation is slower than Barnett relaxation, but it increases slower with the grain size as $\tau_{\rm iER}\sim a^{11/2}$, compared to $\tau_{\rm BR,sup}\sim a^{7}$. Moreover, the inelastic relaxation time decreases faster with increasing the suprathermal rotation number $\St$ than the Barnett relaxation. Note that inelastic relaxation does not depend on the magnetic susceptibility of dust, so it can work for non-magnetic (or diamagnetic) grains such as carbonaceous dust.

\subsection{Timescales of internal relaxation}

The efficient internal alignment is only achieved when internal relaxation occurs faster than the grain's orientation randomization by gas collisions, which is described by the gas damping time ($\tau_{\gas}$). Here we calculate the internal relaxation timescales for grains that are suprathermally rotating by RATs with $\Omega=\Omega_{\rm RAT}$ and by METs with $\Omega=\Omega_{\rm MET}$ and compare with the gas damping time. 

We consider a range of parameter space with the density from $n_{\H}\sim 10^{6}-10^{8}\cm^{-3}$ in protostellar cores to $n_{\H}\sim 10^{10}-10^{12}\cm^{-3}$ typical in protostellar disks (see e.g., Figures 2 and 3 in \citealt{Tung:2020ew}). For a given local condition of the gas density, temperature and radiation field, we compute the suprathermal rotation numbers $\St_{\rm RAT}$ and $\St_{\rm MET}$, and plug them into Equation (\ref{eq:tauBar_sup}) and (\ref{eq:tine_LE}) to obtain the timescales of super-Barnett relaxation and inelastic relaxation. The dust temperature is taken as the equilibrium temperature of silicate with $T_{d}=16.4a_{-5}^{-11/5}U^{1/6}\K$ (see \citealt{2011piim.book.....D}), which is valid for large grains. Here, we assume the dust and gas are in thermal equilibrium, i.e., $T_{\gas}=T_{d}$, which is valid for dense regions like protostellar environments. 

\begin{figure*}
\includegraphics[width=0.5\textwidth]{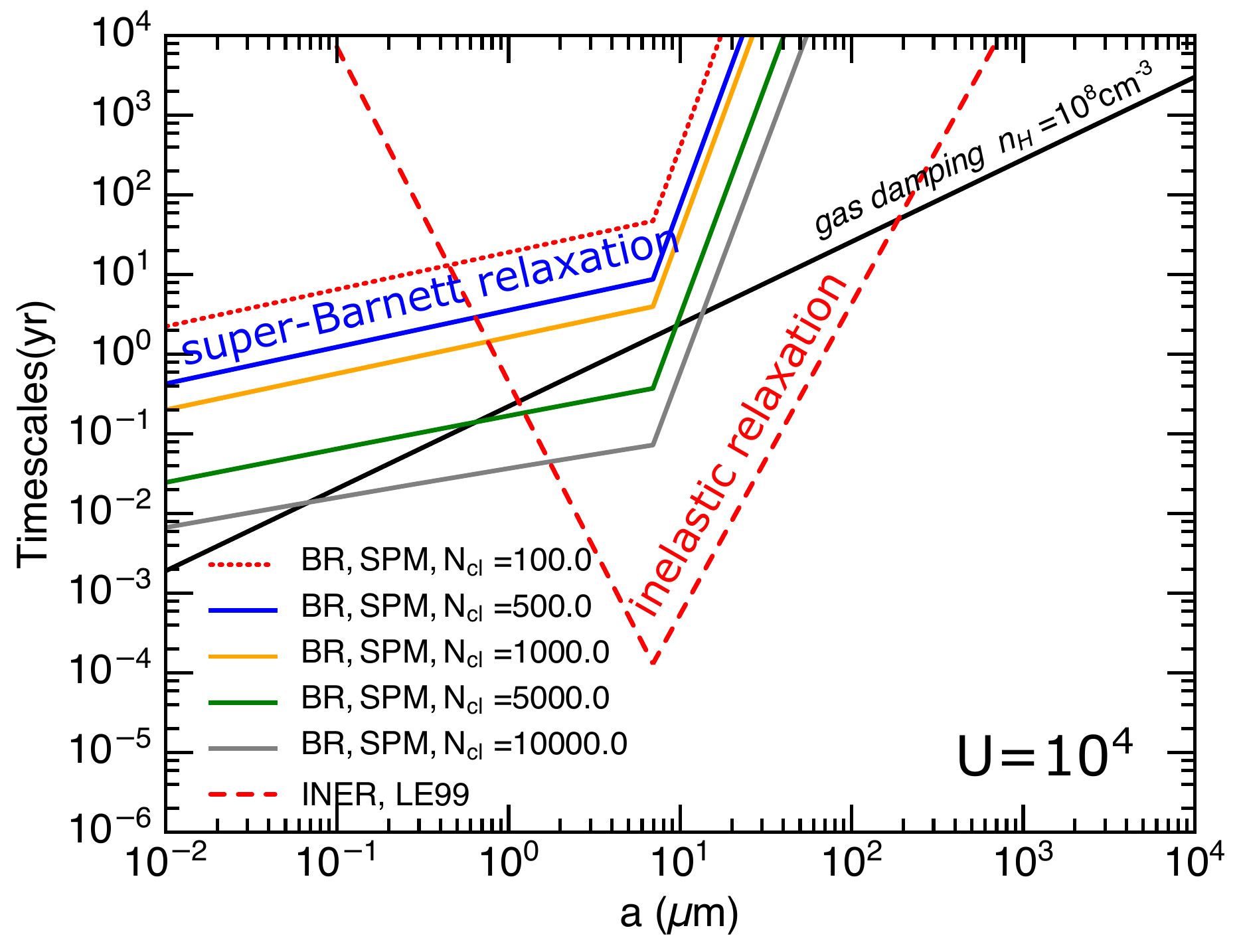}
\includegraphics[width=0.5\textwidth]{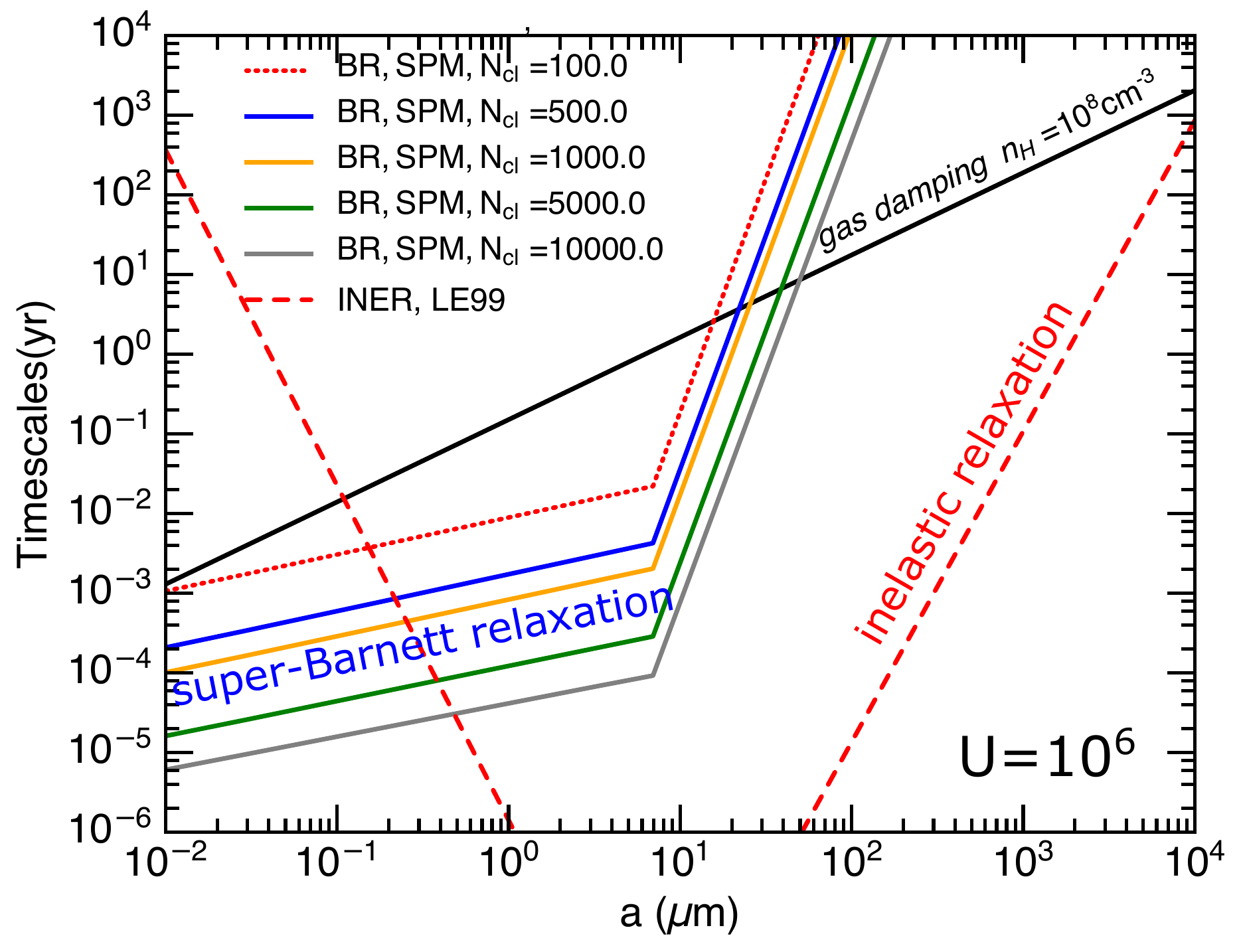}
\caption{Internal relaxation times by inelastic and super-Barnett relaxations for grains spun-up by RATs in a radiation field of $\bar{\lambda}=10\mum, \gamma_{\rm rad}=0.5$ and the gas density $n_{\H}=10^{8}\cm^{-3}$ for $U=10^{4}$ (left panel) and $U=10^{6}$ (right panel). A sharp rise in the timescale at $a\sim \bar{\lambda}/2$ is due to the change in the slope of RATs. Higher radiation strength increases the suprathermal rotation number $\St_{\rm RAT}$ and reduces the relaxation times, extending the range of grains with fast internal relaxation.}
\label{fig:times_a_RAT}

\includegraphics[width=0.5\textwidth]{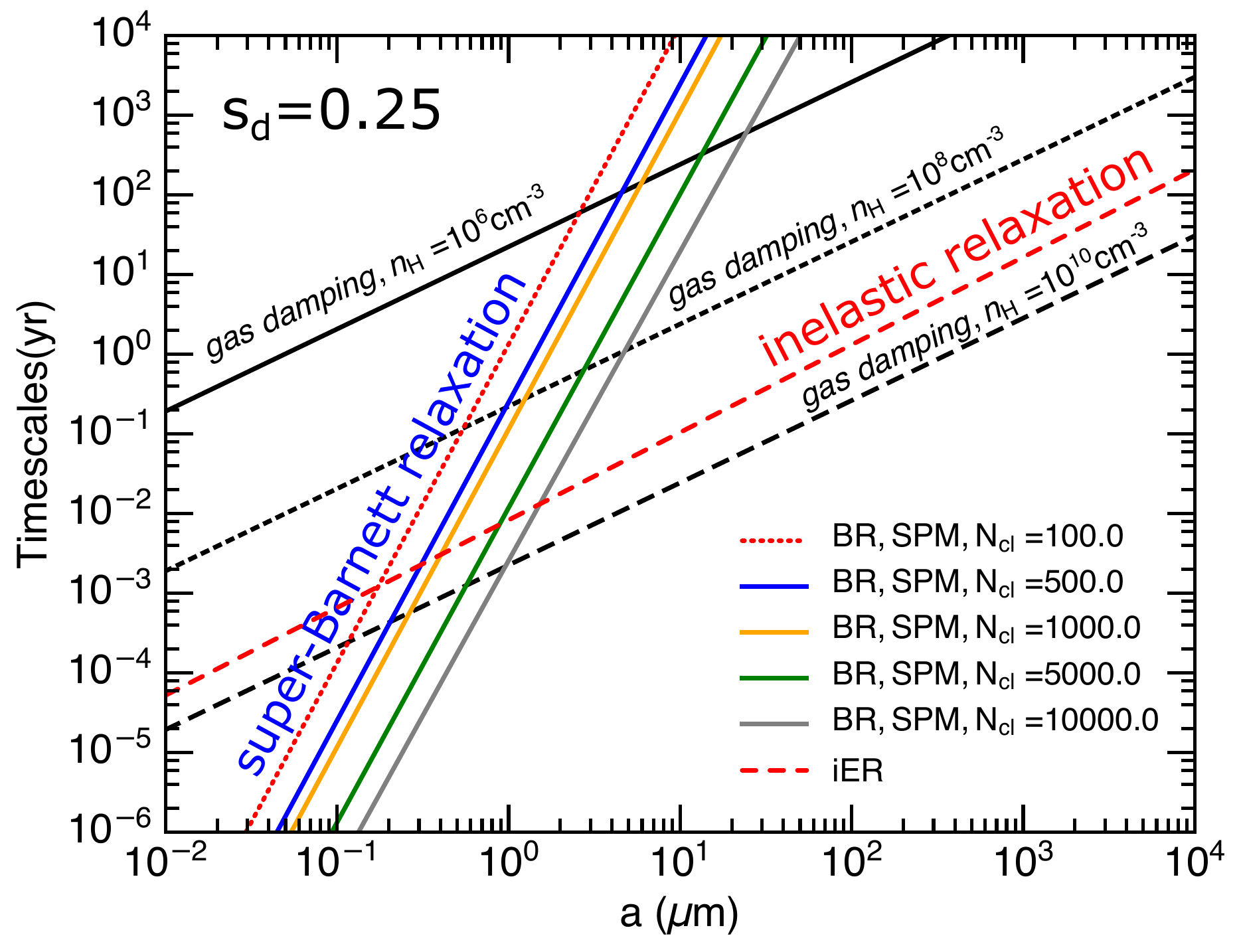}
\includegraphics[width=0.5\textwidth]{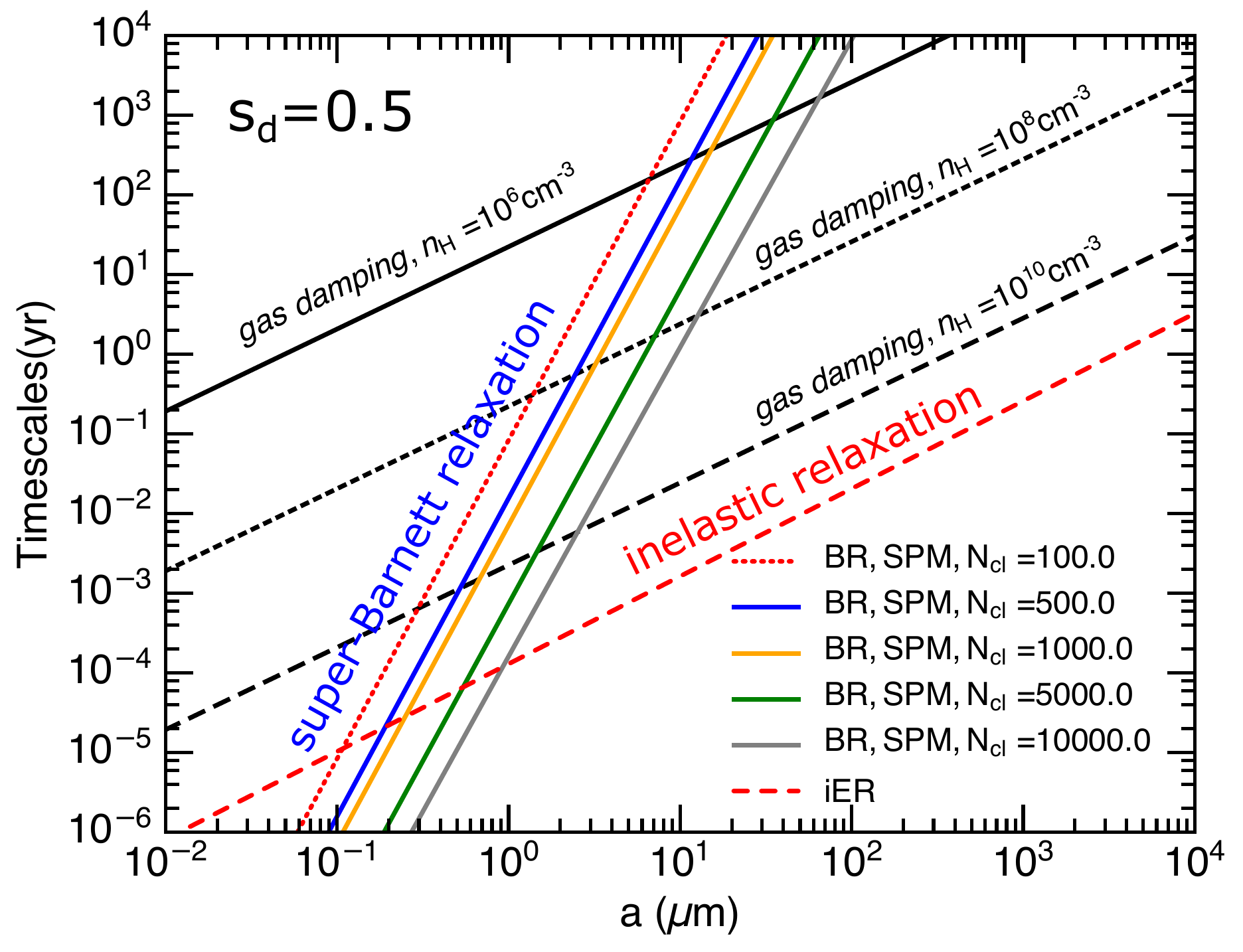}
\caption{Different timescales involved in grain alignment, including super-Barnett relaxation with different iron inclusions ($N_{\rm cl})$, inelastic relaxation, for suprathermally rotating grains by METs with $s_{d}=0.25$ (left) and $s_{d}=0.5$ (right). The gas damping time for three typical densities in protostellar environments is also shown for comparison. Super-Barnett and inelastic relaxation effects are faster for larger $s_{d}$.}
\label{fig:times_a_MET}
\end{figure*}

Figure \ref{fig:times_a_RAT} compares the different timescales by super-Barnett relaxation and inelastic relaxation with the gas damping for the protostellar core of typical density of $n_{\H}=10^{8}\cm^{-3}$, assuming two values of the radiation strength ($U$). For the lower radiation strength, the super-Barnett relaxation is faster than the gas damping for VLGs of size $a\sim 1-20\mum$ when iron inclusions are large of $N_{\rm cl}\gtrsim 5000$, but the inelastic relaxation (dashed line) is fastest for grains of $a\sim 1-100\mum$ (see left panel). For the higher radiation strength, both super-Barnett and inelastic relaxation becomes more efficient. The inelastic relaxation is faster than the gas damping for $a\sim 0.1-10^{4}\mum$, while super-Barnett relaxation is faster for $a<10-50\mum$ for $N_{\rm cl}\sim 100-10^{4}$ (see right panel).

Figure \ref{fig:times_a_MET} shows the different timescales by inelastic relaxation and super-Barnett relaxation for suprathermal rotation by METs for two values of the drift parameters $s_{d}=0.25$ (left panel) and $s_{d}=0.5$ (right panel). The super-Barnett relaxation is faster than inelastic relaxation and gas damping for small grains, but it becomes slower for VLGs of $a>100\mum$. Both Barnett and inelastic relaxation processes are faster for increasing $s_{d}$. Note that the inelastic relaxation has a slightly steeper slope than the gas damping due to its steeper scaling with the gas temperature ($T_{\gas}=T_{d}$) which decreases with the grain size (see Eqs \ref{eq:tau_iER} vs. \ref{eq:tgas}). 
For $s_{d}=0.25$ (left panel), inelastic relaxation is faster than the gas damping for $n_{\H}=10^{6}$ and $10^{8}\cm^{-3}$ for all grain sizes. However, super-Barnett relaxation is faster than the gas damping only for small grains of $a<0.1-10\mum$ at $n_{\H}=10^{10}\cm^{-3}$. For $s_{d}=0.5$ (right panel), thanks to the increased in the suprathermal rotation number, inelastic relaxation becomes faster than the gas damping for all three considered densities. Super-Barnett relaxation becomes also more efficient than the gas damping for grains of $a<1-50\mum$ at $n_{\H}=10^{6}\cm^{-3}$ due to the higher suprathermal rotation number. At $n_{\H}=10^{8}\cm^{-3}$, super-Barnett relaxation is only efficient for small grains of $a<1\mum$ for $s_{d}=0.1$ (left) and $a<10\mum$ for $s_{d}=0.5$ (right), assuming $N_{\rm cl}=10^{4}$.

The total rate of internal relaxation by Barnett and inelastic effects is then $\tau_{\rm INR}^{-1}=\tau_{\rm BR,sp}^{-1}+\tau_{\rm iER}^{-1}$.

\subsection{Critical Sizes of Internal Alignment by Barnett Relaxation}
We now determine the critical sizes for efficient internal alignment by super-Barnett relaxation for grains aligned at high$-J$ attractors by RATs and METs. Using Equations (\ref{eq:tgas}), (\ref{eq:tauBar_sup}), and (\ref{eq:kappa_sp}), we calculate the ratio of the Barnett relaxation to the gas damping times,
\bea
\frac{\tau_{\rm BR,sp}}{\tau_{\gas}}&=& \frac{\gamma_{e}^{2}I_{\|}^{2}4\sqrt{\pi}1.2n_{\H}m_{\H}v_{\rm T}a\Gamma_{\|}}{3Vh^{2}(h-1)J^{2}}\times \frac{[(1+(\omega\tau_{\rm sp}/2)^{2}]^{2}}{\chi_{\rm sp}(0)\tau_{\rm sp}},\nonumber\\
&\simeq & 10^{-3}\frac{\hat{\rho}n_{8}T_{\gas,1}^{1/2}a_{-5}^{6}\Gamma_{\|}}{h^{2}N_{\rm cl,4}\phi_{\rm sp,-2}\hat{p}^{2}}\frac{1}{k_{\rm sp}(\omega)}\left(\frac{J_{d}}{J}\right)^{2}.\label{eq:tBR_tgas}
\ena

Replacing $J/J_{d}=(J/J_{T})(J_{T}/J_{d})=\St\sqrt{(h-1)T_{\rm gas}/T_{\rm d}}$ where $\St=J/J_{T}=\Omega_{0}/\Omega_{T}$, one can determine the critical suprathermal rotation for efficient internal alignment by Barnett relaxation using the criteria $\tau_{\rm BR}/\tau_{\rm gas}<1$ \citep{1997ApJ...484..230L}, yielding
\bea
\St&>&\St_{\rm cri,aJ}({\rm BR})=0.03a_{-5}^{3}\left(\frac{n_{8}T_{\gas,1}^{1/2}\Gamma_{\|}}{h^{2}}\right)^{1/2}\left(\frac{1}{N_{\rm cl,4}\phi_{\rm sp,-2}\hat{p}^{2}}\right)^{1/2}\nonumber\\
&\times&\left(\frac{1}{k_{\rm sp}(\omega)}\right)^{1/2}\left(\frac{(h-1)T_{\rm gas}}{T_{\rm d}}\right)^{1/4},\label{eq:Sp_aJ_BR}
\ena
which implies that for small grains, even thermal rotation of $\St<1$ can have efficient Barnett relaxation. However, to align $a=1,10\mum$ grains, one need $\St_{\rm cri,aJ}({\rm BR})\sim 31$ and $31622$, assuming $n_{8}=1$ and $N_{\rm cl,4}=1$. 

Equation (\ref{eq:tBR_tgas}) implies the steep increase with the grain size as $a^{6}$, so that sufficiently large grains may not have internal alignment. Let $a_{\max, aJ}$ be the maximum grain size for internal alignment between $\ahat_{1}$ and $\bJ$. The maximum size for efficient internal alignment by Barnett relaxation is given by $\tau_{\rm Bar,sp}/\tau_{\gas}=1$, yielding
\bea
a_{\max,aJ}(\rm BR) &\simeq&0.32h^{1/3}\St^{1/3}\left(\frac{N_{\rm cl,4}\phi_{\rm sp,-2}\hat{p}^{2}}{\hat{\rho}n_{8}T_{\gas,1}^{1/2}\Gamma_{\|}}\right)^{1/6}\nonumber \\
&\times&\left(\frac{1}{k_{\rm sp}(\omega)}\right)^{1/6}\left(\frac{(h-1)T_{\rm gas}}{T_{\rm d}}\right)^{1/6}~\mum,~~~~~\label{eq:amax_aJ_BR}
\ena
which implies $a_{\max,aJ}({\rm BR})\approx 0.32\mum$ for $N_{\rm cl}\sim 10^{4}$ and $J/J_{d}=1$ and $n_{8}=1$. One can see that $a_{\max,aJ}$ increases with iron inclusions as $N_{\rm cl}^{1/6}$, with suprathermal rotation as $\St^{1/3}$, but decreases with the gas density as $n_{\H}^{-1/6}$.\footnote{Note that the efficient internal alignment is only achieved when $\tau_{\rm BR,sp}\ll \tau_{\gas}$, but here we derive the critical limit for convenience.}

Plugging $\St=\St_{\rm RAT}$ (Eq. \ref{eq:S_RAT2}) into Equation (\ref{eq:amax_aJ_BR}) one can get $a_{\max,aJ}^{\rm RAT,high-J}({\rm BR})$ due to Barnett relaxation for grains spun up by RATs. Figure \ref{fig:amax_aJ_RAT} shows the critical size of internal alignment by Barnett and inelastic relaxation for grains for the mean wavelength of radiation field of $\bar{\lambda}=10\mum$ and different radiation strengths. The assumed value of $\bar{\lambda}=10\mum$ is typical for the expected radiation field in the protostellar environment, which includes the attenuated radiation field from the central protostar \citep{Hoang.2021} and infrared radiation emitted by the hot dust shell near the protostar \citep{Hoang.2021z5d}. 
For the low radiation strength of $U=1,100$, super-Barnett relaxation is more efficient than inelastic relaxation, but both effects are only effective at $n_{\H}<10^{7}\cm^{-3}$. Increasing the radiation strength to $U=10^{4}, 10^{6}$, inelastic relaxation becomes faster than super-Barnett relaxation, as expected for suprathermal rotation due to its faster increase with the angular momentum and slower variation with the grain size (see Eq. \ref{eq:tine_LE} vs. \ref{eq:tauBar_sup}), which induces efficient internal alignment of large grains even in the region of $n_{\H}\sim 10^{8}-10^{10}\cm^{-3}$.

Similarly, plugging $\St= \St_{\rm MET}$ (Eq. \ref{eq:S_MET}) one can solve for the range of grain sizes with efficient internal relaxation, $a_{\min,aJ}^{\rm MET,high-J},a_{\max,aJ}^{\rm MET,high-J}$ for grains spun up by METs.



\subsection{Critical size for internal alignment by inelastic relaxation}
To estimate the maximum size for internal alignment by inelastic relaxation, we use Equations (\ref{eq:tau_iER}) and (\ref{eq:tgas}),
\bea
\frac{\tau_{\rm iER}}{\tau_{\rm gas}}&&= 0.41a_{-5}^{9/2}\frac{\mu_{8}Q_{3}}{\sqrt{\hat{\rho}}}\frac{n_{8}}{T_{\gas,1}}\frac{g'(s)\Gamma_{\|}}{s}\frac{1}{\St^{3}}.\label{eq:tiner_tgas}
\ena

For a given grain size, the efficient internal alignment occurs when $\tau_{\rm iER}/\tau_{\gas}<1$, which requires the maximum grain size of
\bea
a_{\rm max, aJ}({\rm iER})&=&0.12\left(\frac{\mu_{8}Q_{3}}{\sqrt{\hat{\rho}}}\right)^{-2/9}\left(\frac{T_{\gas,1}}{n_{8}}\right)^{2/9}\nonumber\\
&\times&\left(\frac{s}{g'(s)\Gamma_{\|}}\right)^{2/9}\St^{2/3}\mum,\label{eq:amax_aJ_iER}
\ena
which implies that a suprathermal rotation number of $\St=10^{3}$ can align large grains of $a_{\max,aJ}=12\mum$, assuming the normalized parameters.

Comparing $a_{\max,aJ}({\rm iER})$ with $a_{\max,aJ}({\rm BR})$ (Eq. \ref{eq:amax_aJ_BR}), one can see that at thermal rotation (e.g., low$-J$ attractor), inelastic relaxation is less efficient than super-Barnett relaxation for internal alignment. However, for suprathermal rotation, inelastic relaxation becomes more efficient due to its steeper increase with the suprathermal number as $\St^{2/3}$, compared to $\St^{1/3}$ of super-Barnett relaxation.

For a given grain size, the internal alignment requires the suprathermal rotation of the number,
\bea
\St>\St_{\rm cri,aJ}({\rm iER})&=&0.74a_{-5}^{3/2}\left(\frac{\mu_{8}Q_{3}}{\sqrt{\hat{\rho}}}\right)^{1/3}\left(\frac{g'(s)\Gamma_{\|}}{s}\right)^{1/3}\nonumber\\
&\times&\left(\frac{n_{8}}{T_{\gas,1}}\right)^{1/3},~~~\label{eq:Spcri_aJ_iER}
\ena
which increases rapidly with the grain size and gas density. Here the critical suprathermal rotation number $\St_{\rm cri,aJ}$ corresponds to $\tau_{\rm iER}=\tau_{\gas}$.

Equation (\ref{eq:Spcri_aJ_iER}) implies that inelastic relaxation can be efficient for thermal rotation if $a_{-5}<1$, but for large grains of $a>1\mum$ suprathermal rotation is required at a level of $S_{\rm p}>3.5$, assuming $n_{8}=1$. For a protostellar core of $n_{\H}=10^{8}\cm^{-3}$, one requires $\St_{\rm cri,aJ}(\rm iER)\sim 35$ and $1100$ to align $a=1\mum$ and $10\mum$ grains, respectively. Thus, large grains in denser regions require a high suprathermality to be efficiently aligned by inelastic relaxation.

\subsubsection{Suprathermal rotation by RATs}
For suprathermally rotating grains at high$-J$ attractors by RATs with sizes $a<a_{\rm trans}=\bar{\lambda}/2.5$, using $\St_{\rm RAT}$ from Equation (\ref{eq:S_RAT1}) for Equation (\ref{eq:tiner_tgas}), one obtains
\bea
\frac{\tau_{\rm iER}^{\rm RAT,high-J}}{\tau_{\rm gas}}&\simeq& 7.1\times10^{-11}a_{-5}^{-6}\frac{\mu_{8}Q_{3}}{\hat{\rho}^{2}}\frac{n_{8}}{T_{\gas,1}}\frac{g'\Gamma_{\|}}{s^{5/2}}\left(\frac{\bar{\lambda}}{1.2\mum}\right)^{6}\nonumber\\
&\times&\left(\frac{\gamma_{\rm rad} U_{6}}{n_{8}T_{\gas,1}}\right)^{-3}\left(\frac{1}{1+F_{\rm IR}}\right)^{-3},\label{eq:tiner_tgas1}
\ena
which decreases rapidly with the grain size as $1/a_{-5}^{6}$.

For $a>a_{\rm trans}$, using $\St_{\rm RAT}$ from Equation (\ref{eq:S_RAT2}), one obtains
\bea
\frac{\tau_{\rm iER}^{\rm RAT,high-J}}{\tau_{\rm gas}}&\simeq& 1.1\times 10^{-16}a_{-5}^{3}\frac{\mu_{8}Q_{3}}{\hat{\rho}^{2}}\frac{n_{8}}{T_{\gas,1}}\frac{g'\Gamma_{\|}}{s^{1/2}}\left(\frac{\bar{\lambda}}{1.2\mum}\right)^{-3}\nonumber\\
&\times& \left(\frac{\gamma_{\rm rad} U_{6}}{n_{8}T_{\gas,1}}\right)^{-3}\left(\frac{1}{1+F_{\rm IR}}\right)^{-3},\label{eq:tiner_tgas2}
\ena
which increases rapidly with the grain size as $a^{3}$.

Using the condition for efficient internal alignment as $\tau_{\rm iER}/\tau_{\gas}<1$, Equation (\ref{eq:tiner_tgas1}) yields the range of grain sizes with internal alignment, 
\bea
a&>&a_{\min,aJ}^{\rm RAT,high-J}(\rm iER)\nonumber\\
&=&2.3\times 10^{-3}\left(\frac{\mu_{8}Q_{3}}{\hat{\rho}^{2}}\frac{g'\Gamma_{\|}}{s^{5/2}}\right)^{1/6}\left(\frac{n_{8}}{T_{\gas,1}}\right)^{1/6}\left(\frac{\bar{\lambda}}{1.2\mum}\right)\nonumber\\
&\times&\left(\frac{\gamma_{\rm rad} U_{6}}{n_{8}T_{\gas,1}}\right)^{-1/2}\left(\frac{1}{1+F_{\rm IR}}\right)^{-1/2}\mum,\label{eq:amin_aJ_RAT}
\ena
where $a_{\rm min}$ is the minimum grain size for internal alignment which is the solution of the equation $\tau_{\rm iER}=\tau_{\gas}$.

Similarly, Equation (\ref{eq:tiner_tgas2}) yields the grain sizes for internal alignment as
\bea
a&<&a_{\rm max,aJ}^{\rm RAT,high-J}({\rm iER})\nonumber\\
&=&2.1\times 10^{4}\left(\frac{\hat{\rho}^{2}}{\mu_{8}Q_{3}}\frac{s^{1/2}}{g'\Gamma_{\|}}\right)^{1/3}\left(\frac{T_{\gas,1}}{n_{8}}\right)^{1/3}\left(\frac{\bar{\lambda}}{1.2\mum}\right)\nonumber\\
&\times& \left(\frac{\gamma_{\rm rad} U_{6}}{n_{8}T_{\gas,1}}\right)\left(\frac{1}{1+F_{\rm IR}}\right)\mum,\label{eq:amax_aJ_RAT}
\ena
where $a_{\rm max}$ is the maximum grain size for internal alignment which is the solution of the equation $\tau_{\rm iER}=\tau_{\gas}$. The equation above implies $a_{\rm max,aJ}^{\rm RAT}(\rm iER)\sim 30\mum$ for $U=10^{6}$, $n_{\H}\sim 10^{10}\cm^{-3}$. Therefore, for the conditions of strong radiation fields around a protostar, RATs can induce efficient internal alignment by inelastic relaxation.

Therefore, suprathermal rotation by RATs produces a range of grains that can have efficient (right) internal alignment by inelastic relaxation between $a_{\min,aJ}^{\rm RAT}-a_{\max,aJ}^{\rm RAT}$. 
Equation (\ref{eq:amax_aJ_RAT}) reveals the dependence of $a_{\max,aJ}^{\rm RAT}$ on the values of $\mu_{8}$ and $Q_{3}$, which are determined by the mechanical properties of dust grains. One expects the value of $\mu$ decrease with increasing the grain size due to grain coagulation in protostellar environments that increases the porosity (or increases the filling factor; \citealt{Seizinger:2013ee}). Therefore, for the same local conditions, larger (more porous) grains would have more efficient internal alignment by inelastic relaxation than smaller ones.

\begin{figure*}
\centering
\includegraphics[width=0.35\textwidth]{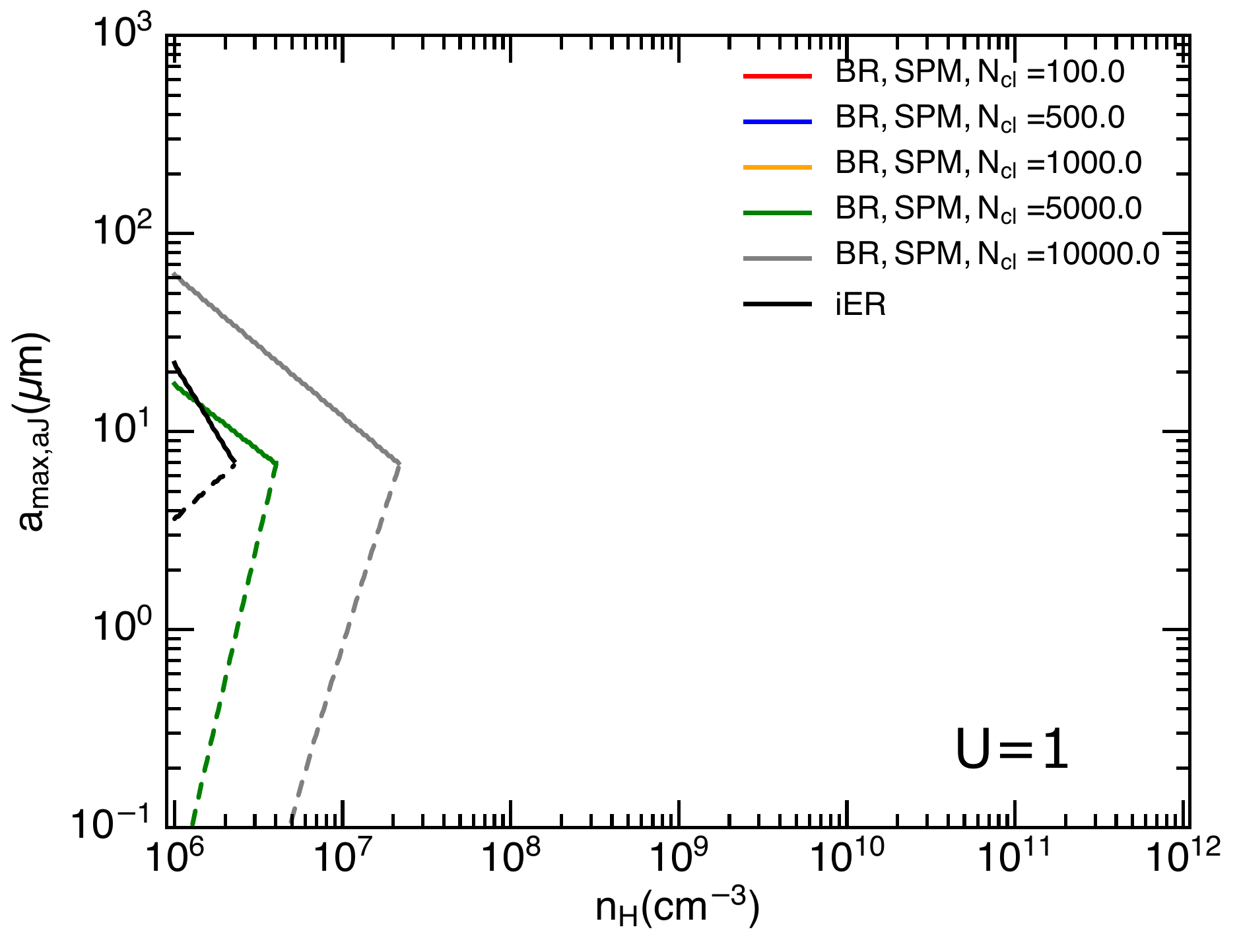}
\includegraphics[width=0.35\textwidth]{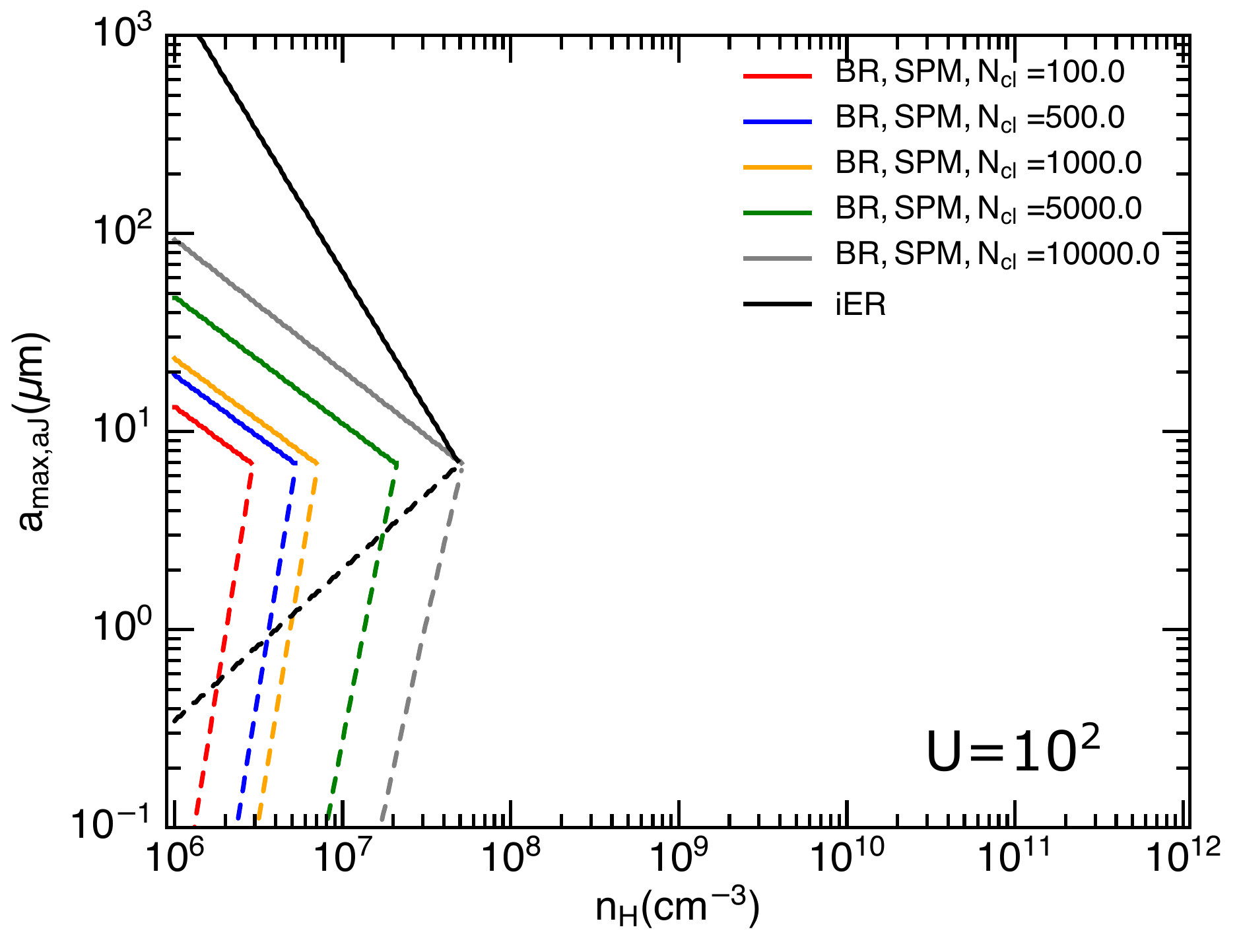}
\includegraphics[width=0.35\textwidth]{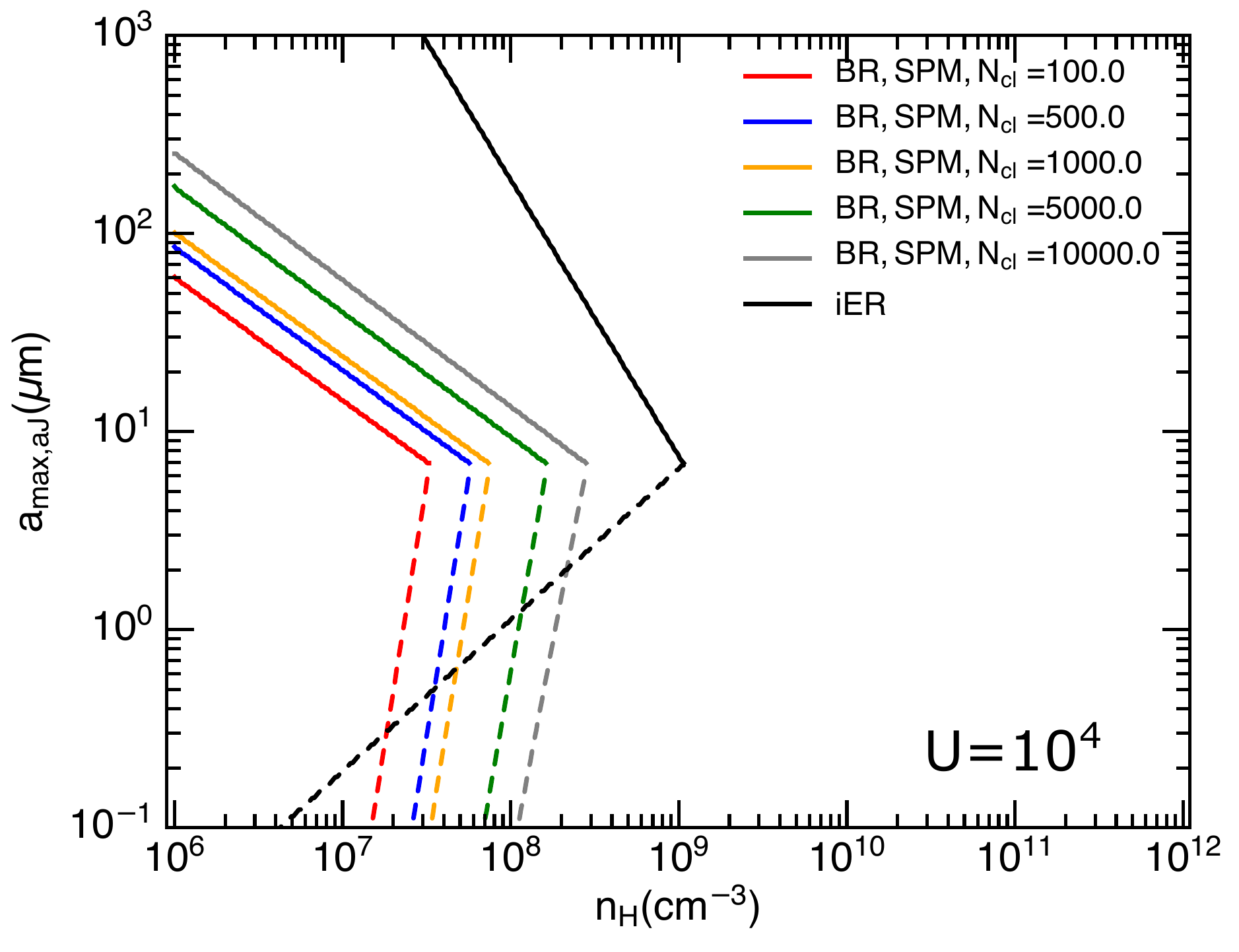}
\includegraphics[width=0.35\textwidth]{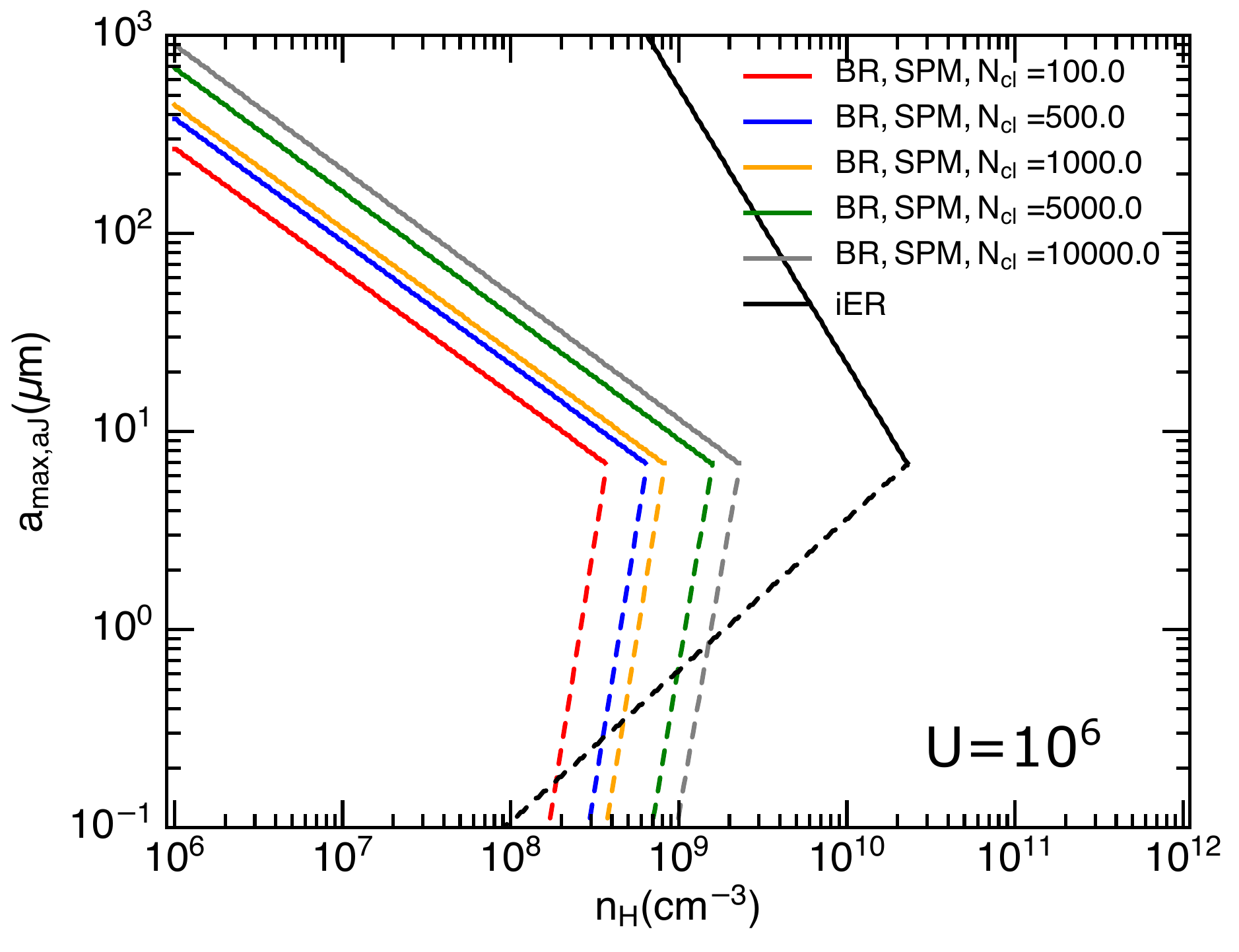}
\caption{Variation of the minimum ($a_{\min,aJ}$, dashed lines) and maximum grain size ($a_{\max,aJ}$, solid lines) for efficient internal alignment with the gas density due to super-Barnett relaxation (BR, SPM) and inelastic relaxation (iER), assuming grains rotating suprathermally by RATs with $\gamma_{\rm rad}=0.5$, $\bar{\lambda}=10\mum$, and four values of $U=1,10^{2},10^{4},10^{6}$. Inelastic relaxation is more efficient than super-Barnett relaxation for $U\gtrsim 10^{4}$. Both effects become inefficient for increasing $n_{\H}$ due to stronger gas damping and lower $\St_{\rm RAT}$.}
\label{fig:amax_aJ_RAT}
\end{figure*}

Figure \ref{fig:amax_aJ_RAT} shows the range of internal alignment sizes by BR and inelastic relaxation due to spin-up by RATs. Inelastic relaxation is more efficient than super-Barnett relaxation for $U\gtrsim 10^{4}$. Both effects become inefficient for increasing $n_{\H}$ due to stronger gas damping and lower $\St_{\rm RAT}$.

\subsubsection{Suprathermal rotation by METs}

For suprathermally rotating grains aligned at high$-J$ attractors by METs, the ratio between the inelastic relation and gas damping timescales becomes
\bea
\frac{\tau_{\rm iER}^{\rm MET,high-J}}{\tau_{\rm gas}}\simeq 0.65s_{d,-1}^{-6}Q_{\rm spinup,-3}^{-3}\frac{\mu_{8}Q_{3}}{\hat{\rho}^{2}}\frac{g'\Gamma_{\|}^{4}}{s^{11/2}}\frac{n_{8}}{T_{\gas,1}},~~~~~\label{eq:tiner_tgas_MET}
\ena
which does not depend on the grain size, but the ratio decreases rapidly with increasing with the drift speed as $s_{d}^{-6}$. Thus, one can determine the critical drift parameter required for internal alignment by inelastic relaxation,
\bea
s_{\rm cri,aJ}(\rm iER)&=&0.09Q_{\rm spinup,-3}^{-1/2}\left(\frac{\mu_{8}Q_{3}}{\hat{\rho}^{2}}\right)^{1/6}\left(\frac{g'\Gamma_{\|}^{4}}{s^{11/2}}\right)^{1/6}\left(\frac{n_{8}}{T_{\gas,1}}\right)^{1/6},\nonumber\\
&\simeq& 0.22Q_{\rm spinup,-3}^{-1/2}\left(\frac{\mu_{8}Q_{3}}{\hat{\rho}^{2}}\right)^{1/6}\left(\frac{n_{8}}{T_{\gas,1}}\right)^{1/6}\nonumber\\
&\times&\left(\frac{g'(\hat{s})\Gamma_{\|}(\hat{s})^{4}}{\hat{s}^{11/2}}\right)^{1/6},\label{eq:sd_cri_ine}
\ena
which implies $s_{d}>0.22$ for $s=0.5$, assuming the normalized parameters. The critical values $s_{\rm cri,aJ}({\rm iER})$ increases in denser/cooler regions of $n_{8}/T_{\gas,1}>1$ and decreases in the less dense/hotter regions of $n_{8}/T_{\gas,1}<1$ (see also Figure \ref{fig:times_a_MET}). For a very dense region of $n_{\H}=10^{12}\cm^{-3}$, the critical drift parameter becomes $s_{\rm cri,aJ}\sim 1.02$.

\begin{figure*}
\includegraphics[width=0.5\textwidth]{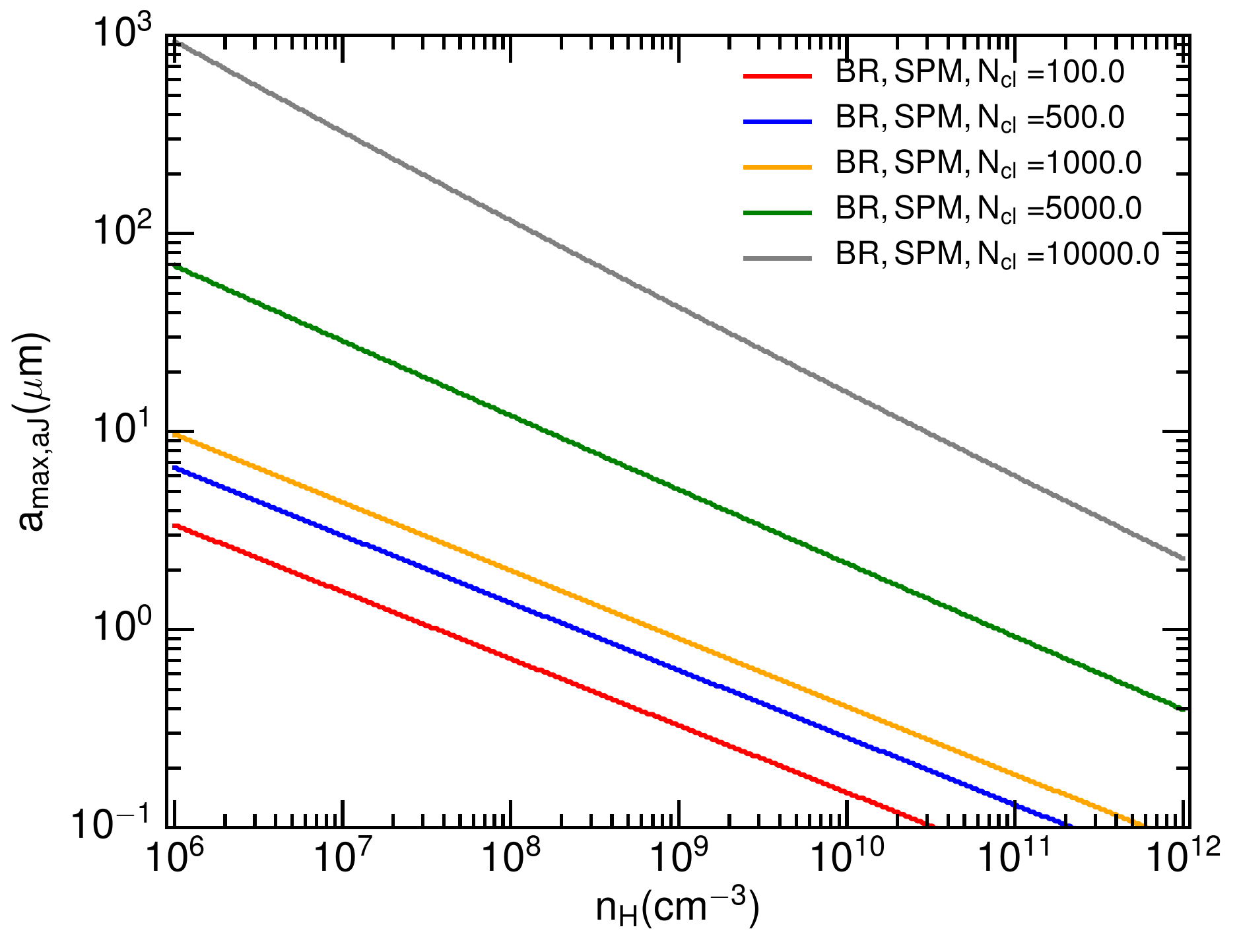}
\includegraphics[width=0.5\textwidth]{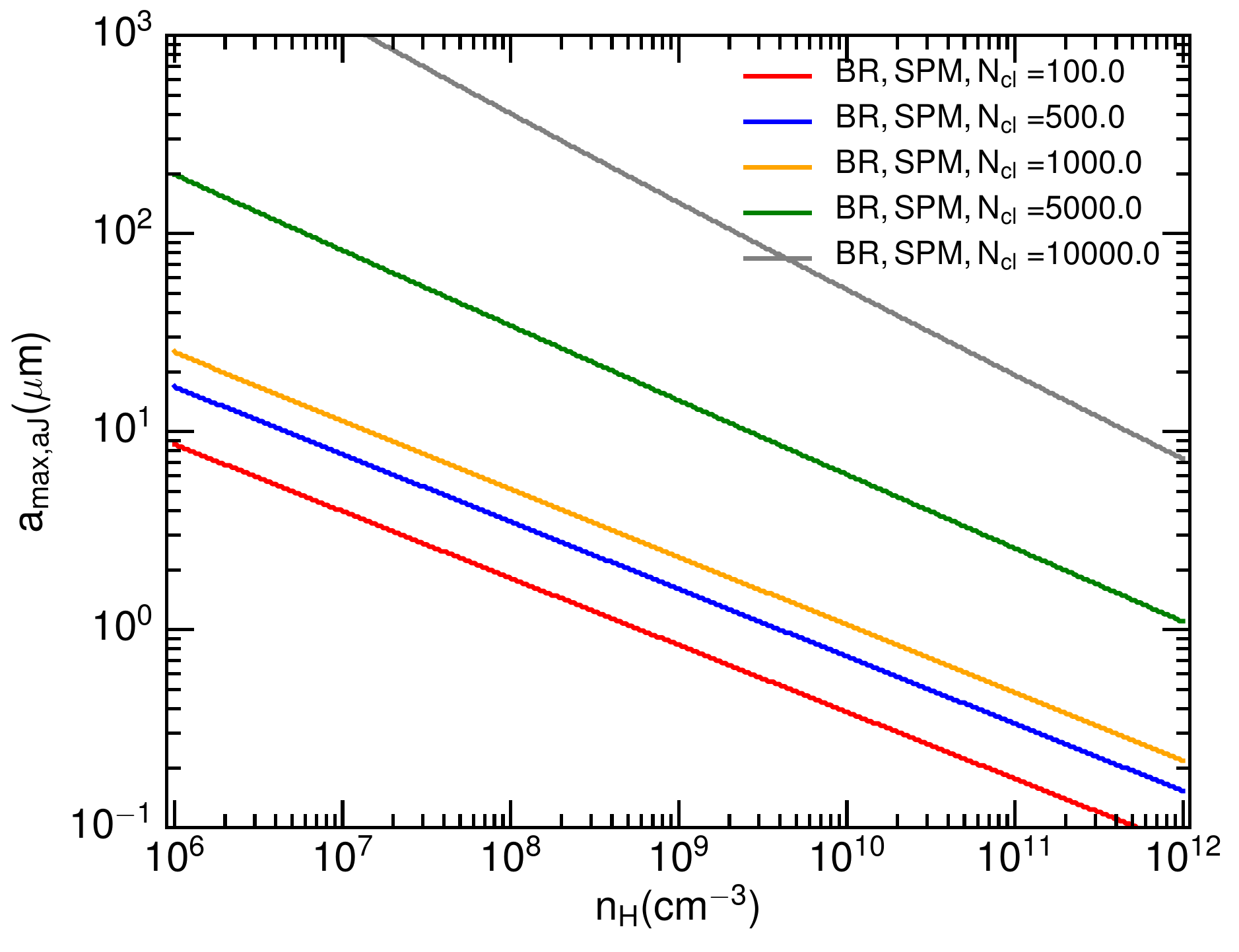}
\caption{Variation of the maximum grain size for efficient internal alignment, $a_{\max,aJ}$, with the gas density due to Barnett relaxation for the typical radiation strength $U=1$, assuming grains rotating suprathermally by METs with $s_{d}=0.25$ (left panel) and $0.5$ (right panel).  Grains with the different level of iron clusters are considered. The internal alignment sizes increases with $N_{\rm cl}$ and the drift parameter $s_{d}$.}
\label{fig:amax_aJ_METs_U1}
\end{figure*}

\begin{figure*}
\includegraphics[width=0.5\textwidth]{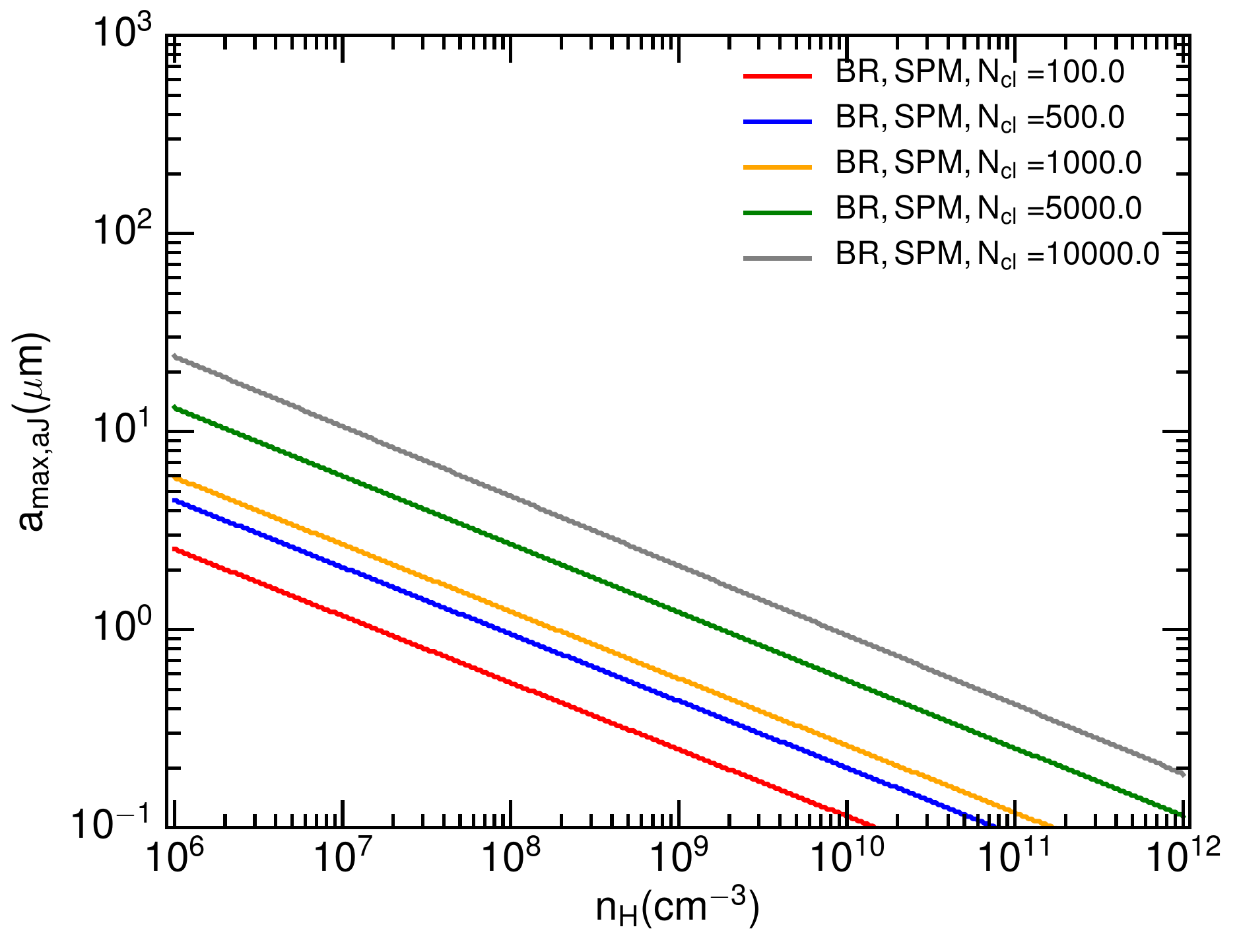}
\includegraphics[width=0.5\textwidth]{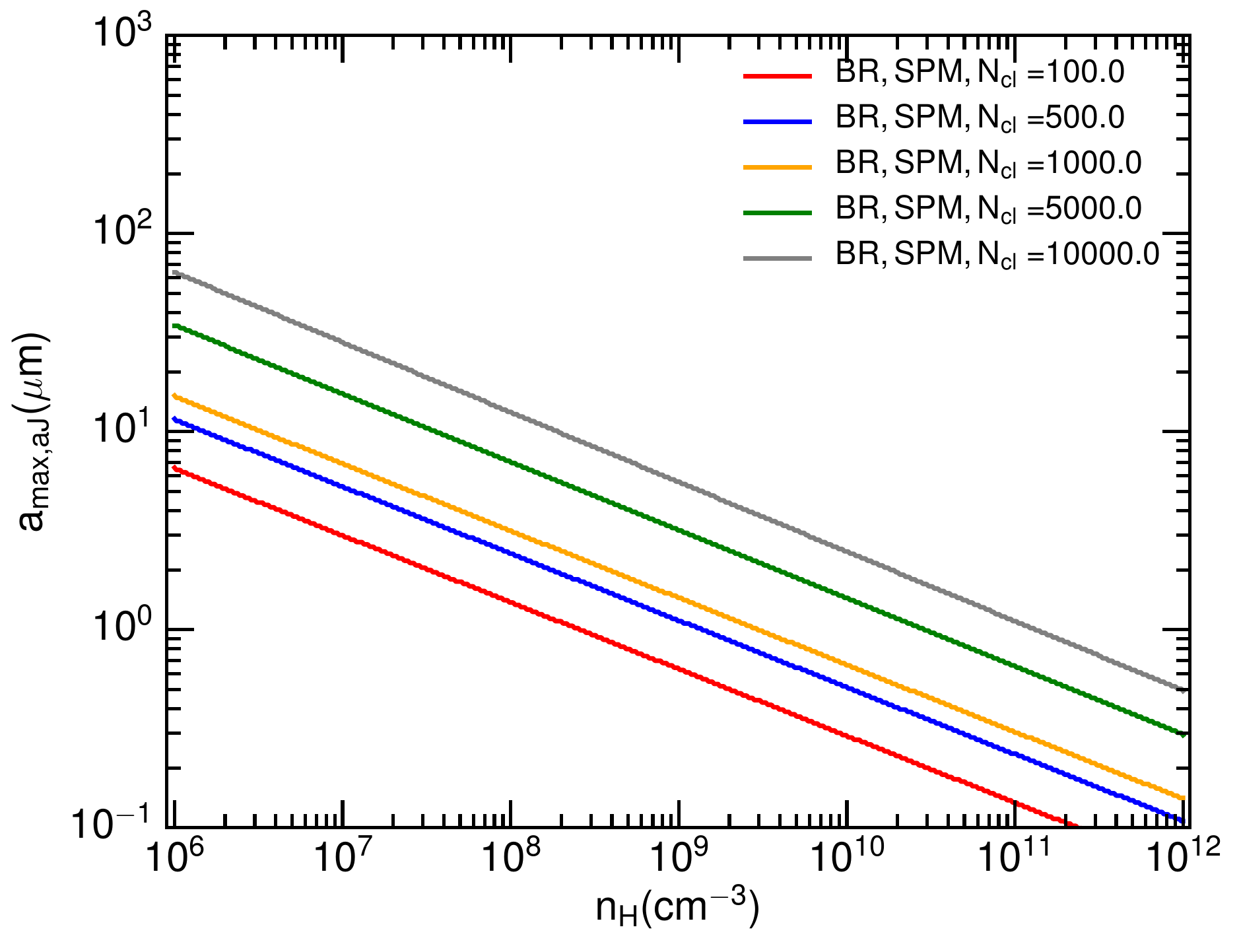}
\caption{Same as Figure \ref{fig:amax_aJ_METs_U1}, but for grains subject to a stronger local radiation field of $U=10^{4}$. The values of $a_{\max,aJ}$ are smaller, especially for the case of large $N_{\rm cl}$, due to the decrease of the magnetic susceptibility with the dust temperature.}
\label{fig:amax_aJ_METs_U1e4}
\end{figure*}

Figures \ref{fig:amax_aJ_METs_U1} and \ref{fig:amax_aJ_METs_U1e4} show the maximum grain size of internal alignment by super-Barnett relaxation due to spin-up by METs, assuming the range of density relevant star-forming regions from $10^{6}-10^{12}\cm^{-3}$, and the local radiation strength of $U=1$ and $U=10^{4}$, respectively. Two values of the drift parameter of $s_{d}=0.25$ (left) and $0.5$ (right). The alignment size increases with $s_{d}$ due to increasing suprathermal rotation.

\section{External Alignment in the presence of magnetic fields}\label{sec:extalign}
Here, we discuss in details the main physical processes involved in the external alignment of the grain angular momentum with a preferred direction by RATs (METs) in the presence of an ambient magnetic field. We will derive general formulae for critical grain sizes for external alignment for protostellar conditions with a high gas density of $n_{\H}\gtrsim 10^{6}\cm^{-3}$. 
 
\subsection{Larmor precession and magnetic alignment}
If the Larmor precession of the grain angular momentum around the ambient magnetic field is very fast compared to the gas damping time, then, the magnetic field becomes an axis of grain alignment, which is called magnetic alignment. From Equations (\ref{eq:tgas}) and (\ref{eq:tauB}), one derives the ratio of the Larmor precession time to the gas damping time for superparamagnetic grains as (see also \citealt{Hoang.2022}):
\bea
\frac{\tau_{B}}{\tau_{\gas}}&=&\frac{2\pi g_{e}\mu_{B}}{\hbar}\frac{1}{\chi_{\sp}(0)B}\left(\frac{1.2n_{\H}m_{\H}v_{\rm T}\Gamma_{\|}a}{\sqrt{\pi}s}\right),\nonumber\\
&\simeq& 0.01a_{-5}\left(\frac{n_{8}T_{\gas,1}^{1/2}T_{d,1}\Gamma_{\|}}{N_{\rm cl}\phi_{\rm sp,-2}\hat{p}^{2}B_{3}s}\right).~~~\label{eq:tauB_taugas}
\ena

The maximum size for the grain alignment with $\bJ$ aligned with the magnetic field ($\Bv$) via Larmor precession, denoted by $a_{\max, JB}$ (Lar), is then determined by $\tau_{B}/\tau_{\gas}=1$, yielding
\bea
a_{\max, JB}^{\rm Lar} \simeq 5.1\times 10^{5} \left(\frac{N_{\rm cl,4}\phi_{\rm sp,-2}\hat{p}^{2}B_{3}\hat{s}}{n_{8}T_{\gas,1}^{1/2}T_{d,1}\Gamma_{\|}}\right) \mum,~~~\label{eq:amax_JB}
\ena
which implies that VLGs of $a\sim 5.1\,\rm mm$ can still be aligned with the magnetic field against gas randomization at $n_{8}\sim 100$, but $a_{\max,JB}^{\rm Lar}\sim 51\mum$ for $n_{8}=10^{4}$, assuming large iron inclusions of $N_{cl,4}=1$. However, for ordinary paramagnetic grains of $N_{\rm cl}\sim 1$, only grains smaller than $a_{\max,JB}^{\rm Lar}\lesssim 51\mum$ can be aligned with $\Bv$ in very dense regions of $n_{8}\gtrsim 1$.

\subsection{Effect of magnetic relaxation on magnetic alignment}
Following \cite{1951ApJ...114..206D}, a rotating paramagnetic grain with the angular momentum $\bJ$ making an angle relative to the ambient magnetic field experiences magnetic dissipation of the grain rotational energy due to the existence of the rotating magnetization with respect to the grain body. This paramagnetic relaxation process eventually leads to the alignment of $\bJ$ with the magnetic field (see Figure \ref{fig:KBE_RAT}), which is known as Davis-Greenstein (DG) alignment mechanism. For grains with iron inclusions, the characteristic time of superparamagnetic relaxation is given by (see e.g., \citealt{2016ApJ...831..159H}) 
\bea
\tau_{\rm mag,sp} &=& \frac{I_{\|}}{VK_{\rm sp}(\Omega)B^{2}}=\frac{2\rho a^{2}}{5K_{\rm sp}(\Omega)B^{2}},\nonumber\\
&\simeq &
0.15\frac{\hat{\rho}a_{-5}^{2}}{N_{\rm cl}\phi_{\rm sp}\hat{p}^{2}B_{3}^{2}}\frac{T_{d,1}}{k_{\rm sp}(\Omega)}
 \yr,\label{eq:tau_DG}~~~
\ena
where $K_{\rm sp}(\Omega)$ from Eq. (\ref{eq:kappa_sp}) has been used.

To see if superparamagnetic relaxation can help align grains against gas collisions, it is convenient to introduce a dimensionless parameter
\bea
\delta_{\rm mag,sp}&=&\frac{\tau_{\rm gas}}{\tau_{\rm mag,sp}}\nonumber\\
&= &56a_{-5}^{-1}\frac{N_{\rm cl,4}\phi_{\rm sp,-2}\hat{p}^{2}B_{3}^{2}}{\hat{\rho} n_{8}T_{\gas,1}^{1/2}}\frac{k_{\rm sp}(\Omega)}{T_{d,1}},\label{eq:delta_m}~~~~
\ena
which implies $\delta_{\rm mag,sp}=5.6$ and $0.56$ for $a=1$ and $10\mum$ with $n_{8}=1$ and $N_{\rm cl,4}=1$. 

As shown in \citep{2016ApJ...821...91H}, grain alignment by superparamagnetic relaxation is inefficient for thermally rotating grains because of internal thermal fluctuations within the dust grain. The joint action of superparamagnetic relaxation and RATs (METs) enhances the alignment degree so that SPM grains can achieve perfect alignment if $\delta_{\rm mag,sp}> 10$ \citep{2016ApJ...831..159H}. Therefore, one can determine the maximum size for which superparamagnetic relaxation is important for external alignment using the condition $\delta_{\rm mag,sp}>1$, yielding
\bea
a_{\max,JB}^{\rm mag}\simeq 5.6\frac{N_{\rm cl,4}\phi_{\rm sp,-2}\hat{p}^{2}B_{3}^{2}}{\hat{\rho} n_{8}T_{\gas,1}^{1/2}}
\frac{k_{\rm sp}(\Omega)}{T_{d,1}} ~{\rm \mu m}.
\label{eq:amax_JB_DG}
\ena

For a protostellar core with a typical magnetic field of $B=10^{3}\,\mu G$ (see \citealt{Hull:2019hw,Pattle.2022} for reviews), one has $a_{\max,JB}^{\rm mag}=55.3\mum$ and $0.553\mum$ for $n_{\H}=10^{5}$, and $10^{7}\cm^{-3}$, assuming $N_{\rm cl}=10^{4}$. Therefore, superparamagnetic relaxation can be important for external alignment of VLGs in prestellar cores or protostellar envelope of density $n_{\H}\lesssim 10^{6}\cm^{-3}$. However, in protostellar cores of very high density of $n_{\H}\gtrsim 10^{7}\cm^{-3}$, super-paramagnetic relaxation is negligible for VLGs.

\subsection{The $k-$RAT vs. $B-$RAT alignment}\label{sec:radalign}
As discussed in Section \ref{sec:paradigm}, the magnetic field is usually the axis of grain alignment in the ISM due to the fact that Larmor precession is much faster than radiative precession and gas collisions. In protostellar environments, grain sizes and radiation field are much larger than the ISM, so that Larmor precession is not likely faster than radiative precession. To determine which direction ($\kv$ or $\Bv$) acts as the axis of grain alignment (see Figure \ref{fig:KBE_RAT}), we need to compare their relative timescales. The $k-$RAT alignment occurs when $\tau_{k}<\tau_{B}$, and $B-$RAT alignment occurs when $\tau_{k}>\tau_{B}$.

Using Equation (\ref{eq:tauk}) one calculate the radiative precession timescale for grains aligned at low$-J$ attractors with $\St=1$, 
\bea
\tau_{k}^{\rm low-J}\simeq 56.8\hat{\rho}^{1/2}T_{\gas,1}^{1/2}\hat{s}^{-1/6}a_{-5}^{1/2}\left(\frac{1.2\mum}{\gamma_{\rm rad}\bar{\lambda}\hat{Q}_{e3}U}\right)
\yr,~~~~~\label{eq:tauk_lowJ}
\ena 
and for grains aligned at high$-J$ attractors with $\St=\St_{\rm RAT}$ from Equation (\ref{eq:S_RAT2}), one has
\bea
\tau_{k}^{\rm high-J}&\simeq&11.7\hat{s}^{-1/3}\left(\frac{\hat{\rho}a_{-5}}{\hat{Q}_{e3}}\right)\left(\frac{1}{n_{8}T_{\gas,1}^{1/2}}\right)\left(\frac{1}{1+F_{\rm IR}}\right)\yr.~~~~~\label{eq:tauk_highJ}
\ena

Comparing $\tau_{k}^{\rm low-J}$ and $\tau_{k}^{\rm high-J}$ with $\tau_{B}$ (Equation \ref{eq:tauB}) yields
\bea
\frac{\tau_{k}^{\rm low-J}}{\tau_{B}}\simeq&&6.96\times 10^{4}\hat{s}^{-1/6} \nonumber \\ 
&&\times \left(\frac{1.2\mum}{\gamma_{\rm rad}\bar{\lambda}\hat{Q}_{e3}U}\right)\frac{N_{\rm cl}\phi_{sp,-2}\hat{p}^{2}B_{3}}{\hat{\rho}^{1/2} a_{-5}^{3/2}}\frac{T_{\gas,1}^{1/2}}{T_{d,1}},~~~\label{eq:tauk_tauB_lowJ}
\ena
for grains at low$-J$ attractors, and
\bea
\frac{\tau_{k}^{\rm high-J}}{\tau_{B}}\simeq1.4\times 10^{4}\frac{\hat{s}^{-1/3}}{a_{-5}\hat{Q}_{e3}}\frac{N_{\rm cl}\phi_{sp,-2}\hat{p}^{2}B_{3}}{n_{8}T_{\gas,1}^{1/2}(1+F_{IR})}\frac{1}{T_{d,1}},~~~\label{eq:tauk_tauB_highJ}
\ena
for grains aligned at high$-J$ attractors.

For grains aligned at low$-J$ attractors, the $k-$RAT alignment occurs when $\tau_{k}^{\rm low-J}<\tau_{B}$. The minimum size for the $k-$RAT alignment, which is also the maximum size for B$-$RAT alignment, is then determined by $\tau_{k}^{\rm low-J}/\tau_{B}=1$, yielding
\bea
a_{\min,Jk}^{\rm RAT,low-J}&&\equiv a_{\rm max,JB}^{\rm RAT,low-J}\simeq 7.85\hat{s}^{-1/9} \left(\frac{1.2\mum}{\gamma_{\rm rad}\bar{\lambda}\hat{Q}_{e3}U_{6}}\right)^{2/3}\nonumber \\
&&\times \left(\frac{N_{\rm cl,4}\phi_{sp,-2}\hat{p}^{2}B_{3}}{\hat{\rho}^{1/2}}\right)^{2/3}\left(\frac{T_{\gas,1}^{1/2}}{T_{d,1}}\right)^{2/3}\mum,~~~~~\label{eq:amin_Jk_lowJ}
\ena
which decreases with increasing the radiation strength but increases with $N_{\rm cl}$. Thus, VLGs aligned at low$-J$ attractors can experience the $k-$RAT alignment even grains have strong iron inclusions of $N_{\rm cl,4}=1$.

For grains aligned at high$-J$ attractors, the $k-$RAT alignment occurs when $\tau_{k}^{\rm high-J}<\tau_{B}$, and one obtains the minimum size for $k-$RAT (maximum size for $B-$RAT) alignment,
\bea
a_{\min,Jk}^{\rm RAT,high-J}&&\equiv a_{\rm max,JB}^{\rm RAT,high-J}\simeq1.4\times 10^{7}\left(\frac{\hat{s}^{-1/3}}{n_{8}T_{\gas,1}^{1/2}\hat{Q}_{e3}}\right)\nonumber\\
&\times&\left(\frac{N_{\rm cl,4}\phi_{sp,-2}\hat{p}^{2}B_{3}}{(1+F_{\rm IR})}\right)\left(\frac{1}{T_{d,1}}\right)\mum.~~~\label{eq:amin_Jk_highJ}
\ena

For protostellar cores of density $n_{8}=1$, $F_{\rm IR}\ll 1$, the value of $a_{\min,Jk}^{\rm RAT,high-J}$ becomes independent on the radiation strength $U$. Dust grains at high$-J$ attractors all have $B-$RAT alignment even with a low iron inclusions of $N_{\rm cl}\sim 10$. In the very high density of $n_{\H}>10^{12}\cm^{-3}$, $a_{\min,Jk}^{\rm RAT,high-J}\sim 1.4(N_{\rm cl}/10)\mum$, so VLGs with small iron clusters of $N_{\rm cl}<100$ can have the $k-$RAT alignment, assuming $T_{\gas,1}=1, B_{3}=1$. In the low density and strong radiation, $F_{\rm IR}\gg 1$, and $a_{\min,Jk}^{\rm RAT,high-J}\propto 1/F_{\rm IR}\sim U^{-2/3}$. The $k-$RAT alignment can become important for smaller grains.

In summary, the difference in the radiative precession timescales at low$-J$ and high$-J$ attractors has important implication for the grain alignment via $k-$RAT vs. $B-$RAT. Suppose that initially a grain is rotating thermally due to gas collisions. One turns on a radiation source and the grain is simultaneously spun up and driven to precess around $\kv$. At the early stage, the radiative precession can be very fast, and $k-$RAT alignment is possible. As $J$ increases due to spin-up by RATs, its precession rate decreases, which will establish the $B-$RAT alignment when grains are stably aligned at high$-J$ attractors. In protostellar cores, grains at low$-J$ attractors most likely have the $k-$RAT alignment. However, grains at high$-J$ attractors can have both the $B-$RAT alignment and $k-$RAT alignment, depending on the gas density, grain magnetic susceptibility, and magnetic field strength. 

\subsection{Mechanical precession and $v-$MET vs $B-$MET alignment}

For grains to be aligned with the gas flow, one must have $\tau_{v}\ll \tau_{\rm gas}$. For rotating grains with a suprathermal rotation number $\St$, one obtains
\bea
a_{\rm min,Jv}^{\rm MET}>0.035\left(\frac{\St}{Q_{\rm prec,-1}s_{d,-1}^{2} \Gamma_{\|}}\right)^{2/3}\mum,
\label{eq:amin_Jv_MET}
\ena
which implies that even for a moderate suprathermal number of $\St\sim3$, mechanical precession is faster than the gas randomization. 

For grains aligned at high$-J$ attractors by METs, i.e., $\St=\St_{\rm MET}$ (Eq.\ref{eq:S_MET}), the $v-MET$ alignment requires $\tau_{v}<\tau_{\rm gas}$ which yields
\bea
\frac{Q_{\rm prec,-1}}{Q_{\rm spinup,-3}}>0.2,\label{eq:Qprec_Qspinup}
\ena
which implies that a small magnitude of the precession component of METs can establish the $v-MET$ alignment.

Similar to RATs, in the presence of magnetic fields, the grain experiences both the Larmor precession and mechanical precession. To determine which direction ($\Bv$ vs. $\bv_{d}$) is the axis of grain alignment, one needs to compare the timescales of mechanical precession with the Larmor precession. Using $\tau_{v}$ and $\tau_{B}$ from Equations (\ref{eq:tau_v}) and (\ref{eq:tauB}), one obtains
\bea
\frac{\tau_{v}}{\tau_{B}}\simeq 0.21\frac{\St}{Q_{\rm prec,-1}s_{d,-1}^{2}n_{8}T_{\gas,1}^{1/2}}\frac{N_{\rm cl}\phi_{sp,-2}\hat{p}^{2}B_{3}}{\hat{\rho} a_{-5}^{5/2}}.~~~\label{eq:tauv_tauB}
\ena

For grains aligned at low$-J$ attractors, $\St=1$, the minimum size for the $v-$MET alignment is determined by $\tau_{v}^{\rm low-J}/\tau_{B}<1$, yielding
\bea
a_{\min,Jv}^{\rm MET,low-J}\simeq13.6\left(\frac{N_{\rm cl,4}\phi_{sp,-2}\hat{p}^{2}B_{3}}{\hat{\rho}Q_{\rm prec,-1}s_{d,-1}^{2}n_{8}T_{\gas,1}^{1/2}}\right)^{2/5}\mum,~~~\label{eq:amin_Jv_lowJ}
\ena
which implies that VLGs can have the $v-$MET alignment.

For grains aligned at high$-J$ attractors by METs, by plugging $\St=\St_{\rm MET}$ into Equation (\ref{eq:tauv_tauB}) one obtains
\bea
\frac{\tau_{v}^{\rm high-J}}{\tau_{B}}\simeq 2.1\times 10^{4}\frac{Q_{\rm spinup,-3}}{Q_{\rm prec,-1}n_{8}T_{\gas,1}^{1/2}}\frac{N_{\rm cl,4}\phi_{sp,-2}\hat{p}^{2}B_{3}}{\hat{\rho}^{1/2} a_{-5}},~~~
\ena
such that the condition $\tau_{v}^{\rm high-J}/\tau_{B,sp}<1$ yields
\bea
a>a_{\rm min,Jv}^{\rm MET,high-J}&\simeq&2.1\times 10^{4}\left(\frac{Q_{\rm spinup,-3}}{Q_{\rm prec,-1}}\right)\left(\frac{1}{n_{8}T_{\gas,1}^{1/2}}\right)\nonumber\\
&\times& \left(\frac{N_{\rm cl,4}\phi_{sp,-2}\hat{p}^{2}B_{3}}{\hat{\rho}^{1/2}}\right)\mum,~~~\label{eq:amin_Jv_highJ}
\ena
which decreases with increasing $n_{\H}$ and decreasing $N_{\rm cl}$. For the protostellar core of density $n_{\H}=10^{8}\cm^{-3}$ and assuming the typical parameters, the mechanical precession is faster than the Larmor precession for VLGs with $a>a_{\rm min,Jv}\sim 0.21\,\rm cm$. At a higher density of $n_{\H}=10^{12}\cm^{-3}$, the $v-$MET alignment can occur for smaller sizes of $a>2.1\mum$ for $N_{\rm cl,4}=1$. In the case of grains with small iron inclusions of $N_{\rm cl}\lesssim 10$, small grains with $a>0.21\,\mu$m can have the $v-$MET alignment in protostellar cores.

In summary, similar to RATs, large grains at low$-J$ attractors can have the $v-$MET alignment, whereas large grains at high$-J$ attractors likely have the $B-$MET alignment. Since the net degree of grain alignment at low$-J$ attractors is rather low, its contribution to the observed polarization is subdominant if $f_{\rm high-J}$ is considerable (e.g, $f_{\rm high-J}\sim 0.1-0.2$).

\begin{table*}
\begin{center}
\caption{Alignment of VLGs in Protostellar Environments.}\label{tab:intalign}
\begin{tabular}{|l| l| l|} \hline
{\it Internal Alignment (IA)} & {\it Fast internal relaxation} & {\it Slow internal relaxation} \cr
\hline
{\it High$-J$}  & Right IA ($\hat{\ba}_{1}\parallel \bJ$) & Right IA \cr
\hline
{\it Low$-J$}  & Right IA & Right IA or wrong IA ($\hat{\ba}_{1}\perp \bJ$) \cr
\hline
{\it External Alignment} & {\it Superparamagnetic grains} & {\it Paramagnetic grains}  \cr
\hline
{\it High$-J$} & $B-$RAT ($B-$MET) & $k-$RAT ($v-$MET) \cr

\hline
{\it Low$-J$}  & $k-$RAT ($v-$MET) & $k-$RAT ($v-$MET) \cr

\hline

\end{tabular}
\end{center}
\end{table*}

Table \ref{tab:intalign} summarizes different regimes of internal and external alignment for VLGs in protostellar environments by RATs due to stellar radiation and METs due to gas flows grain drift for two cases of fast internal relaxation (i.e., $\tau_{\rm INR}<\tau_{\rm gas}$) and slow internal relaxation ($\tau_{\rm INR}>\tau_{\rm gas}$). For internal alignment, grains aligned at high$-$ attractors have right IA in both cases of fast and slow relaxation, whereas grains aligned at low-$J$ have two possibilities of right IA and wrong IA. For external alignment, superparamagnetic grains can be aligned with $\bJ\|\Bv$ (B-RAT/B-MET) at high-$J$ attractors, while paramagnetic grains likely have the alignment along the radiation or gas flow (k-RAT/v-MET). Both superparamagnetic and paramagnetic grains aligned at low$-J$ attractors likely have the alignment along the radiation or gas flow.

\subsection{Minimum size for grain alignment by RATs and METs}
Regardless of the axis of grain alignment ($\Bv,\bk$ or $\bv$), efficient alignment of dust grains with a preferred direction in space is only achieved when grains rotate suprathermally because the grain orientation is randomized by gas collisions when they rotate thermally (\citealt{2016ApJ...821...91H}). Therefore, one can determine the minimum size for grain alignment by RATs by using the suprathermal condition of $\St_{\rm RAT}=3$ (e.g., \citealt{2014MNRAS.438..680H}). Using Equation (\ref{eq:S_RAT1}) yields
\bea
a_{\rm align}^{\rm RAT}&\simeq&0.016s^{-5/21}\hat{\rho}^{-1/7}\left(\frac{\bar{\lambda}}{1.2\mum}\right)^{4/7}\left(\frac{\gamma_{\rm rad} U_{6}}{n_{8}T_{\gas,1}}\right)^{-2/7}\nonumber\\
&\times&\left(\frac{1}{1+F_{\rm IR}}\right)^{2/7} \mum,\label{eq:aali_RAT}
\ena
which implies that small grains of $a>0.017\mum$ can be aligned in high-density regions of $n_{8}=1$ but strong radiation field of $U_{6}=1$.

Similarly, the minimum size for grain alignment by METs is defined by $\St_{\rm MET}=3$ \citep{2018ApJ...852..129H}, which yields
\bea
a_{\rm align}^{\rm MET}=0.013\hat{\rho}^{1/3}s_{d,-1}^{-4/3}Q_{\rm spinup,-3}^{-2/3}\mum,\label{eq:aali_MET}
\ena
which implies that small grains of $a>0.013\mum$ can be aligned by METs with $s_{d}\sim 0.1$, assuming $Q_{\rm spinup,-3}=1$.

\section{Application for a Protostellar disk}\label{sec:disk}
Now, we apply our general analysis from the previous section for a typical case of a protostellar disk and study the various grain alignment mechanisms within the disk mid-plane.

\subsection{Disk model assumption}
We consider a flared disk model with the mid-plane surface density at the disk radius $r$ from the central star given by (see, e.g., \citealt{Tung:2020ew})
\bea
\Sigma(r)=\Sigma_{0}\left(\frac{r}{\AU}\right)^{-\alpha},\label{eq:sigmaR}
\ena
where $\Sigma_{0}$ is the mass surface density at $r=1\AU$, and we assume $\alpha=3/2$. The typical values of $\Sigma_{0}\sim 100-1000\g\cm^{-2}$ for different disks (see \citealt{Williams:2011js}). 

We adopt the flared disk from \cite{1997ApJ...490..368C} with the pressure-scale height increasing with the radius as
\bea
\frac{H_{\rm p}}{r}=0.19\left(\frac{r}{\AU}\right)^{1/7}.\label{eq:H_R}
\ena

The gas number density at radius $r$ is calculated as
\bea
n_{\H}(r)&=&\frac{1}{2m_{\H}}\frac{\Sigma(r)}{\sqrt{2\pi}H_{\rm p}}=\frac{1}{2m_{\H}}\frac{\Sigma_{0}}{\sqrt{2\pi}H_{\rm p}}\left(\frac{r}{\AU}\right)^{-\alpha}\nonumber\\
&=&4\times 10^{13}\left(\frac{\Sigma_{0}}{10^{3}\g\cm^{-2}}\right)\left(\frac{r}{\AU}\right)^{-\alpha-8/7}\cm^{-3},\label{eq:nH}
\ena
and we denote $n_{0}= 4\times 10^{13}(\Sigma_{0}/10^{3}\g\cm^{-2})\cm^{-3}$ and $\alpha_{n}=\alpha+8/7$.

The gas temperature in the disk plane can be described by \cite{1997ApJ...490..368C}
\bea
T_{\rm gas}(r)=T_{0}\left(\frac{r}{\AU}\right)^{-\beta},\label{eq:Tgas_disk_R}
\ena
where $\beta=3/7$ and and $T_{0}\approx 150\K$ for $r<84\AU$ and $T_{\rm gas}(r)=21\K$ for $84\AU<r<209\AU$ (see e.g., \citealt{Tung:2020ew}). 

The magnetic field in the disk consists of the toroidal and poloidal components (see e.g., \citealt{2017ApJ...839...56T}). Here, we assume the strength of the magnetic field can be approximated as 
\bea
B(r)=B_{0}\left(\frac{\dot{M}}{10^{-8}M_{\odot}\yr^{-1}}\right)\left(\frac{r}{\AU}\right)^{-11/8},\label{eq:BR}
\ena
where $\dot{M}$ is the mass accretion rate and $B_{0}=10^{6}\mu$G (see \citealt{Yang.20212a5}).

As noted above, the gas temperature has a piecewise profile with $T_{\rm gas}\sim r^{-\beta}$ for $r<84\,\rm au$, while $T_{\rm gas}=21\K=\rm const$ for $r>84\,\rm au$. In the following sections, we only show our relevant variables with $T_{\rm gas}\sim r^{-\beta}$ as a general expression. These variables are indeed easily converted to the case of $r>84\,\rm au$ by replacing $T_{0}=21\K$ and $\beta=0$.
\subsection{Magnetic alignment by superparamagnetic relaxation and fast Larmor precession}

Plugging $n_{\H}(r), T_{\rm gas}(r)$, and $B(r)$ into Equation (\ref{eq:amax_JB_DG}), one obtains the maximum size for alignment by super-paramagnetic relaxation, 
\bea
a_{\rm max,JB}^{\rm mag}(r)&\simeq& 17.7\frac{N_{\rm cl,4}\phi_{\rm sp,-2}\hat{p}^{2}B_{0,6}^{2}}{\hat{\rho} n_{0,13}T_{0,2}^{1/2}}
\left(\frac{k_{\rm sp}(\omega)}{T_{d,1}}\right)
\nonumber\\
&\times&\left(\frac{\dot{M}}{10^{-8}\,M_{\odot}\yr^{-1}}\right)^{2}\left(\frac{r}{\AU}\right)^{\alpha_{n}+\beta/2-11/4}\mum,~~~~~
\label{eq:amax_JB_DG_disk}
\ena
where $B_{0,6}=B_{0}/(10^{6}\,\mu G)$, $n_{0,13}=n_{0}/(10^{13}\cm^{-3})$, $T_{0,2}=T_{0}/(100\K)$, and $\alpha_{n}+\beta/2-11/4=0.107$, which implies $a_{\rm max,JB}^{\rm mag}\sim 17.7,24.6, 30.1$, and $34.2\mum$ at $r=1,10, 50,$ and $100\AU$, respectively, assuming $N_{\rm cl,4}=1$.

Similarly, using Equation (\ref{eq:amax_JB}), one obtains the maximum size for magnetic alignment of $\bJ$ with $\Bv$ at disk radius $r$,
\bea
a_{\rm max,JB}^{\rm Lar}(r)&\simeq&7.9 \left(\frac{N_{\rm cl,4}\phi_{\rm sp,-2}\hat{p}^{2}B_{0,6}s}{n_{0,13}T_{0,2}^{1/2}T_{d,1}\Gamma_{\|}}\right)\nonumber\\
&\times& \left(\frac{r}{\AU}\right)^{\alpha_{n}+\beta/2-11/8} \mum,~~~\label{eq:amax_JB_disk}
\ena
which yields $a^{\rm Lar}_{\rm max,JB}\sim 25, 59, 108$, and $139\mum$ at $r=1, 10, 50,$ and $100\AU$, respectively, assuming the high level of iron inclusions with $N_{\rm cl,4}=1$. 

\begin{figure*}
\includegraphics[width=0.5\textwidth]{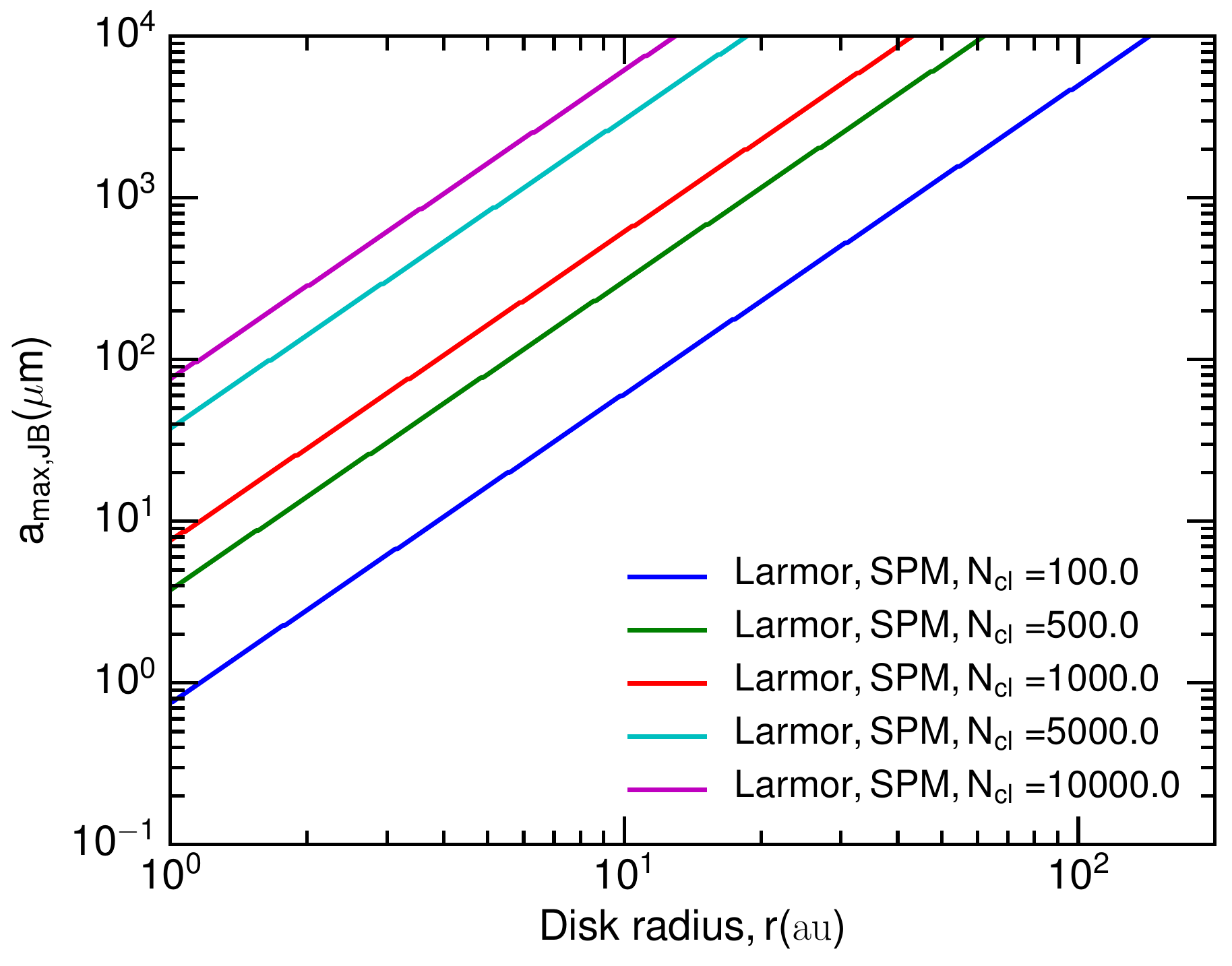}
\includegraphics[width=0.5\textwidth]{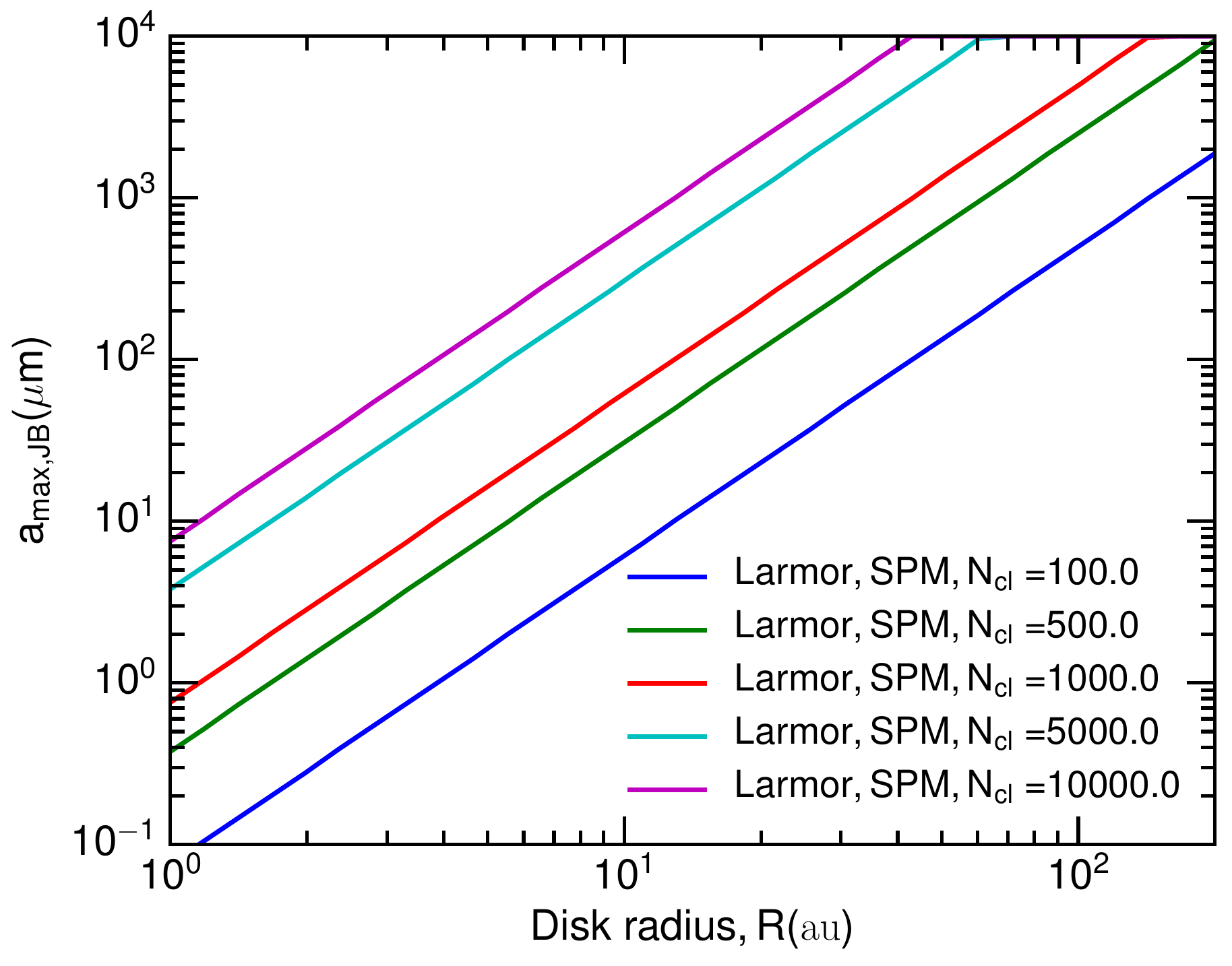}
\caption{Maximum size for magnetic alignment vs. disk radius for grains with different iron inclusions of $N_{\rm cl}\ge 100$, assuming the disk surface mass density $\Sigma_{0}=100$ (left) and $\Sigma_{0}=10^{3}\g\cm^{-2}$ (right).}
\label{fig:amax_JB}
\end{figure*}

Figure \ref{fig:amax_JB} shows the maximum size for the external alignment of $\bJ$ with $\Bv$ as a function of the disk radius for grains with different iron inclusions of $N_{\rm cl}\ge 100$ and for the two disk models. For $\Sigma_{0}=100\,\g\cm^{-2}$ (left panel), VLGs with the maximum number of iron atoms per cluster of $N_{\rm cl}=10^{4}$ can be magnetically aligned throughout the disk. Beyond the disk radius $r>10\AU$, VLGs of sizes $a>10\mum$ can be aligned with the magnetic field. As $a_{\rm max, JB}^{\rm Lar}\sim \Sigma^{-1}_{0}$, a 10 times higher gas surface density ($\Sigma_{0}=10^{3}\g\cm^{-2}$) results in a magnitude lower in $a_{\rm max,JB}^{\rm Lar}$ (right panel).

\subsection{Radiation field, RATs, and grain alignment}
Dust grains in the disk mid-plane are illuminated by attenuated stellar radiation and thermal emission from hot dust in the surface layer. Radiative transfer modeling in \cite{2017ApJ...839...56T} obtains the mean wavelength of the radiation field in the disk interior is $\bar{\lambda}\sim 50-150\mum$ and anisotropy degree $\gamma_{\rm rad}\approx 0.3-1$. Assuming the gas-dust thermal equilibrium (i.e., $T_{d}(r)=T_{\gas}(r)$), the radiation strength in the disk-plane can be estimated using the gas temperature from Equation (\ref{eq:Tgas_disk_R}):
\bea
U(r)=\left(\frac{T_{\gas}(r)}{16.4\K}\right)^{6}=U_{0}\left(
\frac{r}{\AU}\right)^{-6\beta},\label{eq:Ur}
\ena
where $U_{0}=(T_{0}/16.4\K)^{6}\simeq 5.9\times 10^{5}(T_{0}/150\K)^{6}$ for silicate grains, and the weak dependence of the grain temperature on its size is disregarded for simplicity (see \citealt{2011piim.book.....D}).

Plugging $n_{\H}(r)$ and $U(r)$ into Equation (\ref{eq:amin_aJ_RAT}), one obtains the minimum size for efficient internal alignment by inelastic relaxation for grains at high$-J$ attractors by RATs,
\bea
a_{\min,aJ}^{\rm RAT,high-J}(\rm iER)&\simeq&889.6\left(\frac{\mu_{8}Q_{3}}{\hat{\rho}^{2}}\frac{g'\Gamma_{\|}}{s^{5/2}}\right)^{1/6}\left(\frac{n_{0,13}}{T_{0,2}}\right)^{1/6}\nonumber\\
&\times&\left(\frac{\bar{\lambda}}{100\mum}\right) \left(\frac{\gamma_{\rm rad} U_{0,6}}{n_{0,13}T_{0,2}}\right)^{-1/2}\nonumber\\
&\times&\left(\frac{r}{\AU}\right)^{2(-\alpha_{n}+4\beta)/3}\mum,\label{eq:amin_aJ_RAT_disk}
\ena
where $2(-\alpha_{n}+4\beta)/3=-0.62$ for $\alpha_{n}=3/2+8/7$ and $\beta=1/2$, which decreases with increasing the disk radius. 

Similarly, the maximum size of internal alignment by inelastic relaxation is (see Equation \ref{eq:amax_aJ_RAT})
\bea
a_{\rm max,aJ}^{\rm RAT,high-J}(\rm iER)&\simeq&0.081\left(\frac{\sqrt{\hat{\rho}}}{\mu_{8}Q_{3}}\frac{s^{1/2}}{g'\Gamma_{\|}}\right)^{1/3}\left(\frac{T_{0,2}}{n_{0,13}}\right)^{1/3}
\nonumber\\&\times&\left(\frac{\bar{\lambda}}{100\mum}\right) \left(\frac{\gamma_{\rm rad} U_{0,6}}{n_{0,13}T_{0,2}}\right)\nonumber\\
&\times&\left(\frac{r}{\AU}\right)^{4(\alpha_{n}-4\beta)/3}\mum,\label{eq:amax_aJ_RAT_disk}
\ena
where $4(\alpha_{n}-4\beta)/3=1.24$, indicating an increase with the disk radius $r$.

The above equations reveal that there exists no satisfactory range of grain sizes that can have fast inelastic relaxation at $r=1\AU$ because $a_{\max,aJ}<a_{\min,aJ}$, assuming the normalized parameters. Reducing the values of $\mu Q$ can increase the efficiency of inelastic relaxation, as implied by Equations (\ref{eq:amin_aJ_RAT_disk}) and (\ref{eq:amax_aJ_RAT_disk}).

\begin{figure}
    \includegraphics[width=0.5\textwidth]{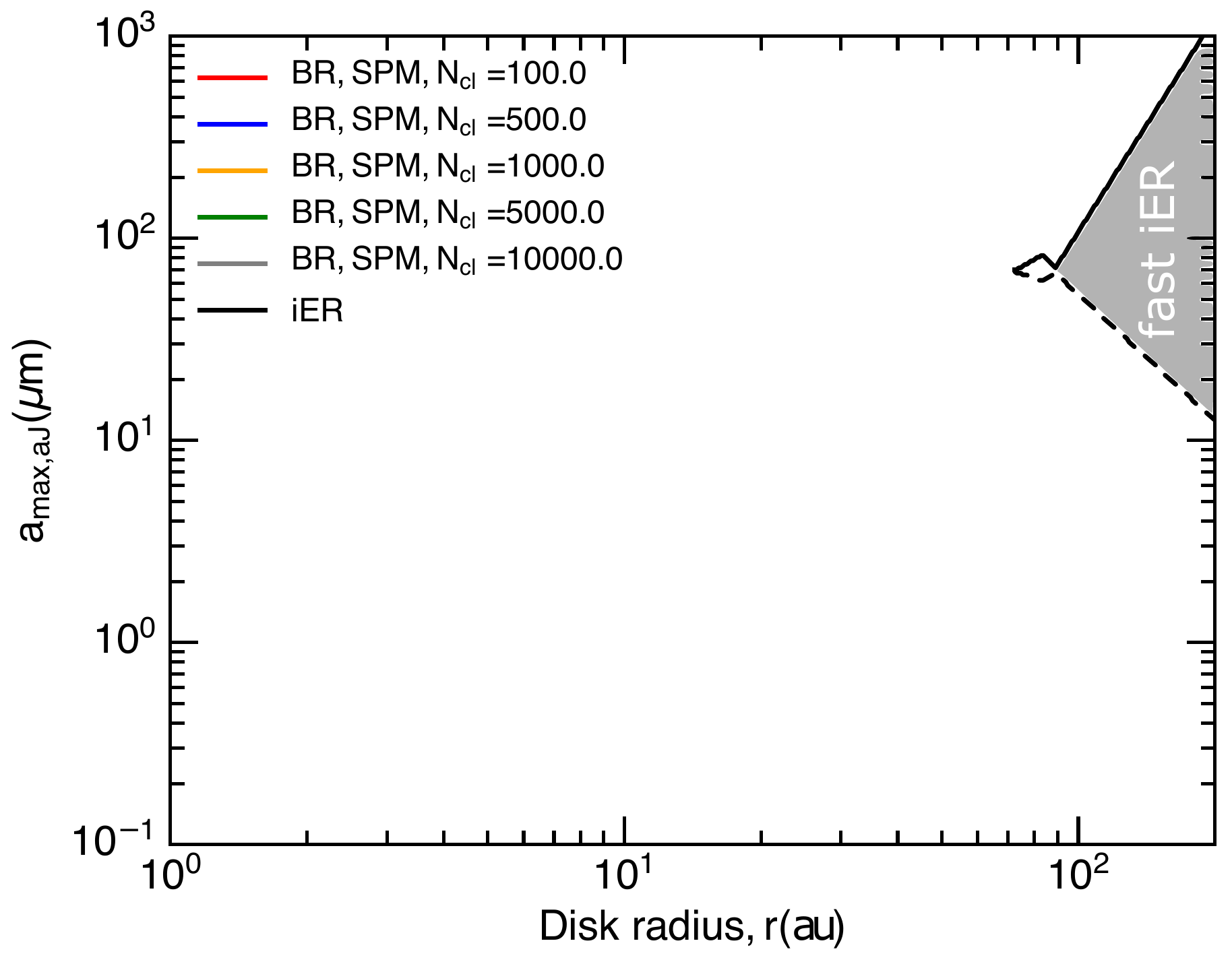}
\caption{Variation of the minimum (dashed line) and maximum size (solid line) for internal alignment by super-Barnett and inelastic relaxation for grains aligned at high-J attractor by RATs for a protostellar disk with the surface density of $\Sigma_{0}=10^{2}\g\cm^{-2}$ and $\mu Q=10^{8}\erg\cm^{-3}$. Inelastic relaxation is efficient for $r>100\AU$ (shaded area), whereas super-Barnett relaxation is inefficient in the entire disk midplane.}  
\label{fig:amax_aJ_RAT_disk}
\end{figure}

We can now calculate the maximum size for internal alignment by super-Barnett relaxation using Equation (\ref{eq:amax_aJ_BR}). For a more massive disk of $\Sigma_{0}=10^{3}\g\cm^{-3}$, we found that both super-Barnett relaxation and internal relaxation is inefficient in the whole disk. For a thinner disk with the surface mass density $\Sigma_{0}=100\g\cm^{-3}$ and low values of $\mu Q=10^{8}\erg\cm^{-3}$, inelastic relaxation can be efficient for VLGs beyond $r>100\AU$, as shown in Figure \ref{fig:amax_aJ_RAT_disk}, but super-Barnett relaxation is still inefficient. 

For external alignment, plugging $U$ and $n_{\H}(r)$ into Equation (\ref{eq:aali_RAT}), one obtains the minimum alignment size by RATs,
\bea
a_{\rm align}^{\rm RAT}(r)&\simeq&5.4\left(\frac{\bar{\lambda}}{100\mum}\right)^{4/7} \left(\frac{\gamma_{\rm rad} U_{0,6}}{n_{0,13}T_{0,2}}\right)^{-2/7}\nonumber\\
&\times&\left(\frac{r}{\AU}\right)^{-2(\alpha_{n}-5\beta)/7}\mum,~~~\label{eq:aali_RAT_disk}
\ena
where $-2(\alpha_{n}-5\beta)/7=-0.14$, which implies the decrease of grain alignment by RATs with increasing the radius.

To understand if grain alignment occurs via $k-$RAT or $B-$RAT, plugging $n_{\H}(r)$, $T_{\rm gas}(r)$ and $B(r)$ into Equation (\ref{eq:amin_Jk_lowJ}), one the minimum size for the $k-RAT$ (or maximum size for $B-$RAT) alignment at low$-J$ attractors,
\bea
a_{\min,Jk}^{\rm RAT,low-J}(r)&\equiv&a_{\rm max,JB}^{\rm RAT,low-J}\simeq49.2 \hat{s}^{1/9}\left(\frac{100\mum}{\gamma_{\rm rad}\bar{\lambda}\hat{Q}_{e3}U_{0,6}}\right)^{2/3}\nonumber\\
&\times&\left(\frac{N_{\rm cl,4}\phi_{sp,-2}\hat{p}^{2}B_{0,6}}{\hat{\rho}}\right)^{2/3}\left(\frac{r}{\AU}\right)^{(4\beta-11/12)}\mum,~~~\label{eq:amin_Jk_lowJ_disk}
\ena
which follows that VLGs can have the $k-$RAT alignment, especially grains with small iron clusters of $N_{\rm cl}\lesssim 10$.

Similarly, using Equation (\ref{eq:amin_Jk_lowJ_disk}), one obtains the minimum size for the $k-RAT$ (or maximum size for $B-$RAT) alignment at high$-J$ attractors,
\bea
a_{\min,Jk}^{\rm RAT,high-J}(r)&\equiv&a_{\rm max,JB}^{\rm RAT,high-J} \simeq4.3\times 10^{5} \hat{s}^{1/6}\left(\frac{1}{n_{0,13}T_{0,2}^{1/2}\hat{Q}_{e3}}\right)\nonumber\\
&\times&\left(\frac{N_{\rm cl,4}\phi_{sp,-2}\hat{p}^{2}B_{0,6}}{\hat{\rho}}\right)\left(\frac{r}{\AU}\right)^{\alpha_{n}+\beta/2-11/8}
\mum,~~~\label{eq:amin_Jk_highJ_disk}
\ena
where $\alpha_{n}+\beta/2-11/8=1.48$, which increases rapidly with increasing the disk radius.

The above equations suggest that for grains with iron inclusions, for grains aligned at the low$-J$ attractor, the $k-$RAT alignment can occur for $a>a_{\rm min,Jk}^{\rm RAT,low-J}\sim 10.6, 49.4\mum$, even with large iron clusters of $N_{\rm cl}=10^{3}$ and $10^{4}$, respectively. However, the $k-$RAT alignment at high$-J$ attractors requires huge grains of $a>40\cm$ for $N_{\rm cl}=10^{4}$ and $a>430\mum$ for $N_{\rm cl}\sim 10$. Large paramagnetic (i.e., $N_{\rm cl}\sim 1$) grains of $a>43\mum$ can have the $k-$RAT alignment. The difference arises from the fact that the radiation precession is much slower when the grain rotates faster (see Eq.\ref{eq:tauk}). Therefore, VLGs with large iron inclusions aligned at high-J attractors can experience the $B-$RAT alignment, while those at low-J attractors experience the $k-$RAT alignment. 

\subsection{Grain drift, METs, and grain alignment}
In protostellar disks, dust grains have Keplerian azimuthal motion while gas follows sub-Keplerian pressured-supported disk because the gas experiences the gas pressure but the dust does not (\citealt{Takeuchi.2002}). Therefore, dust moves faster than the gas and experiences the head-wind.

For a disk with the central protostar of mass $M_{\star}$, the Keplerian angular velocity is $\Omega_{K}=(GM_{\star}/r^{3})^{1/2}$, and the Keplerian tangential velocity becomes
\bea
v_{K}&=&\Omega_{K} r=\left(\frac{GM_{\star}}{r}\right)^{1/2}
\nonumber\\
&=&2.98\times 10^{6}\left(\frac{M_{\star}}{M_{\odot}}\right)^{1/2} \left(\frac{r}{\AU}\right)^{-1/2}\cm\s^{-1}.\label{eq:vK}
\ena

Dust grains orbit the central protostar at the Keplerian velocity $v_{K}$. The gas moves at a sub-Keplerian velocity so that its velocity relative to the dust can be written as $v_{\rm gas}=\eta v_{K}$ with the coefficient $\eta<1$. The gas relative velocity has the radial and azimuthal components,
\bea
v_{r}=\frac{2{\rm Stk}}{1+{\rm Stk}^{2}}\eta v_{K}\label{eq:v_r_disk}\\
v_{\phi}=\frac{{\rm Stk}^{2}}{1+{\rm Stk}^{2}}\eta v_{K},\label{eq:v_phi_disk}
\ena
where ${\rm Stk}$ is the Stokes number (\citealt{kk7}), and the total dust-gas relative velocity is $v_{d}=(v_{r}^{2}+v_{\phi}^{2})^{1/2}$. 

For high Stokes numbers of ${\rm Stk}\gg 1$, the azimuthal component dominates, and $v_{d}\approx v_{\phi}\approx \eta v_{K}=v_{d0} (r/\AU)^{-1/2}$. The grain drift parameter is then 
\bea
s_{d}(r)=\frac{v_{d0}}{v_{T}}\left(\frac{r}{\AU}\right)^{-1/2}=s_{d0}\left(\frac{r}{\AU}\right)^{(\beta-1)/2},~~~\label{eq:sd_R}
\ena
where $s_{d0}$ is the drift parameter at $r=1\AU$, and $s_{d}$ decreases with $r$.
Comparing $s_{d}(r)$ with the critical drift for inelastic relaxation from Equation (\ref{eq:sd_cri_ine}), 
\bea
s_{\rm cri,aJ}&\simeq& 1.02 Q_{\rm spinup,-3}^{-1/2}\left(\frac{\mu_{8}Q_{3}}{\hat{\rho}^{2}}\right)^{1/6}\left(\frac{n_{13}}{T_{2}}\right)^{1/6}\nonumber\\
&\times&\left(\frac{g'(\hat{s})\Gamma_{\|}(\hat{s})^{4}}{\hat{s}^{11/2}}\right)^{1/6},~~~\label{eq:sd_cri_aJ_disk}
\ena
which implies that the grain drift is not enough to induce inelastic relaxation.

Plugging $s_{d}(r)$ into Equation (\ref{eq:tiner_tgas_MET}) one obtains
\bea
\frac{\tau_{\rm iER}}{\tau_{\rm gas}}&\simeq& 6500s_{d0,-1}^{-6}Q_{\rm spinup,-3}^{-3}\frac{\mu_{8}Q_{3}}{\hat{\rho}^{2}}\frac{n_{0,13}}{T_{0,2}}\frac{g'\Gamma_{\|}^{4}}{s^{11/2}}\nonumber\\
&\times&\left(\frac{r}{\AU}\right)^{3-2\beta-\alpha_{n}},~~~\label{eq:tiner_tgas_disk}
\ena
which yields
\bea
r&<&r_{\max,aJ}^{\rm MET}({\rm iER})\nonumber\\
&\simeq&3\times 10^{-6}Q_{\rm spinup,-3}^{6}\left(\frac{s_{d0}}{0.5}\right)^{12}\left[\frac{\hat{\rho}^{2}}{\mu_{8}Q_{3}}\frac{T_{0,2}}{n_{0,13}}\frac{s^{11/2}}{g'\Gamma_{\|}^{4}}\right]^{2}\AU,~~~\label{eq:rmin_iER_MET}
\ena
where $3-2\beta-\alpha_{n}=-0.5$, which implies that METs cannot induce the inelastic relaxation for $s_{d0}=0.5$. Therefore, inelastic relaxation by METs is inefficient to induce efficient internal alignment of grains in the disk.

\begin{figure*}
\includegraphics[width=0.5\textwidth]{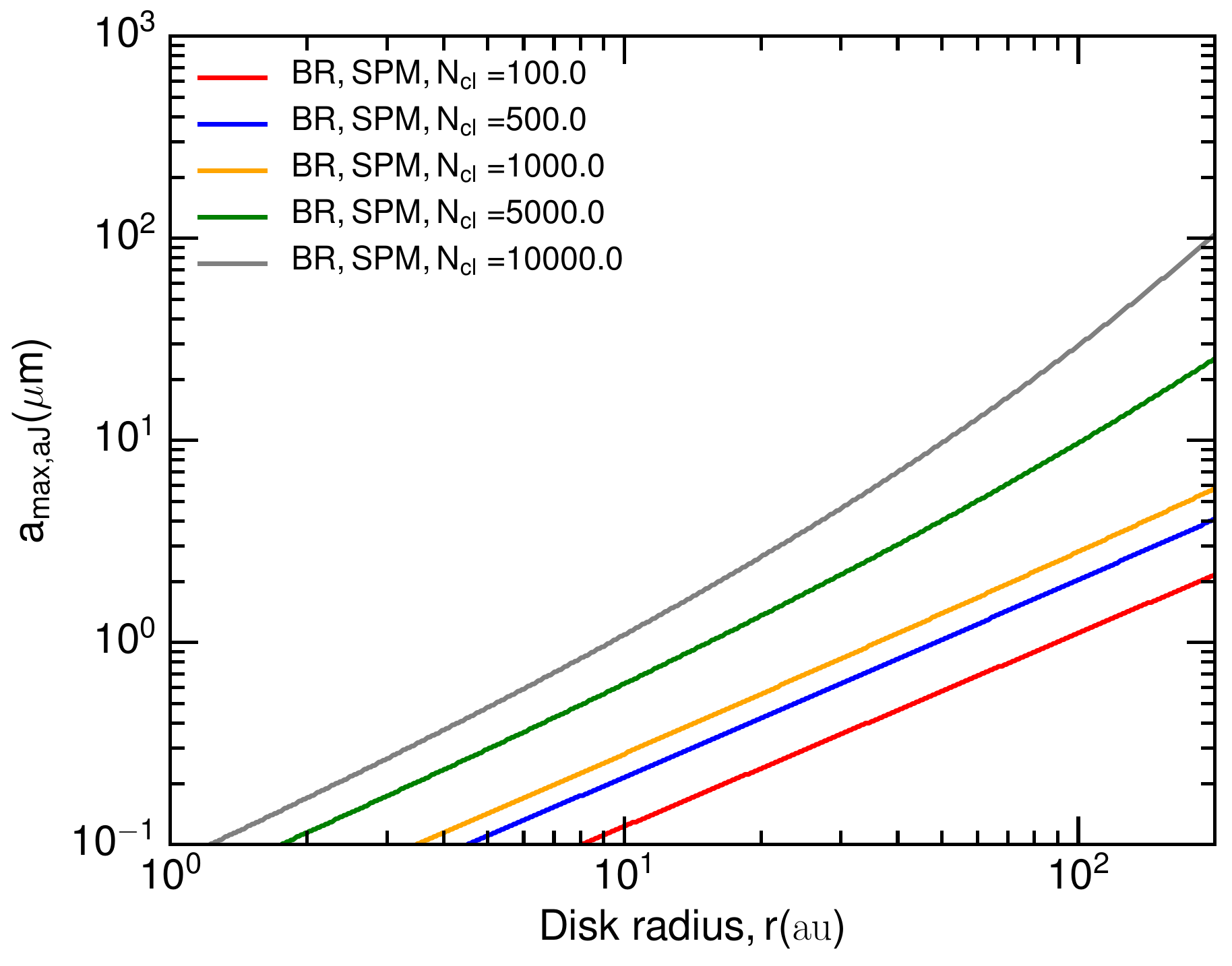}
\includegraphics[width=0.5\textwidth]{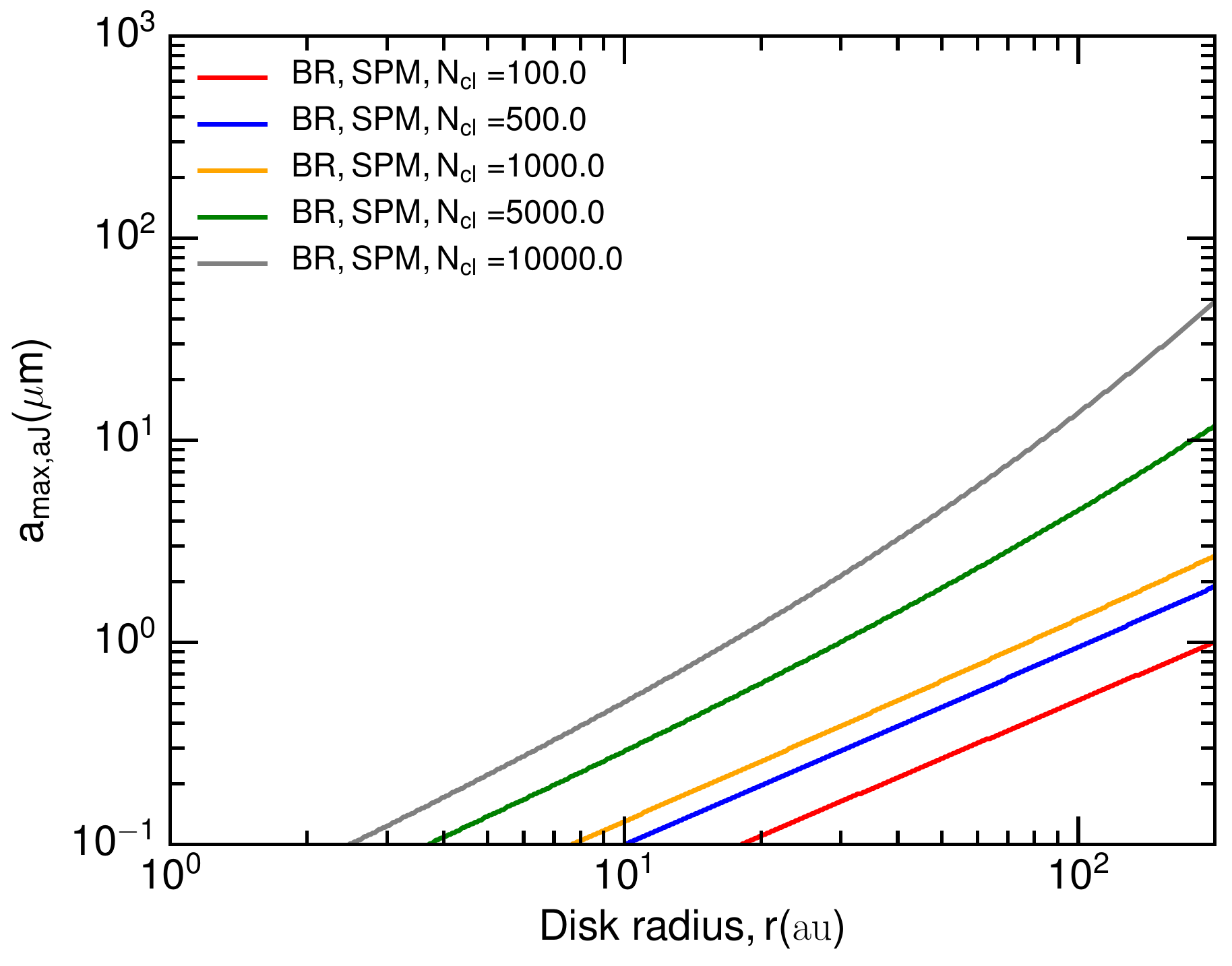}
\caption{Maximum size for internal alignment by super-Barnett relaxation for $\Sigma_{0}=100$ (left) and $\Sigma_{0}=10^{3}\g\cm^{-2}$ (right), assuming grains spun up by METs with $s_{d}=0.25$. The maximum size of internal alignment increases with the size of iron inclusions ($N_{\rm cl}$), but decreases with increasing $\Sigma_{0}$ due to stronger gas damping.}
\label{fig:amax_aJ_sd025}
\end{figure*}

Now, we calculate the maximum size of internal alignment by super-Barnett relaxation effect by first plugging $n_{\H}(r), T_{\gas}(r), s_{d}(r)$ into Equation (\ref{eq:S_MET}) to obtain $\St_{\rm MET}$. Then, plugging $\St_{\rm MET}$ and $B(r)$ into Equation (\ref{eq:tBR_tgas}) and solve for the solution of the equation $\tau_{\rm BR}/\tau_{\rm gas}=1$. The results are shown in Figure \ref{fig:amax_aJ_sd025} for two values of the disk surface mass $\Sigma_{0}$. The maximum size of internal alignment is larger for the lower $\Sigma_{0}$ due to weaker gas damping. VLGs of $a>10\mum$ can be internally aligned at $r>50\AU$, and only smaller grains can be aligned in the inner disk, assuming grains of large iron inclusions with $N_{cl,4}=1$.

\begin{figure*}
\includegraphics[width=0.5\textwidth]{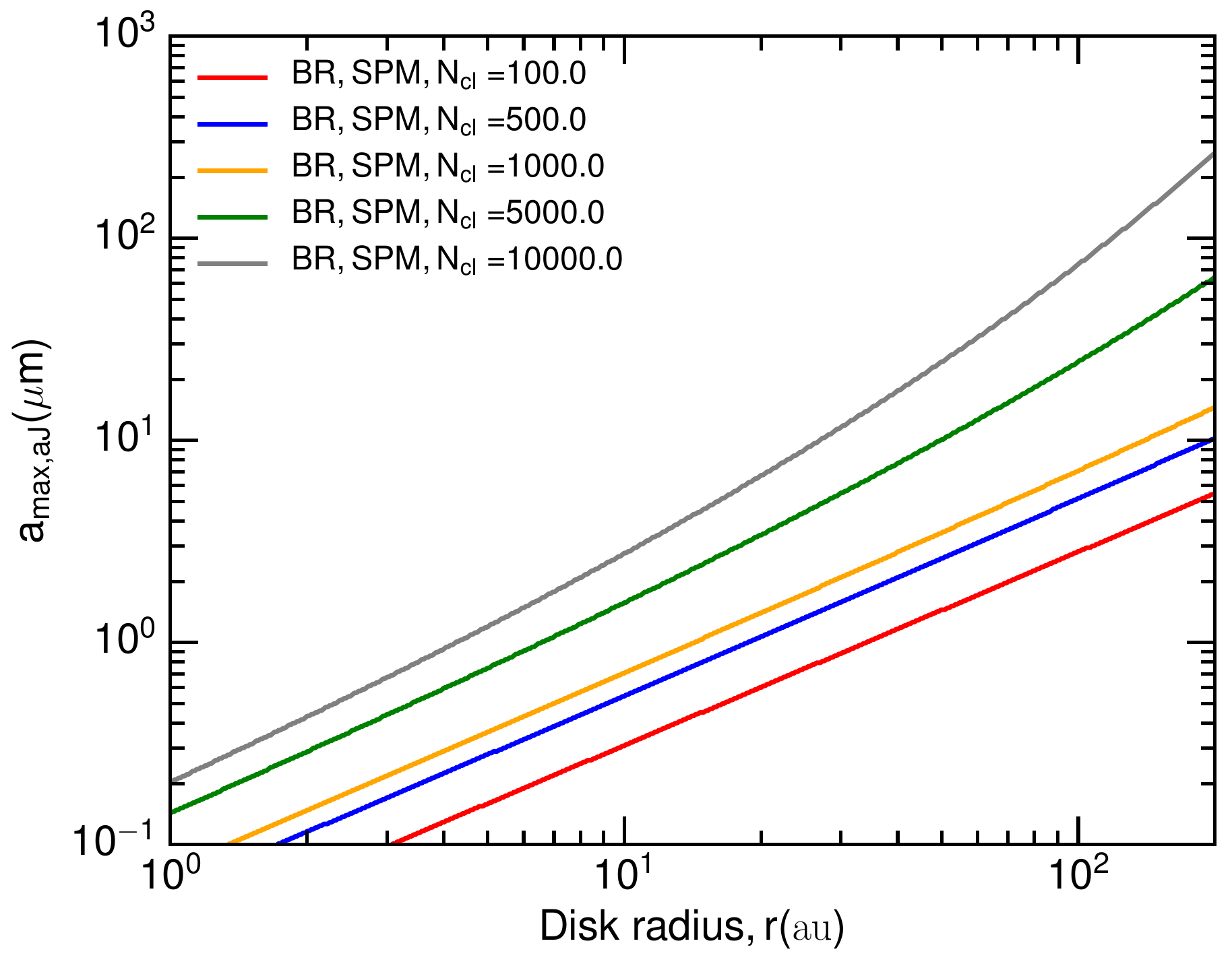}
\includegraphics[width=0.5\textwidth]{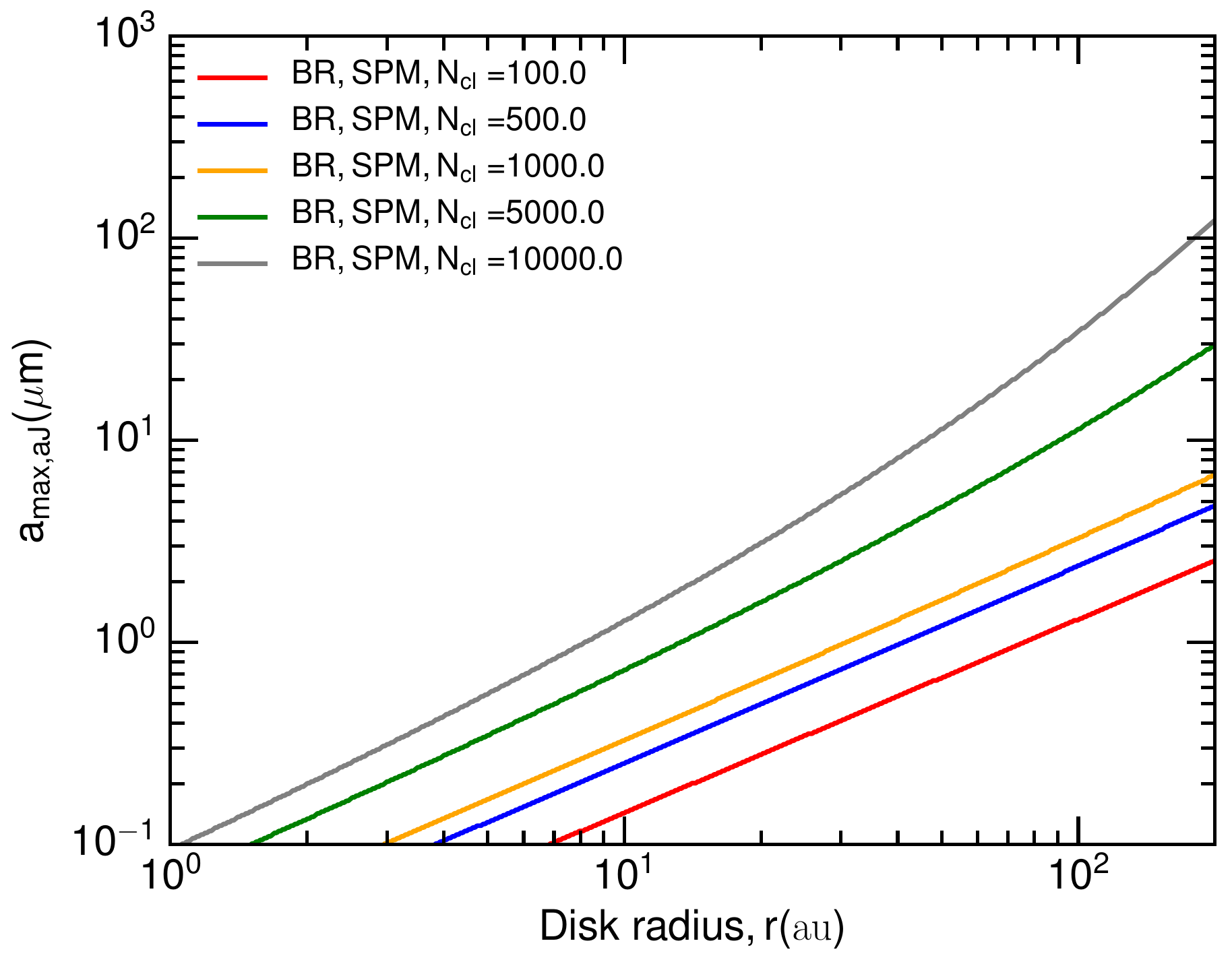}
\caption{Same as Figure \ref{fig:amax_aJ_sd025} but for $s_{d}=0.5$. The maximum size for internal alignment is increased significantly due to stronger METs, and the $\gtrsim 100\mum$ grains with large iron clusters can be internally aligned at $r>100\AU$.}
\label{fig:amax_aJ_sd05}
\end{figure*}

Figure \ref{fig:amax_aJ_sd05} shows the same results as Figure \ref{fig:amax_aJ_sd025} but for a larger drift parameter ($s_{d}=0.5$). The stronger METs increase the maximum size of internal alignment because of the increase of the super-Barnett relaxation with $\St_{\rm MET}$. VLGs of large iron clusters with size $a>10\mum$ can have efficient internal relaxation beyond $r>20\AU$, and VLGs of $a>100\mum$ can be aligned at $r>100\mum$. However, mm-sized grains with large iron inclusions still have slow internal relaxation, although they can be spun up by METs to suprathermal rotation.

To understand whether grain alignment occurs with $\bJ$ along the magnetic field ($B-$MET) or the drift direction ($v$-MET), we plug $n_{\H}, T_{\rm gas}$ and $B$ into Equation (\ref{eq:amin_Jv_highJ}) and obtain the critical size for the $v$-MET alignment,
\bea
a_{\rm min,Jv}^{\rm MET,high-J}(r)&=&4230\hat{\rho}^{-1/2}\hat{s}^{1/2}\frac{Q_{\rm spinup,-3}}{Q_{\rm prec,-1}}\frac{N_{\rm cl,4}\phi_{sp,-2}\hat{p}^{2}B_{0,6}}{n_{0,13}T_{0,2}^{1/2}}\nonumber\\
&\times&\left(\frac{r}{\AU}\right)^{\alpha_{n}+\beta/2-11/8}\mum,~~~\label{eq:amin_Jv_disk}
\ena
where $\alpha_{n}+\beta/2-11/8=1.48$, so that VLGs of $a>423,4230 \mum$ can experience the alignment with $\bJ$ along the drift direction for $N_{\rm cl}=10^{3}, 10^{4}$, respectively. For grains with small iron inclusions of $N_{\rm cl}<10$, the $v-$MET alignment can occur for $a>4.23\mum$. Comparing to $a_{\rm min, Jk}$, one can see that $v-$MET can occur for smaller grains than the $k-$RAT.

We now estimate the minimum alignment size by METs using Equation (\ref{eq:aali_MET}),
\bea
a_{\rm align}^{\rm MET}(r)\simeq0.013\hat{\rho}^{1/3}s_{d0,-1}Q_{\rm spinup,-3}^{-2/3}\left(\frac{r}{\AU}\right)^{-2(\beta-1)/3}\mum.~~~~\label{eq:aali_MET_disk}
\ena
Comparing with $a_{\rm align}^{\rm RAT}$ (Eq. \ref{eq:aali_RAT_disk})
one can see that METs are more efficient than RATs in aligning grains in the disk mid-plane for the typical parameters of $Q_{\rm spinup,-3}=1$.

\section{Discussion}\label{sec:discuss}

\subsection{Comparison of RATs and METs in protostellar environments}
Modern grain alignment theory implies that both RATs (METs) can induce three fundamental effects on the grain dynamics, including grain alignment, spin-up/spin-down, and precession of the grain angular momentum around the radiation (gas flow)\citep{2007MNRAS.378..910L,2007ApJ...669L..77L,2018ApJ...852..129H}. For protostellar environments containing large grains and strong radiation field (gas flow), strong RATs (METs) can rapidly align the grain angular momentum along the radiation direction, $\kv$ (gas flow $\bv$), i.e., axis of grain alignment, with a fraction of grains at a high$-J$ attractor and the rest at the lower$-J$ attractor \citep{Hoang:2008gb,2018ApJ...852..129H}. In the presence of an ambient magnetic field, the grain experiences also Larmor precession around $\Bv$. Therefore, the axis of grain alignment is then determined by the direction ($\kv$, $\bv$ or $\Bv$) around which the grain precession rate is greatest \citep{2019ApJ...883..122L}. We have derived the general formulae for the critical sizes for grain alignment with the magnetic field, radiation direction, and gas flow, for grains with different iron inclusions in protostellar environments (see Section \ref{sec:extalign}).

When the grain is aligned at a low$-J$ attractor, the radiation (mechanical) precession is then determined by the radiation intensity (gas flux) and the precession component of RATs (METs) efficiency. However, when the grain is rotating at a high$-J$ attractor, the radiation (mechanical) precession rate is independent of the radiation intensity (flow speed) and only determined by the ratio of the magnitude of precession component to spin-up components, $Q_{\rm prec}/Q_{\rm spinup}$, depends on the grain properties and radiation field. 
Following the Analytical MOdel (AMO), the component $Q_{e3}$ is determined by the interaction of the photon/gas flow with the massive spheroidal body, which then weakly depends on the whole grain properties. However, the component $Q_{\rm spinup}$ depends on the grain properties. For RATs, the variation of $Q_{\rm spinup}$ with grain shapes is rather small, within an order of magnitude. However, for METs, $Q_{\rm spinup}$ can vary by three to four orders of magnitudes, and are about two orders of magnitude lower than the precession component \citep{2016MNRAS.457.1958D,Reissl.2022}. Therefore, the ratio of the MET components is $Q_{\rm prec}/Q_{\rm spinup}\gg 1$. This has an important implication for external alignment of $\bJ$ along the drift direction instead of along the B-field because the axis of alignment depends on the mechanical precession rate, which is proportional to $Q_{\rm prec}/Q_{\rm spinup}$ (see Eq. \ref{eq:tauv_tauB}) when grains are rotating suprathermally by METs. Moreover, the radiative suprathermal number increases with the grain size as $\St_{\rm RAT}\sim a^{1/2}$ for large grains of $a>\bar{\lambda}/2$, whereas the mechanical suprathermal number increases rapidly with the grain size as $\St_{\rm MET}\propto a^{3/2}$. Furthermore, the radiative suprathermal number decreases with the gas density (Eqs. \ref{eq:S_RAT1} and \ref{eq:S_RAT2}), but the mechanical suprathermal number does not depend on the gas density (Eq. \ref{eq:S_MET}). As a result, METs can be more important than RATs and dominate the inelastic relaxation and internal alignment of VLGs in very dense regions like protostellar disks. Detailed modeling of dust polarization by the MET alignment will be presented elsewhere. 


\subsection{Effects of iron inclusions on internal and external alignment}\label{sec:f_high_J}
Observations reveal that about $90\%$ of the Fe abundance is locked in dust \citep{2009ApJ...700.1299J,2016ApJ...825..136D}. Although the form of Fe in dust is unclear, one expects that a fraction of Fe is in the forms of metallic (Fe) and iron oxide (FeO, Fe$_2$O$_3$, Fe$_{3}$O$_{4}$) nanoparticles. The existence of iron inclusions is expected for dust in protostellar environments. Interestingly, iron nanoparticles are reported to be present in the local interstellar dust grains captured by the {\it Cassini} mission \citep{Altobelli:2016dl} and in primitive interplanetary dust \citep{Hu.2021yit}. 

Iron inclusions in dust increase significantly the grain magnetic susceptibility, which increases the rate of Barnett relaxation (see Eq.\ref{eq:tauBar_sup}) and IA. Therefore, the range of grain sizes with efficient IA is determined by the level of iron inclusions locked in the dust (see Eq. \ref{eq:amax_aJ_BR}). 

Iron inclusions also increase the efficiency of external alignment of $\bJ$ with the magnetic field through two following effects. First, the enhanced magnetic susceptibility increases the Larmor precession (wrt. radiative precession) and thus extends the range of grain sizes that have the $B-$RAT ($B-$MET) alignment. Second, iron inclusions increase the rate of magnetic relaxation, which enhances the fraction of grains that can be aligned at high-J attractors, $f_{\rm high-J}$ \citep{2016ApJ...831..159H}. Therefore, the degree and pattern of thermal dust polarization depends sensitively on the level of iron locked in the dust. 

\cite{2016ApJ...831..159H} quantified the effect of iron inclusions embedded in dust grains on external alignment and found that iron inclusions can increase $f_{\rm high-J}$ to $100\%$ for $\delta_{\rm mag,sp}\gtrsim 10$, which is known as super-RAT (SRAT) and super-MET (SMET) alignment (see also \citealt{Lazarian.2020}). For the ISM, the gas damping rate is low due to a low gas density ($n_{\H}<10^{5}\cm^{-3}$), so that a small level of iron inclusions can help grains to reach $\delta_{\rm mag,sp}>10$ and produce universal high$-J$ attractors \citep{2016ApJ...831..159H}. However, for protostellar environments of higher gas densities, the value of $\delta_{\rm mag,sp}$ is greatly reduced, assuming the same level of iron inclusions as in the ISM. As a result, the maximum size that grain alignment is affected by superparamagnetic relaxation is reduced to $a_{\rm max,JB}^{\rm mag}\sim 0.56(N_{\rm cl,4}\phi_{\rm sp,-2}\hat{p}^{2}B_{3}^{2})/(\hat{\rho} n_{8}T_{\gas,1}^{1/2})$ for $n_{\H}=10^{8}\cm^{-3}$ (see Eq. \ref{eq:amax_JB_DG}). Therefore, the value $f_{\rm high-J}$ for VLGs is only determined by RATs/METs, as opposed to SRAT or SMET mechanisms. Based on calculations of RATs for an ensemble of Gaussian random shapes, \cite{Herranen.2021} found that the value of $f_{\rm high-J}$ spans from $10-70\%$ for grains aligned only by RATs, in the absence of magnetic relaxation.

\subsection{Internal alignment of VLGs with slow internal relaxation}\label{sec:IA}
Our study in this paper focuses on the determination of the range of grain sizes that have internal relaxation faster than the gas damping that induces the right IA with the grain long axis perpendicular to the magnetic field ($\Bv$), $\kv$ or the gas flow ($\bv_{d}$). Our detailed calculations for protostellar environments show that large grains aligned at low-J attractors likely have slow internal relaxation. VLGs aligned at high-J attractors have high probability of having fast internal relaxation due to the increase of the relaxation rate with the angular momentum. However, some VLGs may have slow internal relaxation, depending on the grain magnetic susceptibility and the angular momentum. The question now is how the axis of grains with slow internal relaxation ($\tau_{\rm INR}>\tau_{\gas}$) align? Addressing this question is important for modeling of dust polarization and interpretation of observational data.

\cite{2009ApJ...697.1316H} first studied the alignment of grains without internal relaxation by RATs and found that strong RATs can rapidly bring $\bJ$ to be aligned with $\Bv$ at high$-J$ attractors on a timescale much shorter than the gas damping $\tau_{\gas}$ ({\it fast} alignment; \citealt{2007MNRAS.378..910L,Lazarian.2020}), and the IA occurs with $\ahat_{1}\|\bJ$ (right IA). However, for grains aligned at low$-J$ attractors, they found that IA can occur with two possible configurations of $\ahat_{1}\|\bJ$ (right IA) or $\ahat_{1}\perp \bJ$ (wrong IA). The discussion is then extended for both RATs and METs by \cite{Lazarian.2020bwy} for carbonaceous grains that have no internal relaxation. However, \cite{2009ApJ...697.1316H}'s study did not account for the effect of gas random collisions on the grain orientation. Here, we discuss the potential effect of gas randomization on alignment of VLGs with slow internal relaxation.

Grains aligned at low$-J$ attractors with (sub)thermal rotation are obviously sensitive to strong randomization by gas collisions \citep{Hoang:2008gb,2016ApJ...831..159H}. In the absence of external torques (e.g., RATs, METs or paramagnetic torques), the grain axis is only affected by gas collisions. Physically, thermal dust-gas collisions induce energy equipartition with the grain rotational energy $\langle J_{\|}^{2}\rangle/2I_{\|}=\langle J_{\perp}^{2}\rangle/2I_{\perp}=kT_{\rm av}/2$ with the average temperature $T_{\rm av}=(T_{d}+T_{\gas})/2$, following the Equipartition Theorem. This corresponds to $\langle J_{\|}^{2}\rangle=h\langle J_{\perp}^{2}\rangle$ with $h>1$, revealing that the angular momentum tends to align with $\ahat_{1}$, i.e., the degree of alignment of the grain axis of maximum inertia ($\ahat_{1}$) with $\bJ$ is of $Q_{X}>0$.
Quantitatively, thermal gas collisions will establish the Maxwellian distribution for the angle $\theta$ between $\ahat_{1}$ and $\bJ$, $f_{\rm MW}(\theta)=h/(4\pi)\times (\cos^{2}\theta+h\sin^{2}\theta)^{-3/2}$ (\citealt{Jones:1967p2924}; \citealt{Hoang:2010jy}). The net degree of IA is $Q_{X,\rm MW}\propto \int f_{\rm MW}(\theta)1/2(3\cos^{2}\theta-1)\sin\theta d\theta$ is still positive due to "inertial non-sphericity", i.e., on average, $\ahat_{1}$ is still parallel to $\bJ$ (right IA), although with the IA degree is rather small of $<10\%$ for grains of axial ratio $s>0.4$ (\citealt{1997ApJ...484..230L}). 

However, the effect of external torques (RATs or METs) can change dramatically the distribution of the angle $\theta$ from the Maxwellian distribution. As shown in \citep{2009ApJ...697.1316H}, if the driving RATs are strong enough, grains can be stably aligned at low$-J$ attractors with two possible configuration of the right or wrong IA. The wrong IA can also be sustained with the help of pinwheel torques \citep{2009ApJ...695.1457H,Lazarian.2020}. On the other hand, for grains at high$-J$ attractors, RATs can rapidly stabilize the alignment of right IA even the grains have no internal relaxation \citep{2009ApJ...697.1316H} if the RAT alignment is faster than the gas randomization ({\it fast} alignment case). Let $f_{\rm right-IA}$ be the fraction of grains at low-J attractors that have the right IA. Therefore, if RATs/METs are very strong, the torques can maintain a fraction $f_{\rm high-J}$ of grains with right IA at high-J attractors and a fraction $(1-f_{\rm high-J})f_{\rm right-IA}$ with the right IA, and $(1-f_{\rm high-J}(1-f_{\rm right-IA})$ with the wrong IA. 

Nevertheless, in protostellar environments with a high gas density, the case of {\it slow} alignment is more common than the {\it fast} alignment, which implies that grains must go through a long period of low-J rotation before reaching high-J attractors \citep{Hoang:2008gb,2016ApJ...831..159H,Lazarian.2020}. Therefore, by the time grains reach a high$-J$ attractor, gas collisions have already significantly alter the IA due to the inefficient internal relaxation. Thus, perfect (right) IA at high$-J$ attractors is only satisfied when the internal relaxation is much faster than the gas damping, which was used in this paper to derive the range of grain sizes that have efficient IA. Grains with slow internal relaxation are expected to have {\it right} IA, but its degree of IA is low due to the gas collisions. The exact value of IA degree depends on the timescale required for slow alignment, which can be from several to $\sim 100\tau_{\gas}$ \citep{Hoang:2008gb,2016ApJ...831..159H,Lazarian.2020}. Comparison of dust polarization from detailed numerical modeling that take into account the above effects of IA to observational data would help to constrain the physics of IA.

\subsection{Grain alignment and tracing magnetic fields with dust polarization toward protostellar cores}
Grains in protostellar cores are subject to both protostellar radiation field and gas-grain drift due to ambipolar diffusion as well as strong rotational damping by gas collisions. We here summarize our mains results of grain alignment by RATs and METs for protostellar cores and discuss implications for tracing magnetic fields via dust polarization. 

\subsubsection{$k-$RAT vs $B-$RAT alignment and $B-$field tracing}
Assuming that grains can be aligned at the low$-J$ and high$-J$ attractors by RATs, we derived the analytical formulae for the critical sizes for IA and external alignment (including the $k-$RAT and $B-$RAT alignment) as functions of the local physical parameters (see Section \ref{sec:radalign}). 

For a protostellar core of typical density $n_{\H}\sim 10^{7}-10^{8}\cm^{-3}$, the interstellar radiation field is heavily reddened \citep{Hoang.2021}. For the typical reddened interstellar radiation field strength of $U\lesssim 1$ and $\bar{\lambda}=10\mum$, our results in Figure \ref{fig:amax_aJ_RAT_disk} show that RATs are insufficient to spin grains up to highly suprathermal rotation and induces efficient inelastic relaxation. However, grains with iron inclusions can have fast super-Barnett relaxation and achieve efficient IA for grain sizes of $a\lesssim 10\mum$ (see left panel of Figure \ref{fig:amax_aJ_RAT_disk}). 

The presence of a central protostar can increase the radiation strength to $U\gtrsim 100$, increasing the suprathermal rotation number $\St_{\rm RAT}\gg 1$ (see Eq. \ref{eq:S_RAT1}). As a result, inelastic relaxation becomes faster, inducing the efficient IA of VLGs upto $a\sim 10^{3}\mum$, whereas super-Barnett relaxation is also efficient for large grains of $a\lesssim 50\mum$ (see the right panels of Figure \ref{fig:amax_aJ_RAT_disk}), as found in \cite{Hoang.2022}. Because mm-sized grains are not expected to be present in protostellar cores, the efficient IA with $\ahat_{1}\perp \bJ$ is expected for all dust grains due to the combination of the super-Barnett and inelastic relaxation effects. 

For external alignment, we found that grains at the low$-J$ attractors can have $k-$RAT alignment in the region with strong protostellar radiation fields (e.g., near the protostar). Since grains aligned at low$-J$ attractors are expected to have a low degree of alignment due to thermal fluctuations, the dust polarization by the $k-$RAT alignment is low. However, large grains aligned at high$-J$ attractors have radiative precession much slower than the Larmor precession because of the increase of $\tau_{k}$ with $J$, which results in two possibilities of $k-$RAT and $B-$RAT alignment, but grains with iron inclusions most likely experience $B-$RAT due to enhanced magnetic susceptibility (see Eq. \ref{eq:amin_Jk_highJ}). Since grains at high$-J$ attractors have a higher degree of alignment due to fast rotation, dust polarization produced by grain alignment at high-J attractors is dominant. For an ensemble of grains with $f_{\rm high-J}$, the dust polarization properties are mostly determined by grains at high$-J$ attractors due to its dominance. Therefore, thermal dust polarization toward protostellar cores can reliably trace magnetic fields thanks to its efficient IA and $B-$RAT alignment.

\subsubsection{$v-$MET vs $B-$MET alignment and B-field tracing}
It is known that ambipolar diffusion is ubiquitous in dense prestellar cores due to low ionization fraction of the gas. This effect can produce the drift velocity to $v_{\rm drift}\sim 0.2-0.3 \km\s^{-1}$ calculated in (\citealt{1995ApJ...453..238R}), corresponding to or $s_{d}\sim 0.4-0.6$ for $v_{T}\sim 0.4T_{\gas,1}^{1/2}\km\s^{-1}$ for which METs can spin-up grains to suprathermal rotation. One can see that the range of available drift parameter $s_{d}$ is greater than the critical value required for the efficient inelastic relaxation for the density of dense cores (see Eq. \ref{eq:sd_cri_ine}). Note that Barnett relaxation still is important for grains of $a<10\mum$. As a result, thanks to METs, both super-Barnett and inelastic relaxation become efficient and induce the efficient IA for grains in protostellar cores. 

For external alignment, in the protostellar cores, METs can align small grains with size $a_{\rm align}^{\rm MET}\sim 0.01\mum$ as given by Equation (\ref{eq:aali_MET}). Grains aligned at high$-J$ attractors can still be efficiently aligned with the magnetic field due to the Larmor precession, i.e., $B-$MET alignment (see Eq. \ref{eq:amax_JB}). Same as RATs, grains at low$-J$ attractors can experience the $v-$MET alignment. Therefore, the net alignment configuration is the right IA of $\ahat_{1}\| \bJ$ and external alignment of $\bJ\|\Bv$, which implies the dominance of $B-MET$ alignment over $v-MET$. As a result, dust polarization from grains aligned by METs can most likely trace B-fields in protostellar cores.

\begin{figure}
\includegraphics[width=0.5\textwidth]{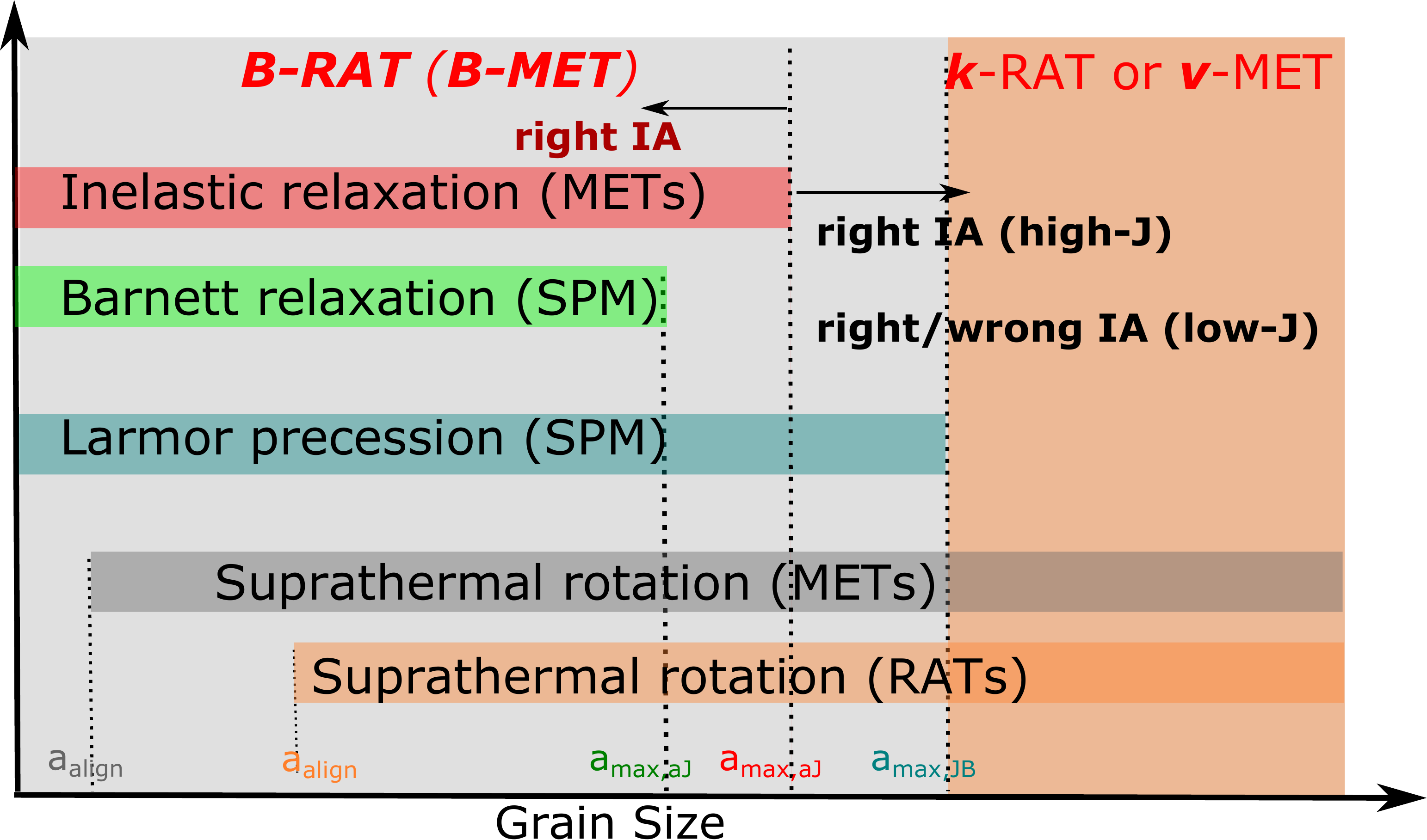}
\caption{Schematic illustration of internal alignment (IA) and external alignment of grains as a function of the grain size expected for protostellar environments, assuming grains are aligned by RATs (METs). Largest grains can experience the $k-$RAT alignment (orange area) with two possible states of right and wrong IA. Smaller grains tend to have the $B-$ RAT (-MET) alignment due to fast Larmor precession, with right IA (gray area) by inelastic relaxation (pink bar) and Barnett relaxation (green).}
\label{fig:alignment_protostar}
\end{figure}

Figure \ref{fig:alignment_protostar} summarizes our understanding of grain alignment for superparamagnetic grains in protostellar environments (see Sections \ref{sec:IntAlign} and \ref{sec:extalign} for relevant formulae). In general, small grains of size $a<10\mum$ tend to align with the magnetic fields by RATs/METs due to fast Larmor precession, whereas VLGs of $a>10\mum$ can align along the radiation direction or the gas flow due to fast radiative/mechanical precession. In such dense protostellar environments, paramagnetic grains, however, cannot align with the magnetic field.

\subsection{Alignment of VLGs in protostellar disks and implications for dust polarization}
Our calculations of critical sizes for grain alignment in Sections \ref{sec:disk} do not require the knowledge on the grain size distribution. However, to discuss the implications of our analysis for observations, we here assume that grains in protostellar disks can be very large of $a\gtrsim 100\mum$ due to grain growth, as revealed by numerous observations \citep{2009ApJ...696..841K,Galametz:2019fj,2014arXiv1402.1354T}.

\subsubsection{$k-$RAT vs. $B-$RAT alignment and dust polarization}
Assuming that grains in the protostellar disk are spun-up by RATs, our results in Section \ref{sec:disk} show that both inelastic relaxation and super-Barnett relaxation are inefficient due to weak radiation field and high gas density for a typical disk of density $\Sigma_{0}=10^{3}\g\cm^{-2}$ (see Eq. \ref{eq:amax_aJ_RAT_disk}). For a thin disk of $\Sigma_{0}=100\g\cm^{-3}$ and low values of $\mu Q$, inelastic relaxation can be efficient only in the outer disk of $r>100\AU$ (see Figure \ref{fig:amax_aJ_RAT_disk}). Therefore, VLGs in protostellar disks do not have efficient internal relaxation, so that both types of right and wrong IA are possible.

For grains at low$-J$ attractors, the RAT alignment can occur with $\bJ$ along $\kv$ for grain size $a>a_{\min,Jk}^{\rm RAT,low-J}$ (Eq. \ref{eq:amin_Jk_lowJ_disk}), even when grains contain large iron clusters (i.e., superparamagnetic material). Large paramagnetic grains of $a>40\mum$ at high$-J$ attractors can have $k-$RAT alignment (see Eq. \ref{eq:amin_Jk_highJ_disk}). However, superparamagnetic grains at high$-J$ attractors cannot align with $\kv$ due to the slow radiative precession, leading to the $B-$RAT alignment. When high$-J$ attractors are absent (i.e., only low$-J$ attractors exist, see Section \ref{sec:RATmodel}), then, the grains have the $k-$RAT alignment and the dust polarization does not trace the magnetic field. Note that the degree of grain alignment at low$-J$ attractors is rather low, leading to the low polarization degree. If high$-J$ attractors are present, the high degree of alignment at high$-J$ attractors dominates the net polarization of dust emission. In this case, dust polarization can trace magnetic fields. 

\begin{figure}
\includegraphics[width=0.5\textwidth]{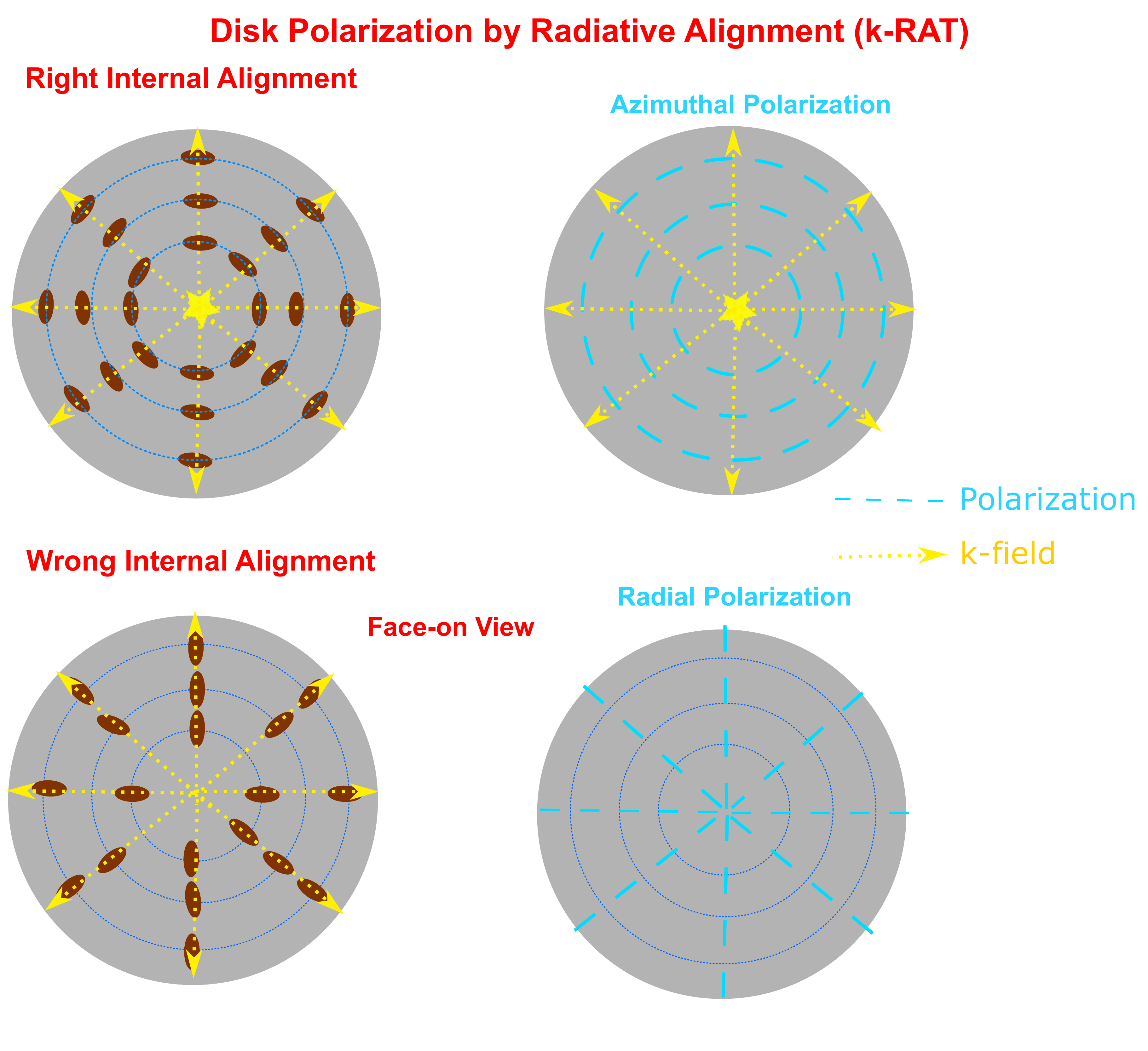}
\caption{Expected polarization pattern of thermal dust emission from aligned grains with $\bJ$ along the radiation direction $\kv$ for the case of right IA (IA, upper) and wrong IA (lower panel).}
\label{fig:pmap_kalign}
\end{figure}

Figure \ref{fig:pmap_kalign} shows the polarization pattern expected from grain alignment in a protostellar disk with $\bJ$ along the radiation direction ($\kv$) for two cases of right IA ($\ahat_{1}\| \bJ$, upper panel) and wrong IA ($\ahat_{1}\perp \bJ$, lower panel) for a disk. The polarization patterns are azimuthal for the right IA and radial for the wrong IA case.

\begin{figure}
\includegraphics[width=0.5\textwidth]{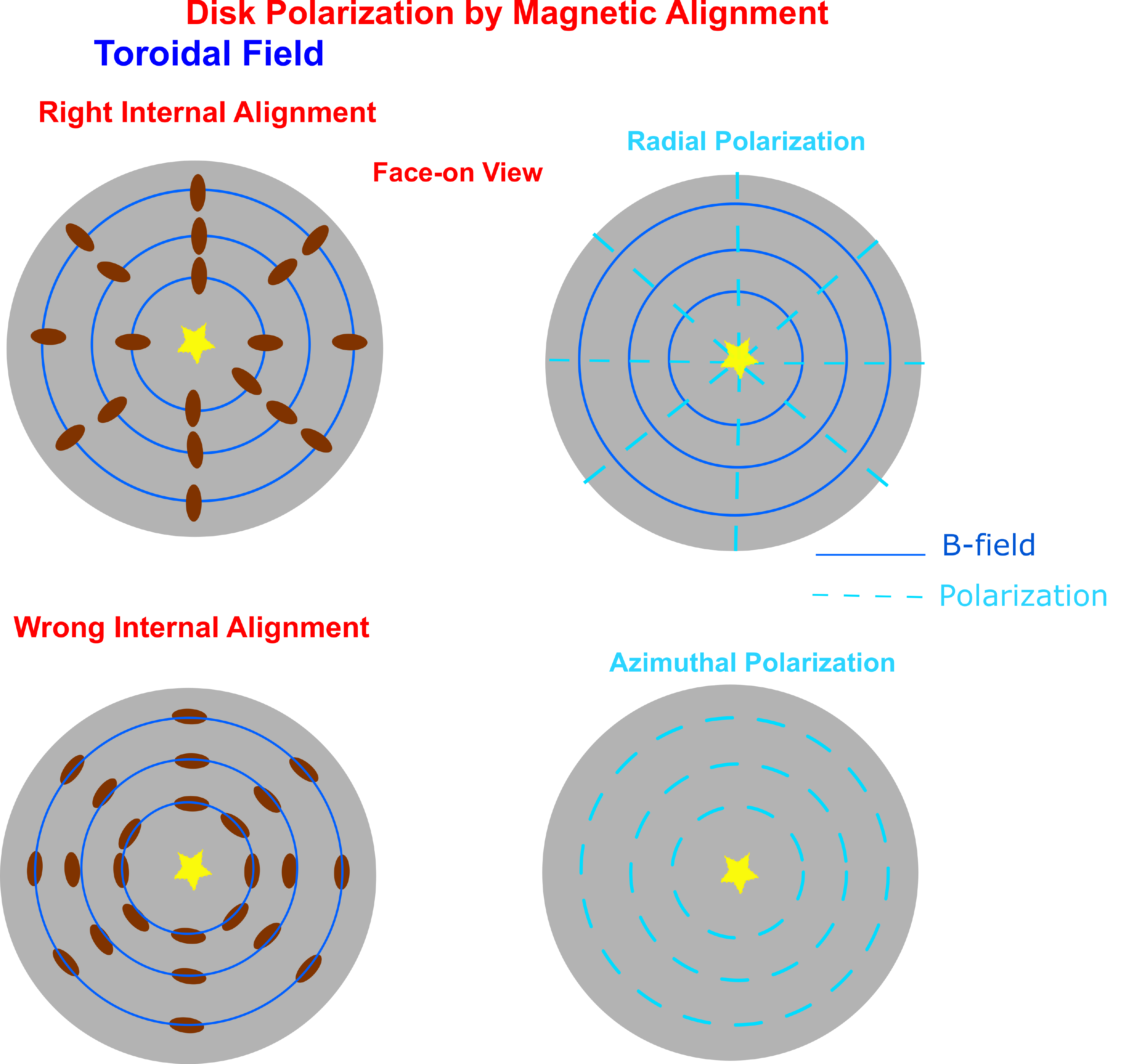}
\caption{Expected polarization pattern of thermal dust emission from grains aligned with $\bJ$ along the magnetic field (magnetic alignment) in a disk with the toroidal magnetic field (blue lines). Both right IA (upper panels) and wrong IA (lower panels) are considered.}
\label{fig:pmap_Balign}
\end{figure}

In Figure \ref{fig:pmap_Balign}, we illustrate the polarization pattern of thermal dust emission induced by grain alignment with $\bJ\|\Bv$ for two cases of right IA (upper panel) and wrong IA (lower panel) for a protostellar disk with the toroidal magnetic field. The polarization patterns are then illustrated in the upper and lower panel respectively. Note that the polarization pattern observed at a given wavelength will be determined by the grain size distribution, iron inclusions, magnetic fields, and radiation and gas flow. If the polarization is dominated by small grains with efficient IA, the polarization degree is high and the pattern is radial (upper right panel). On the other hand, if the dust polarization is dominated by VLGs with inefficient IA, the polarization degree is rather low and the pattern is azimuthal if the wrong IA is present (lower right panel). A detailed modeling of multiwavelength dust polarization from protostellar disks will be presented elsewhere.

\subsubsection{$v-$MET vs. $B-$MET alignment and dust polarization}
Assuming grains in the disk are spun-up by METs, our results in Section \ref{sec:disk} suggest that inelastic relaxation is slower than the gas damping because the gas-grain drift parameter $s_{d}$ is below the critical value required for METs (see Eq. \ref{eq:sd_cri_aJ_disk}). However, VLGs of size $a\sim 10-50\mum$ aligned at high-J attractors can have fast super-Barnett relaxation and thus efficient IA at disk radius $r>50\AU$, provided $N_{cl,4}\sim 1$. However, VLGs of $a>100\mum$ have slow internal relaxation throughout the disk, leading to inefficient IA (see Figures \ref{fig:amax_aJ_METs_U1} and \ref{fig:amax_aJ_METs_U1e4}). For grains aligned at low$-J$ attractors with thermal rotation, internal relaxation is much slower than the gas damping, leading to inefficient IA and possible alignment with long axis parallel to the spinning axis (i.e., wrong IA). Since the net alignment degree for thermal rotation at low$-J$ attractors is rather small, the decisive parameter for describing the dust polarization is the fraction of grains aligned at high$-J$ attractors, $f_{\rm high-J}$ \citep{2016ApJ...831..159H,2019ApJ...883..122L,Herranen.2021}.

For external alignment, our results reveal that grains at high$-J$ attractors can align with $J$ along the drift velocity $\bv_{d}$ if grains have a low level of iron inclusions or ordinary paramagnetic material. On the other hand, grains with large iron inclusions have the $B-$MET alignment due to the fast Larmor precession. Grains at low$-J$ attractors can have the $v-$MET alignment. Therefore, dust polarization can trace magnetic fields if $f_{\rm high-J}$ is considerable (e.g, $f_{\rm high-J}\sim 0.1-0.2$).

\begin{figure}
\includegraphics[width=0.5\textwidth]{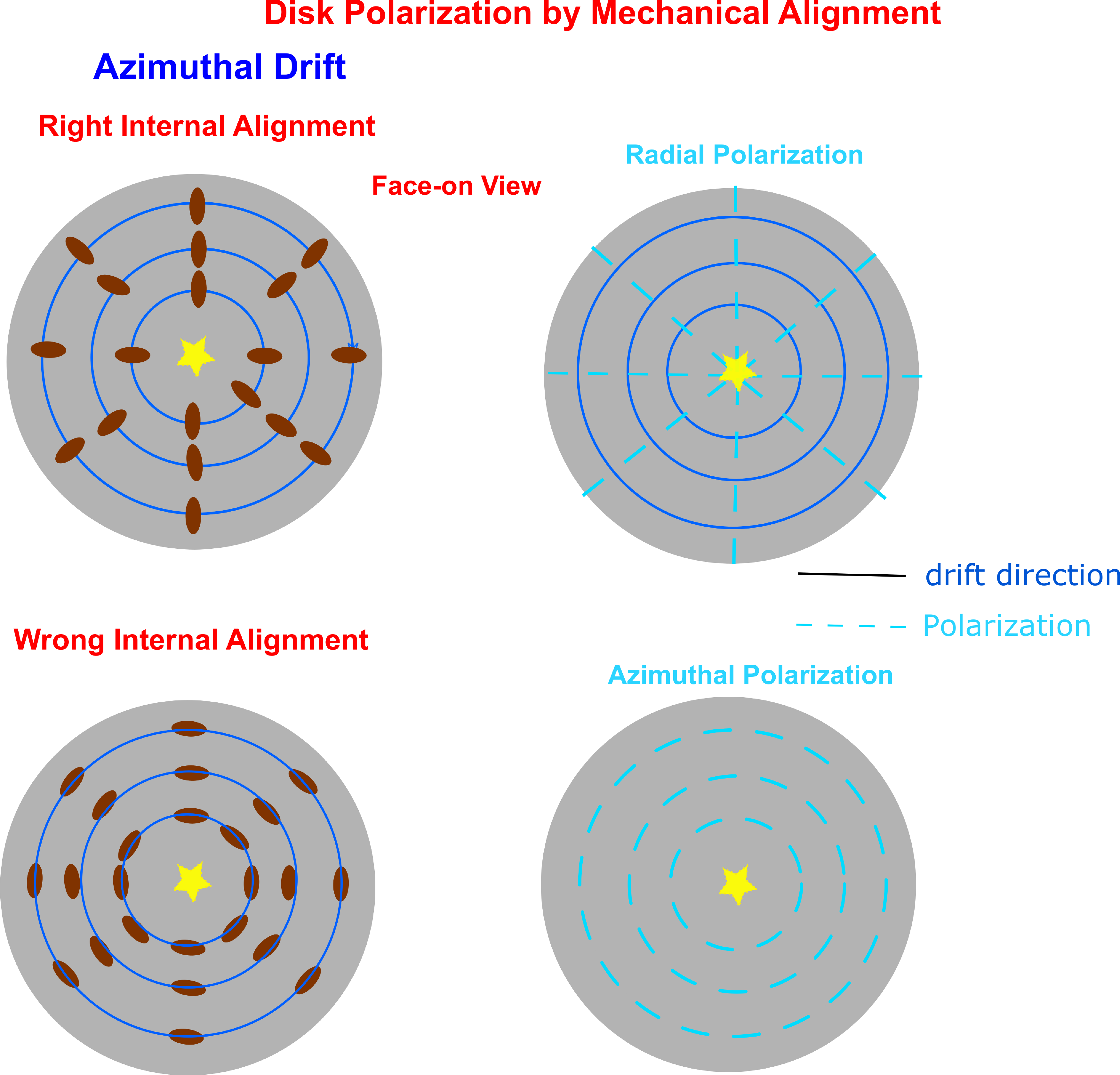}
\caption{Same as Figure \ref{fig:pmap_Balign} but for the alignment with $\bJ$ along the drift direction $\bv_{d}$, which is assumed to be azimuthal due to the Keplerian motion. The polarization pattern is similar to Figure \ref{fig:pmap_Balign}.}
\label{fig:pmap_valign}
\end{figure}

In Figure \ref{fig:pmap_valign}, we show the polarization pattern for $\bJ$ aligned with the drift direction ($\bv_{d}$) assumed to be azimuthal. For the case of right IA, the polarization pattern is radial (upper right panel). For the case of wrong IA, the polarization vector is parallel to the drift $v_{d}$, producing the azimuthal polarization pattern. One can see that in the disk, the polarization pattern by mechanical alignment is similar to magnetic alignment because both the drift direction and magnetic field are azimuthal. The contribution of the inward drifting will causes the drift direction to be spiral, leading to the difference in the polarization pattern from the magnetic alignment. 

Those grains can produce the polarization map as seen in the upper panel of Figure \ref{fig:pmap_valign}. For grains of thermal rotation (grains at low$-J$ attractors), they experience wrong IA and the polarization map is like the lower panel of Figure \ref{fig:pmap_valign}.

We note that the magnetic field in the disk is very uncertain, but it is expected to be dominantly toroidal because of the disk rotation (see e.g., \citealt{Flock.2015}). Thus, the magnetic field aligns with the drift direction if the azimuthal drift component is dominant. In this case, $B-$MET and $v-$MET are the same, and the alignment with $\bJ$ directed along the azimuthal direction. The net polarization angle of dust emission is only characterized by internal relaxation.

Our results in this paper reveal that dust polarization can still trace magnetic fields in the protostellar disk due to the reduction of radiative precession at high$-J$ attractors.

\subsubsection{Previous numerical modeling of grain alignment and dust polarization from disks}
The first detailed modeling of polarized dust emission from circumstellar disks using the early RAT theory \citep{1996ApJ...470..551D,1997ApJ...480..633D} is presented in \cite{2007ApJ...669.1085C} where the perfect IA and external alignment of grains with the magnetic field are assumed for all grains larger than $a_{\rm align}$. Later, using the modern RAT alignment theory \citep{2007MNRAS.378..910L,Hoang:2008gb,2009ApJ...697.1316H,2014MNRAS.438..680H,2016ApJ...831..159H}, \cite{2017ApJ...839...56T} performed numerical modeling of dust polarization and demonstrated the importance of the azimuthal polarization due to the $k-$RAT alignment, assuming the perfect IA (i.e., efficient internal relaxation) for grains at high$-J$ attractors (i.e., {\it right} $k-$RAT) and the right IA for grains at low-J attractors. However, we find that Barnett relaxation is inefficient for large grains without iron inclusions that can experience the $k-$RAT alignment. Inelastic relaxation is also inefficient due to weak radiation fields in the disk. Therefore, the assumption of perfect IA for the $k-$RAT alignment is not valid for all grains, but only valid for small grains. Moreover, \cite{2017ApJ...839...56T} calculated the radiative precession time for grains at low-J attractors with the thermal rotation of $J=J_{T}$ (i.e., $\tau_{k}^{\rm low-J}$) and then compared that timescale with the Larmor precession to determine the critical size for $k-RAT$ alignment even for grains aligned at high$-J$ attractors. However, as shown in this paper, the radiation precession rate at high$-J$ attractors is much slower due to faster rotation, so that large grains with iron inclusions can still have the $B-$RAT alignment. This results in an overestimate of the importance of the $k-$RAT alignment in protostellar disks because $a_{\rm min,Jk}^{\rm RAT,high-J}\gg a_{\rm min,Jk}^{\rm RAT,low-J}$. Therefore, numerical modeling of dust polarization taking our detailed physics presented in this paper is required to accurately predict the polarization map and spectrum and reliably interpret polarization data. A detailed modeling of dust polarization from grains with iron inclusions in protostellar environments will be presented elsewhere (Giang et al. in prep).

\subsection{Implications for interferometric polarization observations to protostellar disks}

The physical condition of protostellar disks is suitable for multiple alignment mechanisms to be active. However, due to the rapid decrease of the gas density with increasing disk radius, each alignment mechanism can be efficient in some specific region. 

In the outer region of radius $r>100\AU$, the gas density is lower, so that the efficient IA can be achieved by super-Barnett relaxation due to spin-up by METs. Moreover, $B-$RAT alignment can occur in this outer region, as given by Figure \ref{fig:amax_JB}. In this case, the polarization pattern is like the upper panel in Figure \ref{fig:pmap_Balign}. In the inner region of $r<100\AU$, IA can be efficient for small grains by super-Barnett relaxation, and MET alignment can occur due to high densities. As the result, the polarization pattern would not change from the outer region to inner region. However, in the inner disk, the polarization does not trace B-fields, but trace the infalling motion of the gas flow. Therefore, the transition from outer to inner disks can be produced. In the intermediate disk radius $10\AU<r<100\AU$, grain alignment may be the contribution of both RATs and METs.

In the disk condition where the gas-dust drift is azimuthal, which is similar to the magnetic toroidal field, the joint action of METs and the Larmor precession can most likely induce the right IA and external alignment with the shortest axis $\ahat_{1}$ parallel to $\bJ\|\Bv$ for small grains. For VLG, one can have wrong IA due to inefficient internal relaxation, such that the alignment can be the long axis parallel to $\bJ\|\Bv$ (see Figure \ref{fig:alignment_protostar}).

Submillimeter Array (SMA) observations to a number of high-mass protostars on sub-parsec scales of $<0.1$ pc are also reported \citep{2014ApJ...792..116Z}. ALMA polarization observations have been conducted toward a large sample of low-mass protostars (Class 0 YSOs) in Perseus (\citealt{2018ApJ...855...92C}) and Ophiuchus clouds \citep{Sadavoy.2019}, showing the different polarization patterns from the envelope to the central region (disk of $\sim$ 100 au scales). 
In particular, ALMA observations toward class I/II YSOs (protoplanetary disks) show different polarization patterns at different wavelengths \citep{Harrison.2019,Aso.2021}. For example, multiwavelength ALMA observations toward HL Tau reveal a striking difference in polarization properties at three wavelengths (875$\mu$m, 1.3 mm, and 3.1 mm; \citealt{Stephens:2017ik,Kataoka:2017fq}), where the polarization pattern changes from being along short disk axis at 870$\mu$m to azimuthal pattern at 3.1 mm. Such an azimuthal polarization was suggested due to polarized emission from aligned grains with $\ahat_{1}\|\bJ\|\kv$ ({\it right} $k-$RAT) instead of $B-$RAT \citep{Kataoka:2017fq}). The ALMA polarization observed toward HD 142527 by \cite{2016ApJ...831L..12K} revealed the variation of the polarization pattern from the north to the south, which is suggested due to the difference in the grain size distributions \citep{Ohashi:2018gl}. 

Our detailed analysis here indicates that the alignment mechanisms are far more complicated in protostellar disks because the case of the $B-$RAT alignment with the wrong IA can also reproduce an azimuthal polarization pattern (see Figure \ref{fig:pmap_Balign}). Furthermore, the case of $k-$RAT with the wrong IA instead produces the radial polarization pattern (see Figure \ref{fig:pmap_kalign}). Moreover, \cite{Guillet.20204q3} noticed that the effect of scattering on Rayleigh regime for VLGs of $a>100\mum$ can cause the polarization parallel to $B$-field. Therefore, detailed modeling of dust polarization accounting for detailed dust physics discussed in this paper and comparison with multiwavelength observational data would help to clarify the main polarization mechanisms. Moreover, circular polarization of dust emission from protostellar disks would be valuable to disentangle the different grain alignment mechanisms, as previously suggested \citep{2007JQSRT.106..225L,2016ApJ...831..159H,2019ApJ...883..122L} and modeled in \cite{Draine.2022}.

\subsection{Constraining iron locked in dust and magnetic fields using k-RAT vs. B-RAT dust polarization}
Our detailed analysis show that the critical sizes of aligned grains by k-RAT depends on the iron level and magnetic field strength. The observed polarization pattern and fraction then will depend on these parameters. Observations of dust polarization can help us to constrain the magnetic properties using k-RAT and B-RAT \citep{2019ApJ...883..122L}.

Our results also show that the minimum size for k-RAT at low-J attractors depends on the radiation strength $U_{0}$. Therefore, the spike in the protostar luminosity due to episodic accretion (see \citealt{Francis.2022}) can change the polarization pattern from B-RAT to k-RAT if the polarization is dominated by grains aligned at low-J attractors. Moreover, the increase of radiation strength extend the range of grain sizes with efficient IA at high-J attractors and facilitate the B-RAT alignment. Observations toward episodic protostars would be valuable to test this transient transition from the RAT alignment paradigm and constrain the fraction of grains with high-J attractors $f_{\rm high-J}$ and grain irregularity.\footnote{The timescale of episodic accretion is much longer than the various timescales involved in grain alignment (see Eqs. \ref{eq:tgas}, \ref{eq:tauk_lowJ}, \ref{eq:tauB}).}

\subsection{Implications for grain growth of aligned grains}
Our results show that VLGs in protostellar disks can still have efficient IA by inelastic relaxation if they rotate suprathermally by METs. This has important effect of grain growth and the internal structure of dust aggregates \cite{Hoang.2022}. If grains are aligned with the short axis along the drift velocity, grain growth by grain-grain collisions would form dust aggregates containing embedded individual grains of parallel short axes. If the alignment with the long axis parallel to the drift velocity, resulting dust aggregates will contain individual grains having parallel long axes. Such features may leave imprints in the structure of cometary dust and interplanetary dust.

We note that, in addition to grain alignment, RATs and METs can induce rotational disruption when grains are spun-up to extremely fast rotation \citep{Hoang:2019da,2020Galax...8...52H,Hoang.2021}. Detailed study on rotational disruption by RATs is presented in \citep{Hoang.2021,Hoang.2021z5d} for protostars and in \citep{Tung:2020ew} for protoplanetary disks. A detailed study on the rotational disruption by METs for protostellar environments is presented elsewhere.

\subsection{Selective effect of RAT-D and the increase of the k-RAT alignment}
Following the RAT-D mechanism \citep{Hoang:2019da}, in vicinity of a protostar, grains can be spun-up to extremely fast rotation and disrupted into small fragments when such the resulting centrifugal stress exceeds the grain tensile strength, $S_{\max}$. Following \cite{Hoang.2021}, the minimum size of rotational disruption by the RAT-D mechanism is given by
\bea
a_{\rm disr}&=&\left(\frac{0.8n_{\rm H}\sqrt{2\pi m_{\rm H}kT_{\rm gas}}}{\gamma_{\rm rad} u_{\rm rad}\bar{\lambda}^{-2}}\right)^{1/2}\left(\frac{S_{\rm max}}{\rho}\right)^{1/4}(1+F_{\rm IR})^{1/2}\nonumber\\
&\simeq& 0.17 \left(\frac{\gamma_{\rm rad}U_{6}}{n_{8}T_{1}^{1/2}}\right)^{-1/2}\left(\frac{\bar{\lambda}}{1.2\mum}\right)\left(\frac{S_{\max,7}}{\hat{\rho}}\right)^{1/4}\nonumber\\
&&\times (1+F_{\rm IR})^{1/2}\mum,
\label{eq:adisr_R}
\ena
with $S_{\max,7}=S_{\max}/10^{7}\erg\cm^{-3}$, which implies that large grains can be disrupted in the region of densities of $n_{8}\lesssim 1$ near the protostar (see \citealt{Hoang.2021} for more details).

We now discuss the effect of selective disruption of RAT-D on the dust polarization observed toward a protostar. For an ensemble of grain shapes, a fraction $f_{\rm high-J}$ grains can have high-J attractors points. Thus, the fraction $f_{\rm high-J}$ grains on high-J attractors that have large sizes of $a>a_{\rm disr}$ can be disrupted into smaller ones by RAT-D. However, the fraction $1-f_{\rm high-J}$ grains that are aligned at low-J attractors do not experience RAT-D. Therefore, in the presence of RAT-D, large grains at $a>a_{\rm disr}$ now can only rotate thermally at low$-J$ attractors. This selective disruption by RAT-D decreases the degree of dust polarization \citep{2020ApJ...896...44L} and may change the polarization pattern. The former was observationally tested \citep{LeNgocTram:2021fw,Tram.2021,b6d}.

The potential change of the polarization pattern due to RAT-D arises from the fact that grains at low-J attractors tend to align via k-RAT instead of B-RAT. Indeed, if $a_{\rm disr}<a_{\min,Jk}^{\rm low-J}$, then, all grains above $a_{\rm disr}$ will experience k-RAT. For grains with slow internal relaxation of sizes $a>a_{\max,aJ}^{\rm low-J}\sim 0.1\mum$ for $n_{8}\sim 1$ (see Eq. \ref{eq:amax_aJ_iER}, $\St=1$), there may exist a fraction $f_{\rm right-IA}$ of grains with the right IA and $(1-f_{\rm right-IA})$ with the wrong IA. The dust polarization at longest wavelengths would be dominated by grains at low-J attractors because the population of large grains at high-J attractors is reduced by RAT-D. As a result, the polarization pattern now depends on $f_{\rm right-IA}$. One have the radial polarization pattern if the wrong IA dominates, and the azimuthal pattern if the right IA dominates. Detailed modeling with both RAT-A and RAT-D combined with multiwavelength dust polarization observations would be useful test the selective effect of RAT-D.

\section{Summary}\label{sec:summary}

Using the RAT (MET) alignment framework developed for interstellar grains, we study in detail alignment of very large grains (VLGs) in protostellar environments due to various physical effects. Our main findings are summarized as follows:

\begin{enumerate}

\item For grains aligned at low$-J$ attractors by RATs (METs), both super-Barnett and inelastic relaxation are inefficient for large grains in protostellar cores. The rate of internal relaxation is increased significantly for grains aligned at high$-J$ attractors that have suprathermal rotation. The increase of inelastic relaxation is higher than super-Barnett relaxation due to its faster increase with the angular momentum and slower variation with the grain size.

\item We derived analytical formulae for estimating the range of grain sizes that have the efficient IA by super-Barnett and inelastic relaxation effects for grains aligned at low$-J$ and high$-J$ attractors by RATs and METs. VLGs at high$-J$ attractors can have the efficient IA due to inelastic relaxation and super-Barnett relaxation.

\item For external alignment, we revisit the condition for $k-$RAT ($v-$MET) vs. $B-$ RAT ($B-$MET) by comparing the radiative/mechanical precession time at both low$-J$ and high$-J$ attractors with the Larmor precession. We find that large grains aligned at low$-J$ attractors can have the $k-$RAT/$v-$MET alignment, but large grains with large iron inclusions aligned at high$-J$ attractors have the $B-$RAT/$v-$MET alignment because of the suppression of radiative/mechanical precession at higher angular momentum. The parameter space for $B-$RAT ($B-$MET) is extended due to the decrease of radiative/mechanical precession rate at high$-J$ attractors, enabling a greater parameter space for tracing B-fields via dust polarization in protostellar cores.

\item We studied in detail internal and external alignment of large grains by RATs and METs in a protostellar disk. We find that RATs cannot induce fast internal relaxation, but METs can induce fast super-Barnett relaxation for grains of $a\sim 10-50\mum$ in the outer disk of $r>50\AU$. In the inner disk of $r<50\AU$, VLGs have the inefficient IA due to slow internal relaxation, which results in a low degree of dust polarization at sub/mm wavelengths and decreases toward the central protostar. 

\item The external alignment with the radiation field ($k-$RAT) occurs for VLGs aligned at low$-J$ attractors or grains without iron inclusions. Grains with large iron inclusions aligned at high$-J$ attractors experience the $B-$RAT alignment due to the decrease of radiation precession time with $J$ and increased Larmor precession, enhancing the parameter space for $B-$RAT and tracing the magnetic field via dust polarization.

\item METs are found to dominate over RATs in both internal alignment and external alignment in the disk mid-plane due to its stronger magnitudes. Grains with iron inclusions at high$-J$ attractors can be aligned with the magnetic field ($B-$MET) in the outer regions of the disk. The regions with magnetic alignment increase with the decreasing surface mass density of the disk. Therefore, depending on the disk mass, dust polarization can trace the magnetic field.

\item Thermal dust polarization from aligned grains is expected to change across the disk due to the combination of different alignment mechanisms at different radii. The magnetic alignment by RATs is expected be effective in the outer disk, whereas MET alignment is efficient in the inner disk.

\item The diverse polarization properties observed toward protostellar disks by ALMA may be explained by multiple alignment mechanisms. Detailed modeling of dust polarization based on the grain alignment physics presented in this paper would help to interpret observational data and disentangle different mechanisms and constrain dust properties in protostellar disks.

\item The polarization pattern produced by the $k-$ RAT alignment with the right IA is similarly azimuthal as those produced by $B-$RAT or $v-$MET alignment with the wrong IA. Multiwavelength polarimetric observations can disentangle this degeneracy.
 
\end{enumerate}

\acknowledgments
We thank the referee for a thorough reading and detailed comments that helped improve the presentation of this manuscript. We thank A. Lazarian for valuable comments on the manuscript. T.H is grateful to Bruce T. Draine for discussion on inelastic relaxation of large grains in protostellar disks, which motivated us to conduct this detailed study. T.H. is supported by the National Research Foundation of Korea (NRF) grant funded by the Korea government (MSIT (No. 2019R1A2C1087045). 

\appendix
\section{Review on inelastic relaxation}
\subsection{Acceleration within a rotating grain}\label{sec:apdvisc}

Let ${\bf X}$ be a vector connecting two atoms in the grain. The deformation can induce the change of ${\bf X}$ with time. Taking into account both the intrinsic displacement and rotation, the time variation ${\bf X}$ with respect to the lab frame can described by equation (see \citealt{2002clme.book.....G})
\bea
\left(\frac{d{\bf X}}{dt}\right)_{\rm LF}=\left(\frac{d{\bf X}}{dt}\right)_{\rm GF}+[\bOmega\times {\bf X}],\label{eq:dXdt_LF}
\ena
which returns to the rigid rotation when $(d{\bf x}/dt)_{\rm LF}=0$.

Acceleration of material within a rotating grain includes the intrinsic acceleration and inertial acceleration arising from the grain rotation. To find the total acceleration, we consider the grain angular velocity $\bOmega$. Let $\bu$ be the displacement vector in the body frame. The new position of the atom at the moment $t$ is then
\bea
\br(t) = \br_{0}(t)+\bu(t),
\ena
where $\bu=0$ and $\Rv=\br=\br_{0}$ for without deformation.

The position of the atom at the moment $t+dt$ is then
\bea
\br(t+dt)=\br_{0}(t+dt)+\bu(t+dt).
\ena
Note that even there is no further deformation during $dt$ (e.g., deformation ceases), the vector $\bu(t+dt)$ still changes in the lab frame due to rotation.

The total velocity of the atom in the lab frame is
\bea
\bv_{\rm LF}&=&\frac{d\br}{dt}=\frac{d\br_{0}}{dt}+\frac{d\bu}{dt}=[\bOmega\times \br_{0}]+(\dot{\bu})_{\rm GF}+[\bOmega\times \bu]\nonumber\\
&=&[\bOmega\times \br_{0}]+\bv_{\rm GF}+[\bOmega\times \bu]=\bv_{\rm GF}+[\bOmega\times \br],
\ena
where the first term is the angular velocity of the lattice atoms in the absence of deformation and the atom velocity changes just because of rotation, and $\bv_{\rm GF}=\dot{\bu}$ is the velocity of atoms measured in the GF, and the $du/dt$ is rewritten using Equation (\ref{eq:dXdt_LF}).

The acceleration of the atoms in the lab frame is then
\bea
\ba_{\rm LF}=\frac{d\bv_{\rm LF}}{dt}=\frac{d}{dt}\left([\bOmega\times \br_{0}]\right)+\frac{d}{dt}\bv_{\rm GF}+\frac{d}{dt}\left([\bOmega\times \bu]\right).
\ena

Using Eq. \ref{eq:dXdt_LF} for the time-derivative in the lab frame, one obtains
\bea
\frac{d}{dt}\bv_{\rm GF}&=&\ba_{\rm GF}+[\bOmega\times \bv_{\rm GF}],\\
\frac{d}{dt}\left([\bOmega\times \br_{0}]\right)&=&\dot{\bOmega}\times \br_{0}+[\bOmega\times \dot{\br}_{0}]\nonumber\\
&=&\dot{\bOmega}\times \br_{0}+[\bOmega\times [\bOmega\times \br_{0}],\\
\frac{d}{dt}\left([\bOmega\times \bu]\right)&=&\dot{\bOmega}\times \bu+[\bOmega\times \dot{\bu}]\nonumber\\
&=&\dot{\bOmega}\times \bu+[\bOmega\times [\bOmega\times \bu]+[\bOmega\times \bv_{\rm GF}],
\ena
where $\dot{\br}_{0}=\ddot{\br}_{0}=0$.

Combining the above equations, one then obtains
\bea
\ba_{\rm LF}=\ba_{\rm GF}+2[\bOmega\times \bv_{\rm GF}]+[\dot{\bOmega}\times \br]+[\bOmega\times [\bOmega\times \br]
\ena
where $\br=\br_{0}+\bu$.

As suggested by LE99, the acceleration of the atoms within the grain $a_{\rm GF}\sim (\delta l)^{2}/\tau$ with $\delta l$ the deformation of atoms within the grain, which can be disregarded due to small $\delta l$. The acceleration of the atom in the LF is then just:
\bea
\ba_{\rm LF}=\bOmega\times [\bOmega\times \br]+[\dot{\bOmega}\times \br]+2[\bOmega\times \bv_{\rm GF}],\label{eq:aLF2}
\ena
which implies that even an atom at rest in the GF still has acceleration with respect to the lab $\ba_{\rm LF}=\bOmega\times [\bOmega\times \br]$ due to rotation (i.e., the centripetal acceleration).

According to the equivalence principle, in the rotating frame, atoms will experience a centrifugal force which is opposite to the centripetal acceleration,
\bea
\bF_{\rm cen}=-m \ba_{\rm LF}=\bOmega\times [\br\times \bOmega\times]+[\br\times \dot{\bOmega}]+2[\bv_{\rm GF}\times \bOmega].
\ena 

Therefore, atoms move in the rotating frame can be considered to move relative to an inertial frame with an inertial a centrifugal force $\bF_{\rm cen}$. In other words, the rotating frame can be considered the inertial frame if the object is subject to the centrifugal force, thus all physical laws have the same form. Thus, Equation (\ref{eq:aLF2}) describes the acceleration of an atom within the rotating grain which is co-moving with the grain. If we take into account the gravity acting on the atom, then, the total acceleration is 
\bea
\ba_{\rm LF}=\bOmega\times [\bOmega\times \br]+[\dot{\bOmega}\times \br]+2[\bOmega\times \bv]+{\bf g},\label{eq:aLF3}
\ena
where the first term is due to the non-uniform rotation, the second term is the centrifugal acceleration, and the third term is the Coriolis acceleration exerting on the particle moving in a rotating system (see also \citealt{2018MNRAS.473..728F}). 

\subsection{Stress tensor and strain potential energy}

When the acceleration field in the grain is known, one can calculate the components of the stress tensor using the differential equations $f_{\rm int}=-f_{\rm ext}=\rho a_{i}=\partial \sigma_{ij}/\partial x_{j}$, which yields $\sigma_{ij}=-\int \rho a_{i}dx_{j}+C$, or
\bea
\sigma_{xy}(x,y)=\frac{1}{2}\left(\int_{x}^{a}\rho a_{x}dy+\int_{x}^{b}\rho a_{x}dx\right)+C
\ena
where $C$ is the integral constant determined by the boundary conditions (see \citealt{2001P&SS...49..937E} for a review). 

The stress causes the deformation of material structure within the grain. When the stress tensor is known, one can use the strain-stress relationship to find the strain $\sigma_{ij}=\sum_{kl}C_{ijkl}\epsilon_{kl}$ where $C_{ijkl}$ are the constants dependent on the grain's elasticity. The strain potential energy is calculated as (see LE99)
\bea
W=\frac{1}{2}\sum_{ik}\epsilon_{ij}\sigma_{jk},\label{eq:W}
\ena
which acts to bring the grain to the original state when the stress/external force is turned off.

Due to the time-variation of the centrifugal acceleration induced by the grain's precession, the deformation causes some dissipation of strain energy if the material is inelastic. The alternating stress caused by the precession of $\bOmega$ with $\ahat_{1}$ lags behind the grain material and induces the dissipation of the grain rotation energy into the heat, resulting in the IA of $\bOmega$ and $\bJ$ with $\ahat_{1}$ (\citealt{1973MNRAS.165..403B}). 

\subsection{Inelastic energy dissipation and relaxation time}
The energy loss by inelastic relaxation can occur on a time scale in the order of the precession period $\tau\sim 2\pi/\omega$, which can be written as $\tau=Q/\omega$, the larger $Q$ implies a slower dissipation and the material has better quality. So, $Q$ is the quality factor of the grain material. As a result, one can write the energy loss as
\bea
\frac{dE_{\rm loss}}{dt}= \frac{W}{\tau}=\frac{2\omega W}{Q},\label{eq:dEloss_dt}
\ena
where $W=W^{(\omega)}+2W^{(2\omega)}$ is the total strain energy associated with the principal frequency ($\omega$) and double frequency ($2\omega$) of oscillations, respectively \citep{Efroimsky:2000p5384}.
\newpage
\section{Notations}
\begin{center}
	\renewcommand{\arraystretch}{1.2}
	\label{tab:notation}
    \begin{longtable*}{|p{2cm}|p{3cm}|p{6cm}|p{2cm}|p{3cm}|}	    \caption{Glossary of notations and meanings}
 \\
 		\hline
		Category & Notation & Description & Typical Value (cgs units) & Normalized Notation \\ 
		\hline
        \hline
        \multirow{7}{2cm}{Physical constants} & $k$ & Boltzmann constant & & \\
        \cline{2-5}
        & $\hbar$ & Reduced Planck constant & & \\
        \cline{2-5}
        & $\mu_{B}=e\hbar/2m_{e}$ & Bohr magneton & & \\
        \cline{2-5}
        & $m_{\rm H}$ & Hydrogen atom mass & & \\
        \cline{2-5}
        & $m_e$ & Electron mass & & \\
        \cline{2-5}
        & $g_e$ & Electron spin $g$-factor & & \\
        \cline{2-5}
        & $\gamma_e=-g_e\mu_B/\hbar$ & The electron gyromagnetic ratio & & \\

        \hline
        \hline
        \multirow{9}{2cm}{Environment parameters} & $T_{\rm gas}$ & Gas temperature & & $T_{\rm gas,1}=T/(10\,{\rm K})$ \\
        \cline{2-5}
        & $n_{\rm H}$ & Hydrogen number density & $\gtrsim 10^6$ cm$^{-3}$ for protostellar cores & $n_8=n_{\rm H}/(10^8\,{\rm cm}^{-3})$ \\
        \cline{2-5}
        & ${\bf B}$ & Magnetic field & & \\
        \cline{2-5}
        & $B$ & Magnetic field strength & & {$B_3=B/(10^{3}\mu G)$} \\
        \cline{2-5}
        & $u_{\rm rad}$ & the energy density of the radiation field & & \\
        \cline{2-5}
        & $U=u_{\rm rad}/u_{\rm MMP83}$ & Ratio between the local radiation to that in the solar neighborhood $u_{\rm MMP83}$ & $10^6$ near the protostar & $U_6=U/10^6$ \\
        \cline{2-5}
        & $\gamma_{\rm rad}$ & The anisotropy degree of the radiation field & $0.3$ & \\
        \cline{2-5}
        & $\bar{\lambda}$ & The mean wavelength of the radiation field & $1.2\,\mum$ & \\

        \hline
        \hline
        \multirow{12}{2cm}{Grain mechanical parameters} & $a$ & Semi-major length of grains & $10^{-5}$ cm & $a_{-5}=a/(10^{-5}\,{\rm cm})$\\
        \cline{2-5}
        & $c$ & Semi-minor length of grains & & \\
        \cline{2-5}
        & $s=c/a$ & Ratio between semi-minor and semi-major axis of grains & $0.5$ & $\hat{s}=s/0.5$ \\
        \cline{2-5}
        & $a_{\rm eff}=s^{1/3}a$ & The effective grain size defined as the equivalent sphere of radius $a_{\eff}$ with the same volume as irregular grains & & \\
        \cline{2-5}
        & $\rho$ & Mass density of grain material & $3$ g~cm$^{-3}$ & $\hat{\rho}=\rho/(3$ g~cm$^{-3})$ \\
        \cline{2-5}
        & $I_\parallel$ (Eq. \ref{eq:Ipar}) & Principal moment of inertia for the rotation parallel to the grain symmetry axis & & \\
        \cline{2-5}
        & $I_\perp$ (Eq. \ref{eq:Iparperp}) & Principal moment of inertia for the rotation perpendicular to the grain symmetry axis & & \\
        \cline{2-5}
        & $h=2/(1+s^2)$ & Ratio between $I_\parallel$ and $I_\perp$ & & \\
        \cline{2-5}
        & $T_d$ & Grain temperature & & \\
        \cline{2-5}
        & $n$ & Number density of atoms within the dust material & $10^{23}$ cm$^{-3}$ & $n_{23}=n/10^{23}\,{\rm cm}^{-3}$ \\
        \cline{2-5}
        & ${\hat{\bf a}_1},{\hat{\bf a}_3},{\hat{\bf a}_3}$ & The unit vectors for the grain principal axes; ${\hat{\bf a}_1}$ denotes the axis of maximum inertia  & & \\
        \hline

        \pagebreak
        \caption{(Continued)} \\
        \hline
        Category & Notation & Description & Typical Value & Normalized Notation \\

        \hline
        \hline
        & ${\bf \Omega}$ & The angular velocity of grains & & \\
        \cline{2-5}
        & ${\bf J}$ & The angular momentum of grains & & \\
        \cline{2-5}
        & $J$ & The magnitude of the grain angular momentum & & \\
        \cline{2-5}
        & $J_1=J_\parallel\,{\rm and}\,J_2=J_3=J_\perp$ & Components of ${\bf J}$ projected onto the grain principal axes ${\hat{\bf a}_1},{\hat{\bf a}_3}$, and ${\hat{\bf a}_3}$ & & \\
        \cline{2-5}    
        & $\Omega_{0}=J/I_\parallel$ & The angular momentum defined by the ratio between $J$ and $I_\parallel$ & & \\
        \cline{2-5}
        & $v_{T}=\sqrt{2k T_{\rm gas}/m_{\rm H}}$ & The thermal speed of gas & & \\
        \cline{2-5}
        & $\Omega_{T}=\sqrt{kT_{\rm gas}/I_\parallel}$ & The grain thermal angular speed & & \\
        \cline{2-5}
        & $J_{T}=I_\parallel\Omega_T$ & The grain thermal angular momentum & & \\
        \cline{2-5}
        & ${\rm St}=J/J_T=\Omega_0/\Omega_T$ & The suprathermal rotation number & & \\
        \cline{2-5}
        & $v_d$ & The grain drift speed & & \\
        \cline{2-5}
        & $s_d=v_d/v_T$ & The grain drift parameter & & $s_{d,-1}=s_d/0.1$ \\
        \cline{2-5}
        & $\theta$ & The angle between ${\bf \hat{a}}_1$ and ${\bf J}$ & & \\
                \cline{2-5}
        & $Q_X$ & The degree of internal alignment between ${\bf \hat{a}}_1$ with ${\bf J}$ & & \\
        \cline{2-5}
        & $f_{\rm MW}(\theta)$ & The Maxwellian distribution for the angle $\theta$ established by thermal gas collisions & & \\
        \cline{2-5}
        & $Q_{X,\rm MW}$ & The degree of internal alignment assuming $\theta$ follows $f_{\rm MW}(\theta)$ & & \\

        \hline
        \hline
        \multirow{15}{2cm}{Parameters for paramagnetic and superparamagnetic grains} & $f_p$ & The fraction of paramagnetic (Fe) atoms in the dust grain & $1/7$ for silicate of MgFeSiO${_4}$ structure& \\
        \cline{2-5}
        & $n_p=f_p n$ & Number density of paramagnetic atoms & & \\
        \cline{2-5}
        & $p$ & Coefficient for the effective magnetic moment per paramagnetic atom & $5.5$ & $\hat{p}=p/5.5$ \\
        \cline{2-5}
        & $\mu_p=p\mu_B$ & The effective magnetic moment per paramagnetic atom & & \\
        \cline{2-5}
        & $\chi_p(0)$ (Eq. \ref{eq:curielaw} and \ref{eq:chi_para}) & The zero-frequency susceptibility of paramagnetic grains & & \\
        \cline{2-5}
        & $N_{\rm cl}$ & The number of iron atoms per cluster & $ 20- 10^5$ & $N_{\rm cl,4}=N_{\rm cl}/10^4$ \\
        \cline{2-5}
        & $\phi_{\rm sp}$ & The volume filling factor of iron clusters & $0.01-0.3$ & $\phi_{\rm sp,-2}=\phi_{\rm sp}/10^{-2}$ \\
        \cline{2-5}
        & $\chi_{\rm sp}(0)$ (Eq. \ref{eq:chi_sp}) & The zero-frequency superparamagnetic susceptibility & & \\
        \cline{2-5}
        & $\omega\simeq (h-1)J/(2I_\parallel)\simeq \Omega$ & The precession frequency of the angular momentum ${\bf J}$ around the axis of maximum inertia $\hat{\bf a}_1$ & & \\
        \cline{2-5}
        & $\nu_0$ & Characteristic frequency of thermal fluctuations of iron clusters & $10^9$ s & \\
        \hline

        \pagebreak
        \caption{(Continued)} \\
        \hline
        Category & Notation & Description & Typical Value & Normalized Notation \\

        \hline
        \hline
        & $T_{\rm act}$ & Characteristic temperature required for thermal remagnetization & $0.011$ K & \\
        \cline{2-5}
        & $\tau_{\rm sp}$ (Eq. \ref{eq:tau_sp}) & The timescale of thermally activated remagnetization & & \\
        \cline{2-5}
        & $\chi_2(\omega)$ & The imaginary part of the complex magnetic susceptibility & & \\
        \cline{2-5}
        & $K_{\rm sp}=\chi_2(\omega)/\omega$ (Eq \ref{eq:kappa_sp}) & The ratio between $\chi_2(\omega)$ and $\omega$ & & \\ 
        \cline{2-5}
        & $k_{\rm sp}=\K_{\rm sp}(\omega)\nu_0/\chi_{\rm sp}(0)$ (Eq \ref{eq:gsp}) & Describing the dependence of        
        $K_{\rm sp}(\omega)$ on dust temperature and frequency & & \\

        \hline
        \hline
        \multirow{4}{2cm}{Rotational damping} & $\Gamma_\parallel$ (Eq. \ref{eq:Gamma_par}) & The geometrical factor adopted in the rotational damping time due to gas collisions & $\Gamma_\parallel(s=1/2)\simeq 0.62$ & \\
        \cline{2-5}
        & $\tau_{\rm gas}$ (Eq. \ref{eq:tgas}) & The damping time due to gas collisions & & \\
        \cline{2-5}
        & $F_{\rm IR}$ (Eq. \ref{eq:FIR}) & Coefficient for the rotational damping due to IR emission & & \\
        \cline{2-5}
        & $\tau_{\rm IR}=\tau_{\rm gas}/F_{\rm IR}$ & The damping time due to IR emission & & \\

        \hline
        \multirow{2}{2cm}{External alignment via Larmor precession} & $\mathbf{\mu}_{\rm Bar}$ (Eq. \ref{eq:muBar}) & The grain magnetization obtained via the Barnett effect & & \\
        \cline{2-5}
        & $\tau_B$ (Eq. \ref{eq:tauB}) & The Larmor precession time of ${\bf J}$ around ${\bf B}$ & & \\

        \hline
        \hline
        \multirow{7}{2cm}{External alignment by RATs} & ${\bf k}$ & The radiation direction & & \\
                \cline{2-5}
        & $a_{\rm trans}\simeq \bar{\lambda}/2.5$ & The transition size at which the average RAT efficiency changes the slope & & \\        
        \cline{2-5}
        & $\Omega_{\rm RAT}$ (Eq. \ref{eq:omega_RAT1} and \ref{eq:omega_RAT2}) & The maximum angular velocity of grains spun-up by RATs & & \\
        \cline{2-5}
        & ${\rm St}_{\rm RAT}=\Omega_{\rm RAT}/\Omega_T$ (Eq. \ref{eq:S_RAT1} and \ref{eq:S_RAT2}) & The suprathermal rotation number for grains spun-up by RATs & & \\
        \cline{2-5}
        & $Q_{e3}$ & The third component of RATs that induces the precession of ${\bf J}$ around ${\bf k}$ & $10^{-2}$ & $\hat{Q}_{e3}=Q_{e3}/10^{-2}$ \\
        \cline{2-5}
        & $f_{\rm high-J}$ & The fraction of grains aligned at high-J attractors & & \\
        \cline{2-5}
        & $\tau_k$ (Eq. \ref{eq:tauk}) & The precession time of ${\bf J}$ around ${\bf k}$ & & \\
        \cline{2-5}
        & $\tau^{\rm low-J}_k=\tau_k(\Omega=\Omega_T)$ (Eq. \ref{eq:tauk_lowJ}) & The precession time of ${\bf J}$ around ${\bf k}$ at low-J attractors & & \\
        \cline{2-5}
        & $\tau^{\rm high-J}_k=\tau_k(\Omega=\Omega_{\rm RAT})$ (Eq. \ref{eq:tauk_highJ}) & The precession time of ${\bf J}$ around ${\bf k}$ at high-J attractors & & \\

        \hline
        \hline
        \multirow{4}{2cm}{External alignment by METs} & $Q_{\rm spinup}$ & The spin-up efficiency of METs & $10^{-6}$ - $10^{-3}$ & \\
        \cline{2-5}
        & $\Gamma_{\rm MET}$ (Eq. \ref{eq:Gamma_MET}) & The magnitude of METs induced by the grain drift through ISM gas & & \\
        \cline{2-5}
        & $\Omega_{\rm MET}$ (Eq. \ref{eq:omega_MET}) & The maximum angular rotation velocity of grains spun-up METs & &\\
        \hline  
        \pagebreak
        \caption{(Continued)} \\
        \hline
        Category & Notation & Description & Typical Value & Normalized Notation \\

        \hline
        \hline
        & ${\rm St}_{\rm MET}=\Omega_{\rm MET}/\Omega_{T}$ (Eq. \ref{eq:S_MET}) & The suprathermal rotation number for grains spun-up by METs & & \\
        \cline{2-5}
        & $Q_{\rm prec}$ & The coefficient to describe the MET efficiency to induce the grain precession around the gas flow & 0.1 & $Q_{\rm prec,-1}=Q_{\rm prec}/10^{-1}$ \\
        \cline{2-5}
        & $\Gamma_{\rm MET,prec}$ (Eq. \ref{eq:MET_prec}) & The component of METs inducing the precession of ${\bf J}$ around the drift direction & & \\
        \cline{2-5}
        & $\Omega_{\rm MET,prec}$ (Eq. \ref{eq:omega_MET_prec}) & The precession frequency of ${\bf J}$ around the drift direction & & \\
        \cline{2-5}
        & $\tau_v$ (Eq. \ref{eq:tau_v}) & The precession time of ${\bf J}$ around the drift direction & & \\

		\hline
		\hline
		\multirow{3}{2cm}{External alignment via magnetic relaxation} & $\tau_{\rm mag,sp}$ (Eq. \ref{eq:tau_DG}) & The superparamagnetic relaxation time & & \\
		\cline{2-5}
		& $\delta_{\rm mag,sp}$ (Eq. \ref{eq:delta_m}) & Ratio between the gas damping time and the superparamagnetic relaxation time & & \\
		\cline{2-5}
		& $a^{\rm mag}_{\rm max,JB}$ (Eq. \ref{eq:amax_JB_DG}) & The maximum size for which superparamagnetic relaxation is important for external alignment given by $\delta_{\rm mag,sp}=1$ & & \\

        \hline
        \hline
        \multirow{4}{2cm}{Super-Barnett relaxation for internal alignment} & $J_d=\sqrt{I_\parallel kT_d/(h-1)}$ & The characteristic thermal angular momentum & & \\
        \cline{2-5}
        & $f(\hat{s})=\hat{s}\left[(1+\hat{s}^2)/2\right]^2$ & The geometric factor adopted in the super-Barnett relaxation time & & \\
        \cline{2-5}
        & $\tau_{\rm BR,sp}$ & The Barnett relaxation time for superparamagnetic grains & & \\

        \hline
        \hline
        \multirow{5}{2cm}{Inelastic relaxation for internal alignment} & $\sigma$ & The Poisson ratio taken from \citet{Molina.2003} & $0.5$ & \\
        \cline{2-5}        
        & $\mu$ & The shear modulus & $10^8$ erg/cm$^{3}$ & $\mu_8=\mu/(10^8\,{\rm erg/cm}^{3})$ \\ 
        \cline{2-5}
        & $Q$ & The quality factor of grain material & $100$ (for silicate rocks) & $Q_3=Q/10^3$ \\
        \cline{2-5}
        & $g'(s)=2.2s^{3/2}g(s)$ (see Eq. \ref{eq:gs} for $g(s)$) & The geometrical factor adopted in the inelastic relaxation time & & \\
        \cline{2-5}
        & $\tau_{\rm iER}$ & The inelastic relaxation time & & \\

        \hline
        \hline
        \multirow{4}{2cm}{Critical sizes and drift parameters for internal alignment} & $a_{\rm max,aJ}({\rm BR})$ (Eq. \ref{eq:amax_aJ_BR}) & The maximum grain size for efficient internal alignment by Barnett relaxation (BR) given by $\tau_{\rm BR,sp}=\tau_{\rm gas}$ & & \\
        \cline{2-5}
        & $a_{\rm max,aJ}({\rm iER})$ (Eq. \ref{eq:amax_aJ_iER}) & The critical grain size for efficient internal alignment by inelastic relaxation given by $\tau_{\rm iER}=\tau_{\rm gas}$ & & \\
        \cline{2-5}
        & $a_{\rm min,aJ}^{\rm RAT,high-J}({\rm iER})$ (Eq. \ref{eq:amin_aJ_RAT}) & The minimum grain size for efficient internal alignment by inelastic relaxation given by $\tau_{\rm iER}=\tau_{\rm gas}$ and assuming $\Omega=\Omega_{\rm RAT}$ & & \\
        \hline

        \pagebreak
        \caption{(Continued)} \\
        \hline
        Category & Notation & Description & Typical Value & Normalized Notation \\ 
        
        \hline
        \hline
        & $a_{\rm max,aJ}^{\rm RAT,high-J}({\rm iER})$ (Eq. \ref{eq:amax_aJ_RAT}) & The maximum grain size for efficient internal alignment by inelastic relaxation for grains aligned at high-J given by $\tau_{\rm iER}=\tau_{\rm gas}$ & & \\
        \cline{2-5}
        & $s_{\rm cri,aJ}(\rm iER)$ & The critical drift parameter required for internal alignment by inelastic relaxation given by $\tau_{\rm iER}^{\rm MET,high-J}=\tau_{\rm gas}$ & & \\

        \hline
        \hline
        \multirow{4}{2cm}{Critical sizes for external alignment} & $a^{\rm Lar}_{\rm max,JB}$ (Eq. \ref{eq:amax_JB}) & The maximum size for the grain alignment of ${\bf J}$ with ${\bf B}$ constrained by Larmor precession & & \\
        \cline{2-5}
        & $a_{\rm min,Jk}^{\rm RAT,low-J}$ (Eq. \ref{eq:amin_Jk_lowJ}) & The minimum size for the $k-$RAT alignment at the low$-J$ attractor given by $\tau^{\rm low-J}_k=\tau_k(\Omega=\Omega_{T})=\tau_B$ & & \\
        \cline{2-5}
        & $a_{\rm min,Jk}^{\rm RAT,high-J}$ (Eq. \ref{eq:amin_Jk_highJ}) & The minimum size for the $k-$RAT alignment at the high$-J$ attractor given by $\tau^{\rm high-J}_k=\tau_k(\Omega=\Omega_{\rm RAT})=\tau_B$ & & \\
        \cline{2-5}
        & $a_{\rm min,Jv}^{\rm MET}$ (Eq. \ref{eq:amin_Jv_MET}) & The minimum size for the $v-$MET alignment given by $\tau_v=\tau_{\rm gas}$ & & \\

        \hline
        \hline
        \multirow{2}{2cm}{Critical sizes for external alignment}& $a_{\rm min,Jv}^{\rm MET,low-J}$ (Eq. \ref{eq:amin_Jv_lowJ}) & The minimum size for the $v-$MET alignment at the low$-J$ attractor given by $\tau^{\rm low-J}_v=\tau_v(\Omega=\Omega_T)=\tau_{\rm gas}$ & & \\ 
        \cline{2-5}
        & $a_{\rm min,Jv}^{\rm MET,high-J}$ (Eq. \ref{eq:amin_Jv_highJ}) & The minimum size for the $v-$MET alignment at the high$-J$ attractor given by $\tau^{\rm high-J}_v=\tau_v(\Omega=\Omega_{\rm RAT})=\tau_{\rm gas}$ & & \\

        \hline
        \hline
        \multirow{2}{2cm}{Critical sizes for grain alignment by RATs and METs} & $a^{\rm RAT}_{\rm align}$ (Eq. \ref{eq:aali_RAT}) & The minimum size for grain alignment by RATs given by $\Omega_{\rm RAT}=3\Omega_T$ (or ${\rm St}_{\rm RAT}=3$) & & \\
        \cline{2-5}
        & $a^{\rm MET}_{\rm align}$ (Eq. \ref{eq:aali_MET}) & The minimum size for grain alignment by METs given by $\Omega_{\rm MET}=3\Omega_T$ (or ${\rm St}_{\rm MET}=3$) & & \\

        \hline
        \hline
        \multirow{11}{2cm}{Protostellar disk parameters} & $r$ & Radial distance from the central star & & \\
        \cline{2-5}
        & $\Sigma_0$ & Surface mass density in the mid-plane of the disk at $r=1$ au & $100$ - $1000$ g cm$^{-2}$ & \\
        \cline{2-5}
        & $\Sigma(r)$ (Eq. \ref{eq:sigmaR}) & Surface mass density of the disk as a function of $r$ & & \\
        \cline{2-5}
        & $H_p$ (Eq. \ref{eq:H_R}) & The pressure-scale height for the profile of gas perpendicular to the mid-plane of the disk & & \\
        \cline{2-5}
        & $n_0=4.0\times 10^{13}(\Sigma_0/10^3\,{\rm g\,cm^{-2}})\,{\rm cm^{-3}}$ & The gas density in the mid-plane of the disk at $r=1$ au & & $n_{0,13}=n_0/(10^{13}\,{\rm cm}^{-3})$ \\
        \cline{2-5}
        & $\alpha_n$ & The index for the gas density profile as a function of $r$ & 37/14 & \\
        \cline{2-5}
        & $n_{\rm H}(r)=n_0(r/{\rm au})^{-\alpha_n}$ (Eq. \ref{eq:nH}) & The gas density in the mid-plane of the disk as a function of $r$ & & \\
        \cline{2-5}
        & $T_0$ & The gas temperature in the mid-plane of the disk at $r=1$ au & & $T_{0,2}=T_0/(10^2\,{\rm K})$ \\
        \hline

        \pagebreak
        \caption{(Continued)} \\
        \hline
        Category & Notation & Description & Typical Value & Normalized Notation \\

        \hline
        \hline
        \cline{2-5}
        & $\beta$ & The index for the temperature profile as a function of $r$ & 3/7 & \\
        \cline{2-5}
        & $T_{\rm gas}(r)=T_0(r/{\rm au})^{-\beta}$ (Eq. \ref{eq:Tgas_disk_R}) & The gas temperature in the mid-plane of the disk as a function of $r$ & & \\
        \cline{2-5}
        & $\dot{M}$ & The mass accretion rate of the central star & $10^{-8}\,M_\odot\,{\rm yr}^{-1}$ & \\
        \cline{2-5}
        & $B_0$ & The magnetic field strength in the mid-plane of the disk at $r=1$ au normalized for $\dot{M}=10^{-8}\,M_\odot\,{\rm yr}^{-1}$ & $10^6\,\mu$G & $B_{0,6}=B_0/(10^6\,\mu{\rm G})$\\
        \cline{2-5}
        & $B(r)$ (Eq. \ref{eq:BR}) & The magnetic field strength in the mid-plane of the disk as a function of $r$ & & \\
        \cline{2-5}
        & $M_*$ & Mass of the central protostar & & \\
        \cline{2-5}
        & $\Omega_K$ & The Keplerian angular velocity & & \\
        \cline{2-5}
        & $v_K$ (Eq. \ref{eq:vK}) & The Keplerian velocity & & \\
        \cline{2-5}
        & ${\rm Stk}$ & The Stokes number & & \\
        \cline{2-5}
        & $v_{\rm gas}$ & The gas velocity & & \\
        \cline{2-5}
        & $\eta=v_K/v_{\rm gas}$ & Ratio between $v_{\rm gas}$ and $v_{K}$ & & \\
        \cline{2-5}
        & $v_r$ (Eq. \ref{eq:v_r_disk}) & The radial component of the gas velocity & & \\
        \cline{2-5}
        & $v_\phi$ (Eq. \ref{eq:v_phi_disk}) & The radial component of the gas velocity & & \\
        \cline{2-5}
        & $v_d=(v_r^2v_\phi^2)^{1/2}$ & The dust-gas relative velocity & & \\
        \cline{2-5}
        & $s_d(r)=v_d/v_T$ & The grain drift parameter in the mid-plane of the disk as a function of $r$ & & \\
        \cline{2-5}
        & $s_{d0}$ & The grain drift parameter at $r=1$ au & & $s_{d0,-1}=s_{d0}/0.1$ \\

        \hline
	\end{longtable*}
\end{center}


\bibliography{ms.bbl}

\end{document}